\documentclass[11pt,a4paper]{article}
\usepackage[utf8]{inputenc}
\usepackage[spanish,english]{babel}
\usepackage{amsmath}
\usepackage{amsfonts}
\usepackage{amssymb}
\usepackage{graphicx}
\usepackage{fourier}
\usepackage{color}
\usepackage{caption}
\usepackage{subcaption}
\usepackage{authblk}

\numberwithin{equation}{section}

\usepackage[left=2cm,right=2cm,top=2cm,bottom=2cm]{geometry}

\begin{document}

\title{Thermalization of Holographic Excited States}

\author[a]{Pedro Jorge Martínez\footnote{pedro.martinez@cab.cnea.gov.ar }}
\author[b]{Guillermo A Silva\footnote{silva@fisica.unlp.edu.ar }}
\affil[a]{\it Instituto Balseiro, Centro At\'omico Bariloche 

8400-S.C. de Bariloche, R\'io Negro, Argentina\vspace{2mm}}

\affil[b]{\it Instituto de F\'isica La Plata - CONICET and 

Departamento de F\'isica,
Universidad Nacional de La Plata 

C.C. 67, 1900, La Plata, Argentina}

\maketitle
\thispagestyle{empty}
\begin{abstract}  
We propose a real time holographic framework to study thermalization processes of a family of QFT excited states. The construction builds on Skenderis-van Rees's holographic duals to QFT Schwinger-Keldysh complex-time ordered paths. Thermalization is explored choosing  a set of observables $F_n$ which essentially isolate the excited state contribution. Focusing on theories defined on compact manifolds and with excited states defined in terms of Euclidean path integrals, we identify boundary conditions that allow to avoid any number of modes in the initial field state. In the large conformal dimensions regime, we give precise prescriptions on how to compute the observables in terms of bulk geodesics.  
\end{abstract}

\vspace{12cm}

\pagebreak

\tableofcontents

\section{Introduction}

The first concrete formulation of the holographic correspondence was made in Euclidean signature. This realization proposed the identification of the AdS gravitational partition function with the QFT generating functional. External CFT sources were equated with the asymptotic boundary conditions for bulk fields  \cite{GKP,W}. By using the asymptotic sources as auxiliary tools and considering different topologies, the framework allowed the computation of vacuum and thermal $n$-point correlators \cite{GKP,W,WT}. It soon became clear that keeping non-zero sources at the AdS boundary corresponded to  CFT deformations,  generically triggering RG flows \cite{PW}. However, intrinsic real-time phenomenology was out of reach. In particular, a strong interest in the physics of strongly coupled  quark-gluon plasma \cite{Son:2007vk}, alongside the general quest for a holographic description of QFT hydrodynamics \cite{Glorioso:2018mmw,deBoer:2018qqm} as well as to finding a QFT perspective of black hole interior physics \cite{Shenker02,Shenker03}, revealed the necessity of a real-time formulation of the holographic dictionary. 

From the outset, a Lorentzian formulation of AdS/CFT requires to deal with: (i) the correct prescription for determining   time-ordering in the correlators (Feynmann, Causal, etc.) as well as, (ii) imposing initial/final conditions in time. In Euclidean signature, these issues were absent since only the asymptotic AdS boundary shows up, a manifestation of the uniqueness of the  Euclidean correlator. Important efforts in formalizing the real time scenario \cite{bala,bala2,Mar}  identified timelike (asymptotic) and spacelike (initial/final times) boundaries. The latter of these are tricky to interpret in the holographic setup if we adopt the philosophy of describing everything from the asymptotic boundary. Further generalizations allowed to compute retarded Green functions for thermal systems \cite{SS,HS}. Here, non-trivial chemical potentials on the CFT translated into non-zero bulk gauge fields profiles at the AdS boundary \cite{Hart}. Again, asymptotic boundary conditions on the timelike boundary were used either as auxiliary tools to compute correlation functions \cite{Hart} or as deformations of the CFT \cite{PW}. 
Despite some Euclidean computations 
being successfully carried over to real-time via analytic continuation, applications were fairly restrictive and usually required physical input to obtain the correct result \cite{vaman}.
Moreover, this method did not conceptually explained how initial/final conditions and causality issues were encoded in the holographic map.

A full systematic approach to attack real-time problems addressing the above issues was developed by Skenderis and van Rees (SvR) in a series of works \cite{SvRC,SvRL}. Following original ideas of Schwinger, Keldysh, Hartle and Hawking, SvR proposed to describe Schwinger-Keldysh complex $t$-contours in QFT in terms of glued AdS geometries of mixed signature (see \cite{WitM} for a recent review). For example, the Euclidean AdS-prescription for the standard solid ball was viewed as dual to an ordered straight vertical path (pure imaginary) in the QFT complex $t$-plane. Real-time physics was then obtained by deforming the initial vertical contour to the real axis. General curves in the complex $t$-plane becomes dual to several AdS geometries glued together. In what follows we will collectively refer to the complex  $t$-contours as Schwinger-Keldysh (SK) paths, we will denote them by $\cal C$.
The main advantage of the framework is that the ordering along the SK-contour fixes automatically the correct analytic extension of all real-time correlators  requiring no further input. Although some years have passed since its formulation, the potential of the SvR viewpoint has not been fully explored yet. In this work, we aim to make a step forward in this direction by studying thermalization processes.

In SK formalism, initial/final QFT wavefunctions are described in terms of Euclidean evolution (pure imaginary time segments) with appropriate operator insertions along it. A specific wavefunction arises in standard fashion as a cut open Euclidean path integral. SvR formalism suggests that the operator insertions generating the QFT excited state become dual to asymptotic boundary conditions on the (Euclidean) AdS boundary. This proposal is in line with Hartle-Hawking's wavefunction  \cite{HH}. Since Lorentzian holography permits different kinds of states, a natural question in this context is the characterization of the states generated this way. On the CFT side the operator-state correspondence provides a simple answer in terms of conformal primaries and descendants of the radially quantized theory. On the gravity side, the character of the  states both at zero and finite temperature was elucidated in a series of works \cite{us1,CS,us2,us3,us4,us5,Rabideau}. It was found they behave as holographic coherent states. This means that, in the large $N$ limit, when the CFT becomes generalized free, the dual state is guaranteed to have a geometric interpretation and moreover becomes a coherent state of the bulk field $\Phi$ dual to the inserted $\cal O$ in the CFT. These states have been shown to be an overcomplete basis for the perturvative bulk Hilbert space, often referred to as code subspace \cite{Harlow:2016vwg}. 

Initial studies of these states \cite{Rabideau,MarkChino} revealed that, via a limiting process, one could seemingly create an initial state as localized in the bulk as desired. However, it was noted recently \cite{Belin20} that not all sets of initial data (in particular non-analytic profiles) can be reached via asymptotic sources using Euclidean path integrals. More precisely, it was shown that the problem of finding asymptotic sources for a given general set of initial data is itself ill-posed. A second goal of this work is to explore a related question: is there a precise formula for asymptotic boundary conditions to the bulk path integral such that a single normal (or quasi-normal) state is given as an initial condition? In this context we find an interesting and reassuring answer: by a limiting process, we will build asymptotic sources that avoid any number of normal or quasi-normal modes on the initial state\footnote{Notice that these are necessarily analytic configurations on the initial time slice, so our result is not in conflict with \cite{Belin20}.}. In this fashion, one can build an asymptotic source that avoids all but a single mode. This seems to contradict the common lore stating that high energy eigenstates should not have a simple geometric dual. Tension is resolved by noticing that the source required to obtain a single-mode is actually an infinite superposition of geometric states. 
In a sense, this is the reverse of the celebrated interpretation of the (geometric) BTZ state as a series of (non-geometric) energy eigenstates \cite{VanRaamessay}. 

In this work we will define and perturbatively study a family of observables $F_n$ that are sensitive to the system's response to excited states. These will be described in terms of insertions in mixed signature manifolds dual to SK-contours. In simple scenarios, one may have access to the exact correlators for which analytic continuations (as prescribed by the SK contour) can be made. However, generically, one can only access a geodesic approximation of the correlator. The framework will thus make use of  ``complex geodesics'' in the bulk geometries. 
Our computations share similarities to those in \cite{Balasubramanian:2012tu,Anous:2016kss,Malda20}.

The paper is organized as follows: in Sec. \ref{Sec2} we present all the ingredients to setup the work. Then we consider case studies of increasing complexity in order to show the plethora of possible applications. In Sec. \ref{Sec:AdS} we start tackling pure AdS. i.e. zero temperature. In Sec \ref{BTZ}, in view of its analytic tractability, we consider the BTZ geometry, i.e. a thermal system. Finally, in Sec. \ref{Sec:BH5} we consider an AdS$_5$ BH and contrast our results to those in the BTZ scenario. We conclude in Sec. \ref{Sec:Conc} with a discussion on future directions. We leave to the appendices some technical discussion on the nature of the complex geodesics we consider.

\section{Framework}
\label{Sec2}

In this section we present the main elements of the framework. First, we review the SvR prescription for real-time holography. Second, we review a family of excited states that have a simple holographic dual description and discuss some of its properties. Then, we explicitly build a set of asymptotic sources that generate wavefunctions that lack any arbitrary number normal/quasi-normal modes in its decomposition. Finally, we define a family $F_n$ of observables that isolate the excited state contribution.
Concrete applications are described in the following sections.

\subsection{From SK paths to geometries: SvR prescription}
\label{Sec:SvR}

In \cite{SvRC,SvRL}, Skenderis and van Rees (SvR) developed  a prescription for real-time holography aimed at finding holographic duals to Schwinger-Keldysh (SK) QFT contours. The use of complex-time contours to represent physical systems in QFT is standard and well known to provide correct real-time results \cite{Kamenev,Kamenev2,Kamenev3}. On the gravity side, one should in principle perform the gravitational path integral with complex boundary conditions. This is typically a difficult task and in general out of reach. In practice, however, the SK-path can be often split into several segments of definite signature, either pure imaginary or pure real. Then, candidate saddles can be naturally assigned to each of the SK-segments. These associated saddles possess space-like boundaries (besides the asymptotic ones, which are in one to one correspondence with the SK path segments) that must be then glued together demanding appropriate continuity conditions, i.e. complexified Israel juncture conditions \cite{SvRL,IsraelJ}.
The resulting mixed signature manifold ends up possessing only asymptotic boundaries and serves as a candidate dual to the CFT system\footnote{One may find many bulk duals to a single SK-path. We expect this fact to be interpreted as realizing  phase transitions in the system \emph{a la} Hawking-Page \cite{HP,WT} transition. The problem of finding non-trivial bulk duals to a particular SK-path has not been thoroughly explored yet. We will not pursue this avenue in this work.}. Correlators and observables obtained from these geometries are unambiguous and completely fixed in terms of the asymptotic boundary conditions on the SK path. 

A few comments are in order: 
\begin{itemize}
\item The final manifold may posses several real-time sections. Their physical interpretation stems from the QFT SK-path. As an example, standard (In-In) SK-paths contain a pair of Lorentzian segments moving in opposite directions, often interpreted as the DOF's of the system and bath \cite{Umezawa}. Hence, real time correlators are computed from generating functionals with non-zero asymptotic sources in the real time segments.

\item The bulk manifold will also have a number of asymptotic Euclidean boundaries where sources could  also be turned on. The physical interpretation of non-trivial sources in the  Euclidean sections was elucidated in \cite{us1,CS,Rabideau,us4} as preparing a family of \emph{holographic} excited states whose properties will be described in the next subsection. 

\item  This is a technical comment. As is well known, single Lorentzian sections admit normalizable modes (N-modes) which are not fixed by their corresponding asymptotic  boundary conditions. The SvR-framework  fixes them through the gluing conditions between regions. The outcome is that N-modes coefficients end up depending generically on all the asymptotic sources prescribed in the problem, i.e.: (i) sources on the same region, (ii) on other Lorentzian regions, or (iii) on Euclidean regions. Physically we interpret the dependence on (i) as fixing  the correlator for the theory (i.e. retarded, Feynman, etc), on (ii) as entanglement between different Lorentzian sections and on (iii) as encoding effects of the excited states.

\end{itemize}

\begin{figure}[t]\centering
\begin{subfigure}{0.49\textwidth}\centering
\includegraphics[width=.9\linewidth] {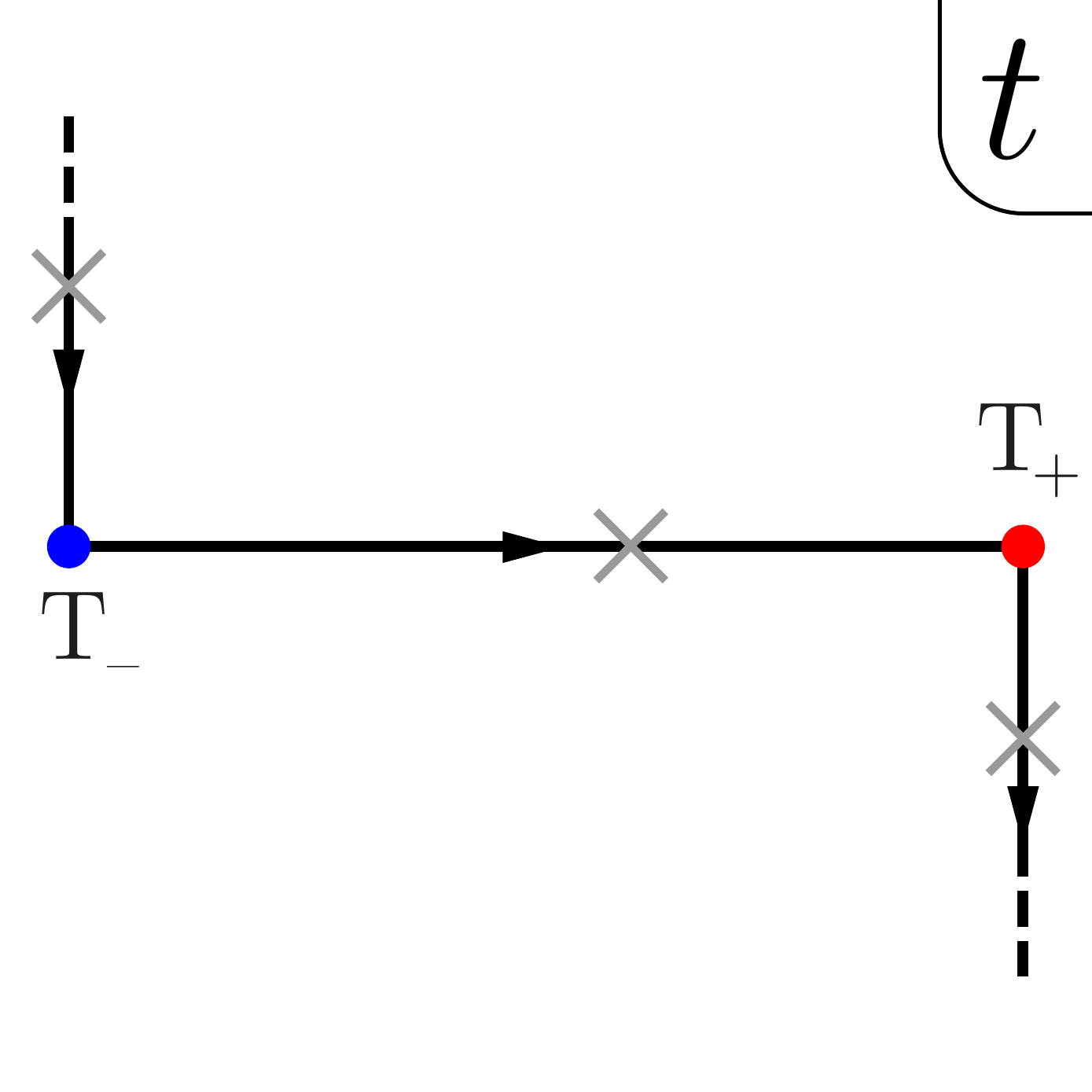}
\caption{}
\end{subfigure}
\begin{subfigure}{0.49\textwidth}\centering
\includegraphics[width=.9\linewidth] {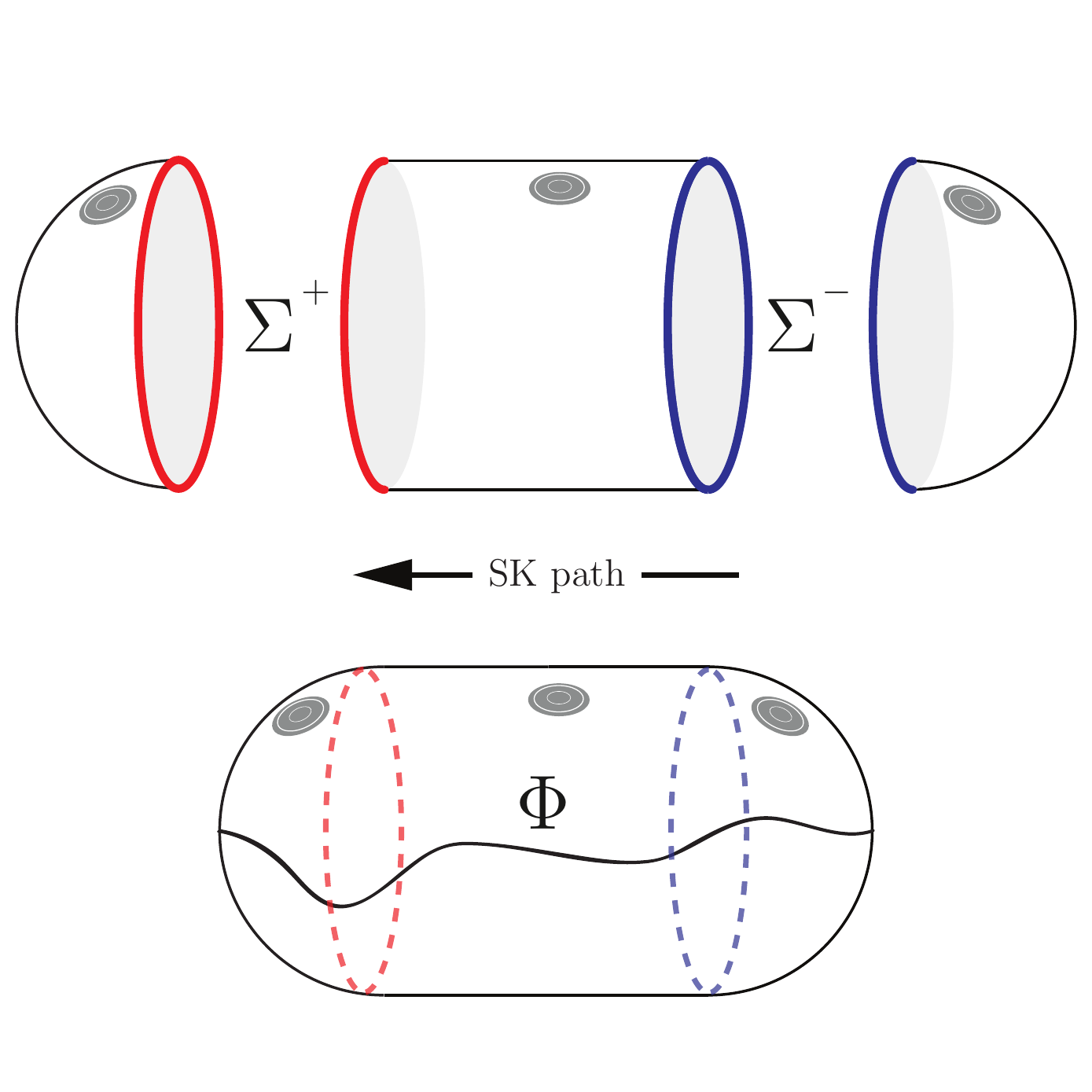}
\caption{}
\end{subfigure}
\caption{(a) In-Out SK-contour in the complex $t$-plane appropriate for describing  a QFT transition amplitude. The  real-time evolution  $\Delta T=T_+-T_-$ is taken to infinity when computing  scattering processes. Grey crosses represent operator insertions. (b) Bulk dual of the In-Out SK path depicted in (a). SvR dictionary associates either half Euclidean AdS spheres or Lorentzian  AdS cylinders to each of the segments in (a). The pieces are ${ C}_1$ glued across $\Sigma^\pm$. The resulting manifold provides the unique classical solution to the bulk fields eom's with prescribed asymptotic boundary conditions represented by grey lumps.}
\label{Fig:In-Out}
\end{figure}

We conclude the SvR-framework review with an application to a simple holographic scattering problem. The  SK-path associated to a traditional QFT scattering process is depicted in Fig. \ref{Fig:In-Out}(a). Initial and final vacuum states are prepared by the vertical (Euclidean) segments at real time coordinates $T_{\mp}$.  
The physical process occurs as we move in the horizontal segment, and real-time $n$-point functions can be computed by inserting, in the QFT path integral, auxiliary sources along this segment. Turning on external sources on the vertical segments is associated to  excitations over the bra/ket vacuum. The bulk dual to a QFT scattering process is represented in Fig. \ref{Fig:In-Out}(b):  Euclidean half sphere sections and Lorentzian AdS cylinders are assigned to each segment and ${C}^1$-glued across $\Sigma^{\pm}$. The final mixed signature manifold is understood as a saddle of the gravitational path integral. On this background, a classical bulk field configuration $\Phi$ can be fully determined in terms of prescribed asymptotic boundary conditions $\phi$. In GKPW spirit, we summarize the relation between dual theories as
\begin{equation*}
Z^{CFT_d}_{0\to 0}[\phi]=\langle 0|e^{-i\int {\cal O}\phi}|0\rangle \equiv  \left(\int_{T_-+i \infty}^{T_-}{\cal D}\Phi \; e^{-I}\right) \left(\int_{T_-}^{T_+}{\cal D}\Phi_{\phi} \;e^{- i I}\right) \left(\int_{T_+}^{T_+-i\infty}{\cal D}\Phi \;e^{-I}\right) =Z^{AdS_{d+1}}_{0\to 0}[\Phi|_{\partial}=\phi]
\end{equation*}
Here $I=I[\Phi]$ are the corresponding bulk actions for each section. We have considered vanishing sources in the Euclidean sections, appropriate for a vacuum to vacuum process, and  denoted by $\cal O$ the QFT operator dual to $\Phi$.  The product of path integrals manifests that the recipe for building bulk duals to SK-paths involves a piece-wise holographic dictionary. 

\subsection{Holographic excited states}
\label{Sec:HES}

In this subsection we elaborate on the nature and properties of holographic excited states \cite{us1,CS,us2,us3,us4,us5,Rabideau}. 
Their bulk state wavefunction is obtained in a Hartle-Hawking fashion as an Euclidean path integral with non-trivial asymptotic boundary conditions for  bulk fields. As we show below, they  have also a precise definition on the CFT side. 
In the semi-classical limit, they posses coherent state properties \cite{us1} so, in particular, they can be used as a (overcomplete) basis for generating $n$-particle states.

In path integral language, excited states $|\phi\rangle$ on each side of the dual pair read
\begin{equation}
\label{exc-state}
\text{\sf CFT:}~~~\langle A(\Sigma)|\phi\rangle\equiv  \int^{A(\Sigma)}{\cal D} A \; e^{-I_{CFT}-\int \phi\; {\cal O}} \qquad\Leftrightarrow\qquad \langle \Phi(\Sigma)|\phi\rangle=\int^{\Phi(\Sigma)} [{\cal D}\Phi]_{\phi} \; e^{-I}
~~~:\,\text{\sf AdS}\end{equation}
Here $A$ denotes CFT fields and $\Phi$ bulk fields dual to CFT primaries ${\cal O}$. We have denoted $A(\Sigma)$ and $\Phi(\Sigma)$ to the field configurations on (boundary and bulk) codim-1 hypersurfaces $\Sigma$ necessary for cut open Euclidean path integrals computed on the vertical segments $t\in(T_-+i\infty,T_-)$ depicted in Fig. \ref{Fig:In-Out}. As usual, external CFT sources $\phi$ translate into boundary conditions for bulk fields $\Phi$ under the holographic map. As mentioned above, excited states are built over fixed  geometric backgrounds, hence, under appropriate circumstances  they will also have a classical profile.

Alternatively, one can write in operatorial CFT language
\begin{equation}
    |\phi\rangle={\cal P}\{ e^{-\int {\cal O}\phi} \}|0\rangle \sim \left(1-\int {\cal O}\phi+\frac 12 \int {\cal P} \{{\cal O} {\cal O}\}\;\phi \;\phi+\dots\right)|0\rangle\;,
\label{CFTex}    
\end{equation}
where the integrals are taken on the vertical segments $t \in (T_- + i\infty , T_- )$. 
In principle, excitations of this sort can be built over any state of the theory. However, in the holographic set-up one is often interested in reference states that have a known bulk dual in the semi-classical limit. Over these, states \eqref{CFTex} will also have a semi-classical bulk dual for any profile of $\phi$ as we now explain.
In the large $N$ limit, single trace operators $\cal O$ become generalized free fields. Then, each term in the series has a $n$-particle state interpretation. In the strict $N\to\infty$ limit the state becomes coherent. From the bulk point of view, the state is constructed out from the ladder operators of canonically quantized bulk fields $\Phi$. As shown in \cite{us2}, $1/N$ corrections deform their coherence property.  

As expected from the coherent nature of the state $|\phi\rangle$, the real and imaginary parts of $\phi\in\mathbb{C}$ have a nice physical interpretation \cite{Thesis,Belin20}: they are related respectively to the vevs $\Phi_\phi$ and $\Pi_\phi$ of the bulk field and its conjugated momentum 
computed at $\tau=t=0$ 
\begin{equation}
\Phi_\phi\equiv\langle\phi|\Phi|\phi\rangle=\int[{\cal D}\Phi]_{\phi} \;\Phi\; e^{-I}, 
\qquad\qquad
\Pi_\phi\equiv\langle\phi|\Pi|\phi\rangle=\int[{\cal D}\Phi]_{\phi} \;\Pi\; e^{-I}.
\label{bcEOM}
\end{equation}
Here bra $\langle\phi|$ is built using Euclidean conjugation \cite{Jackiw}, i.e. conjugation plus time reflection on $\phi$. Since we are computing an expectation value, the path integral is now taken over the whole Euclidean manifold, e.g. the solid AdS sphere for the vacuum. Hence, at the semiclassical level, we can trade boundary conditions $\phi$ into  initial conditions $ \{\Phi_\phi,\Pi_\phi\}$ at $\Sigma$. The explicit map $\{\phi,\phi^*\} \leftrightarrow \{\Phi_\phi,\Pi_\phi\}$ can be found in \cite{Thesis}.
A natural question is whether this map is bijective. This has been discussed in a number of works. In particular \cite{Rabideau,MarkChino} have shown, by a limiting process, that one can create arbitrarily localized initial conditions $\{\Phi_\phi,\Pi_\phi\}$, which would suggest that one can, in a linear approximation, create arbitrary initial conditions. However, \cite{Belin20} has shown that this is not the case. 
Since the construction relies heavily on the analyticity of the fields, non-analytic profiles for $\{\Phi_\phi,\Pi_\phi\}$ cannot be reproduced via this formalism. 

In this work we study a related question: can we refine the AdS/CFT dictionary so that the asymptotic source $\phi$ produces a {\it single} normal or quasi-normal mode of the  system as an initial condition? We will explore this question in the next subsection. 

\subsection{Mode-skipping sources}
\label{Sec:MSS}

Recent works on holographic excited states  \cite{Rabideau,Belin20} analyze CFTs defined on non-compact spatial slices (i.e AdS-Poincaré coordinates) where analytical computations ease. For our purposes, we will work with compact spatial slices, these will imply a discrete energy eigenbasis on both sides of the duality. In this setup, our result is the following: we will give a systematic way to build asymptotic sources $\{\phi,\phi^*\}$ that avoid any number of QN modes in the corresponding initial conditions $\{\Phi_\phi,\Pi_\phi\}$. We will call these boundary conditions {\it mode-skipping sources}. By avoiding all but one frequency one could in principle build a single QN mode initial condition. This will carry some caveats we explore below.

Consider the problem of determining an Euclidean KG field $\Phi$ in an asymptotic AdS bulk from its asymptotic boundary conditions $\phi$. As is well known the solution to the EOMs is unique and can be written as a convolution of the Euclidean bulk to boundary propagator $K_E$ and $\phi$
\begin{equation}
\Phi(\tau,\Omega,r) = \int d\tau' d\Omega'  K_E(\tau,\Omega,r;\tau',\Omega') \phi(\tau',\Omega')= \sum_{l}\int d\omega  e^{i\omega\tau+i l \Omega} K(\omega,l,r)\phi(\omega,l)
\end{equation}
Here $(\tau,\Omega)$ denote boundary coordinates and $r$ is the holographic radial coordinate. 
The Kernel $K_E(\omega,l,r)$ is known to be regular on the real $\omega$-axis. Simple poles arise at $\omega=\pm i\omega_{nl}$ with $\omega_{nl}\in\mathbb{C}$. Normal (stationary) modes have $\omega_{nl}\in\mathbb{R}$, whilst $\omega_{nl}\in\mathbb{C}$ give rise to quasinormal (QN) modes.

Consider a source $\phi(\tau)^*=\phi(-\tau) $, smoothly turning  off at $\tau=0$ as required by \eqref{bcEOM}. Such a source leads to a normalizable field configuration $\{\Phi_\phi,\Pi_\phi\}$ on $\tau=0$, i.e. $\Sigma$. Hence, expanded in terms of the Lorentzian normalizable basis $g_{nl}$ we have
\begin{equation}
\Phi_\phi\equiv\Phi(0,\Omega,r) = \sum_{l} \int d\omega e^{i l \Omega} K_E(\omega,l,r)\phi(\omega,l) = \sum_{nl} C_{\phi;n l} \; g_{n l}(0,\Omega,r)
\label{int}
\end{equation}
here $C_{\phi;nl}$ are the expansion coefficients, $n,l$ being discrete by virtue of the box character of AdS and CFT spatial slice being compact. 
The $\omega$-integral in \eqref{int} can be computed by using residues theorem,  picking contributions from all poles of $K_E(\omega,l,r)$. 

Suppose we are interested in avoiding the contribution from a particular mode  $\tilde \omega\in\{\omega_{nl}\}$. It is immediate to see that the ansatz  
\begin{equation}\label{singlemode}
\phi(\omega,l)=\left(\omega^2+\tilde\omega^2\right)e^{i \omega \epsilon} {\cal F}(\omega,l)=\left(-\partial_\epsilon^2+\tilde\omega^2\right)e^{i \omega \epsilon}{\cal F}(\omega,l)
\end{equation}
will do the job since the factor $\left(\omega^2+\tilde\omega^2\right)$ cancels the pole in $K_E$. The $\epsilon$-factor is inserted to regulate the $\omega$-integral and its  sign  is unimportant, it only determines whether we close the $\omega$-integral through the upper or lower half plane.  Relative weights for the remaining   modes are encapsulated in the smooth function ${\cal F}(\omega,l)$. A source avoiding all but a finite set of   modes will be discussed below. 

For concreteness,  consider ${\cal F}(\omega,l)=f_l $. Transforming back to Euclidean time one finds
\begin{equation}\label{singlemode2}
\phi(\tau,l)=\left(-\partial_\epsilon^2+\tilde \omega^2\right)\delta(\tau+\epsilon)f_l\,.
\end{equation}
This result merits two comments: (i) in the strict $\epsilon\to0$ limit the source sits at $\tau=0$, this may rise concern as $\phi|_{\tau=0}=0$ for convergence \cite{SvRL,us1}. In concrete examples below, we will see that sensible result are obtained if the  $\epsilon\to0$ limit is taken at the end of the computations, and (ii)  \eqref{singlemode2} is non symmetric with respect to $\tau=0$. This makes the proposal inadequate for computing excited expectation values since the source should be manifestly time reflection symmetric \cite{Jackiw}. This is easily solved by defining a reflection symmetric source 
\begin{equation}\label{singlemodereal}
\phi(\tau,l)=\frac{f_l}{2}\left(-\partial_\epsilon^2+\tilde \omega ^2\right)\begin{cases} \delta(\tau + \epsilon) &\tau<0 \\ \delta(\tau-\epsilon) & \tau\geq0\end{cases} \qquad\Leftrightarrow\qquad \phi(\omega,l)=\left(\omega^2+\tilde \omega^2\right)\cos( \omega \epsilon)f_l
\end{equation}
which meets the same relevant properties as \eqref{singlemode}.

It may happen that a given $\tilde \omega=\omega_{\tilde n\tilde l}$ is degenerate, i.e.  $\tilde \omega$   might be reproduced by many combinations of $n$ and $l$. Thus, our anzats \eqref{singlemode}  simply truncates all $\tilde \omega$ modes. Refining our proposal, a source capable of avoiding only the mode $\tilde \omega=\omega_{\tilde n, \tilde l}$, out of a ${\cal F}(\omega,l)$  generating  relative QN modes components, is
\begin{equation}\label{arbmode}
{\cal F}(\omega,l) \quad \to \quad  {\cal F}(\omega,l) 
\left( 1+\delta_{n \tilde n}\delta_{l \tilde l} \left((\omega^2+\omega_{\tilde n, \tilde l}^2)-1\right) \right) = \begin{cases} {\cal F}(\omega,l), &  \omega,l \neq \omega_{\tilde n, \tilde l},\tilde l \\ {\cal F}(\omega_{\tilde n, \tilde l},\tilde l)\left(\omega^2+\omega_{\tilde n, \tilde l}^2\right), & \omega,l = \omega_{\tilde n, \tilde l}, \tilde l \end{cases}
\end{equation}
Of course, we could replace the Kronecker deltas by Gaussians at the expense of small distorsions  of ${\cal F}$.

It now seems natural to build a source that should produce only a single mode excitation by inserting zeroes in all but the desired frequency. 
The proposal for keeping only the $(\tilde n,\tilde l)$-mode is
\begin{equation}\label{manymodereal}
\phi(\omega,l)\propto\prod_{n,l \neq \tilde n, \tilde l}\left(\omega^2+\omega_{nl}^2\right)\cos( \omega \epsilon)
\end{equation}
with a suitable normalization left implicit. However, this result rises a paradox: while asymptotic sources  should always provide a geometric dual, there is a general consensus that energy eigenstates  should not be geometric. The resolution of this tension can be found by noticing that   transforming back   \eqref{manymodereal} to configuration space, our ``single-mode''  configuration on $\Sigma$ actually arises from an infinite number of terms, hence,  an infinite series of smooth  geometric duals. In other words, we conclude that any QN mode  can be decomposed in a basis of holographic excited states\footnote{ A more agnostic reader may also argue that the limiting process $\epsilon\to0$ takes a nicely behaved geometric state into a quench-like i.e. created via an excitation strictly at $\tau=t=0$. Thus, in the strict $\epsilon=0$ limit, we may not regard any state of these mode skipping family as having a good geometric dual.}.
On the other hand, it would be reasonable to expect that removing a single mode out from the set as done in \eqref{singlemode} does not break the state's geometric representation drastically. An interesting direction for research is to quantify the loss of geometric character of states as one increasingly removes more and more modes from the set. We leave further development in this direction for future work.

\subsection{The $F_n$ family of observables}
\label{Sec:Fn}

We now present a family $F_n$, $n\geq1$ of observables which we will explore in this work. 

Start with a general source $\phi(\tau,\Omega)$ defining an excited state $|\phi\rangle$ and consider the expectation value of a product of $n$ operators ${\cal O}_{\Delta_i}(\Omega_i,t)$, with conformal dimension $\Delta_i$, taken to be observables of the theory, inserted at fixed real time $t$. To extract the excited state's features more cleanly, we define $F_n$ as the difference between $\phi$-expectation value and the vev, i.e.
\begin{equation}\label{Fn}
F_n(t,\{\Omega_i\})\equiv \langle \phi |\prod_{i=1}^n {\cal O}_{\Delta_i}(t,\Omega_i )| \phi \rangle-\langle 0 |\prod_{i=1}^n {\cal O}_{\Delta_i}(t,\Omega_i)| 0 \rangle
\end{equation}
In the present work we will perform perturbative holographic computations in the excited state source $\phi$ and in a bulk coupling constant $\lambda$ which, on general grounds, scales as $1/N$. Using \eqref{exc-state} and denoting $\Delta_E$ the conformal dimension of the operator generating the excited state, to leading order we get
\begin{align}\label{Fn2}
F_n(t,\{\Omega_i\}) &= -\int_{0}^\infty d\tau\; \phi(-\tau)^* \langle 0 |{\cal O}_{\Delta_E}(\tau)\prod_{i=1}^n {\cal O}_{\Delta_i}(\Omega_i,t)| 0 \rangle - \int_{-\infty}^0 d\tau\;  \langle 0 |\prod_{i=1}^n {\cal O}_{\Delta_i} (\Omega_i,t) {\cal O}_{\Delta_E}(\tau)| 0 \rangle \phi(\tau)+ O(\phi)^2 \nonumber\\
&= -2\Re \left\{\int_{-\infty}^0 d\tau\; \langle 0 |\prod_{i=1}^n {\cal O}_{\Delta_i}(\Omega_i,t){\cal O}_{\Delta_E}(\tau)| 0 \rangle \phi(\tau) \right\}+ O(\phi)^2 
\end{align} 
Higher orders terms, i.e. $\phi^m$, involve  $(n+m)$-point functions. These are increasingly suppressed by powers of $1/N$. The expression in the second line follows from the source  and operator's analytic properties. We have taken $\prod_{i=1}^n{\cal O}_{\Delta_i}(\Omega_i,t)$ to be an observable, so its expectation value should be a real number in any state of the theory. As such, one can see that $F_n$ is also real by definition.
Since  ${\cal O}_{\Delta_i}$ are inserted at equal times, one can interpret 
$F_n$ as measuring the entanglement between the DOFs at different angular positions.

In this work we will focus on computing $F_1$ and $F_2$, given by
\begin{equation}\label{F1}
F_1(t) = \langle\phi|{\cal O}_\Delta(t)|\phi\rangle \approx  -2\Re\left\{ \int_{-\infty}^0\!\!\! d\tau\, \langle0| {\cal O}_\Delta(t) {\cal O}_{\Delta_E}(\tau) |0\rangle\,\phi(\tau) \right\}+ O(\phi)^2\;,
\end{equation} 
\begin{align}\label{F2}
F_2(t)=  - 2\Re \left\{ \int_{-\infty}^0\!\!\! d\tau \, \langle0|{\cal O}_\Delta(t){\cal O}_\Delta(t){\cal O}_{\Delta_E}(\tau)|0\rangle\,\phi(\tau) \right\}+ O(\phi)^2\;,
\end{align}
as Witten diagrams defined in complexified manifolds, for various scenarios, and explore its properties. We have suppressed the angular dependence for the ease of the notation and   used $\langle0|{\cal O}_\Delta(t) |0\rangle=0$ which follows from translation invariance.   

We now comment on the information we aim to extract from the observables $F_n$. The first relevant information that comes to mind, in thermalizing systems, are of course its QN modes, i.e. the generally complex frequencies in which the system relax to equilibrium. These frequencies, in principle, are already contained in the exact $n$-point vacuum functions, so one may wonder what new information could we extract from our observables. Our main goal is to elucidate the particular properties of the holographic excited states \eqref{exc-state} in a thermalization situation, in particular we will study the thermalization process for mode-skipping sources \eqref{singlemodereal}. We will check that the proposal indeed  avoids completely a particular set of QN modes. An immediate consequence is that if the first $m$ QN modes are known for a given system, one could in principle envisage fine tuning the initial conditions so that all of them are absent. Then, by letting the system evolve in time one should be able to uncover the subsequent exponentially decaying QN modes.

In the examples below, we will also study a number of simple but physically important profiles for $\phi$. To the best of our knowledge, there are no explicit computations in the literature studying the evolution for broad families of sources,  we   aim to fill this gap. In a sense, one can also think of this work as a first application of the framework developed in \cite{us2}.

Finally, besides  the excited states properties, we stress that thermalizing systems may require intrincate SK paths \cite{Ben}. As such, these require  elaborate bulk duals. As mentioned above,  the problem of systematically building gravitational duals to SK paths still remains to be fully explored.

\paragraph{Conformal dimension regimes:}\; 

We now comment on the regimes of $\Delta_i$ we will study. 
To leading order, $F_1$ involves vacuum 2-point functions, which are diagonal in conformal weights, i.e. it will be zero unless $\Delta_1=\Delta_E=\Delta$. Two regimes are then foreseen, i.e. $\Delta\ll1$ or $\Delta\gg1$. The first is usually out of reach for generic backgrounds and requires knowledge of the exact two point function. On the other hand,  the second, equivalent to a heavy bulk particle, is known to be well approximated by a geodesic exploring the bulk. In the following, whenever  the exact and geodesic approximation are available, we will compare the results. 

The case of $F_2$  deals with vacuum 3-pt functions and  provides a richer spectrum of regimes that we call LLL, HHL and HHH. Here L/H stands for Light/Heavy, referring to whether $\Delta \ll1$ or $\Delta \gg1$ respectively. Notice that we will choose the operators ${\cal O}_{\Delta_i}$ in the real time segment to have identical conformal dimensions $\Delta_1=\Delta_2=\Delta$ in order to view $F_2$  as a deformation of a vacuum 2-pt function, hence we actually have 2 free parameters: $\Delta$ and $\Delta_E$. The most general scenario would be LLL, involving the exact 3-pt function, again this is generally out of reach. The scenario HHL, i.e. $\Delta\gg1$ and $\Delta_E\ll1$, involves a geodesic connecting the real time operators, and the excited state source interacting with it via a bulk to boundary propagator. Despite some simplifications, the HHL regime may still be out of reach as it requires knowledge of the full bulk to boundary correlator. Nevertheless, we will discuss an example of this regime in the sections below. Finally,  the HHH regime allows for a full geodesic approximation of the 3-pt correlator. Computing $F_2$ becomes in principle a problem of finding the intersection point, in the complexified bulk, where 3 geodesics meet. The HHH intersecting geodesics problem involves an extremization on the mixed signature manifold. We will be careful below on how to perform such extremization and will find that the proper way to define the problem breaks the  representation of  geodesics as curves in spacetime. 

As a last technical comment, we stress that in all instances and especially in the HHH regime, the deformations generated by the excited state are understood as deformations of a VEV, hence we assume $\Delta_E<2\Delta$. This avoids known effects on 3-pt functions related to OPE mixing when  conformal dimensions get close to each other, see \cite{Malda20}. A treatment for the $\Delta_E \geq 2\Delta$ scenario is beyond the scope of this work.

\section{Case Study I: Pure AdS}
\label{Sec:AdS}

In this section, we start exploring excited staes in the simplest possible scenario, i.e.  pure  AdS in global coordinates at zero temperature. We consider a self interacting real massive scalar field $\Phi$, dual to a CFT primary ${\cal O}_\Delta$, in the regime of no back-reaction. Specifically we will concentrate in AdS$_{2+1}$, but the general case is straightforward to obtain and qualitatively identical. The expected result is simple to envision: pure AdS in global coordinates is dual to a spatially compact CFT system at zero temperature. Thus, any initial conditions we start with will actually persist for all times and recover its initial form in cycles of   $T=2 \pi R_{AdS}$, no thermalization is expected.

\begin{figure}[t]\centering
\begin{subfigure}{0.49\textwidth}\centering
\includegraphics[width=.9\linewidth] {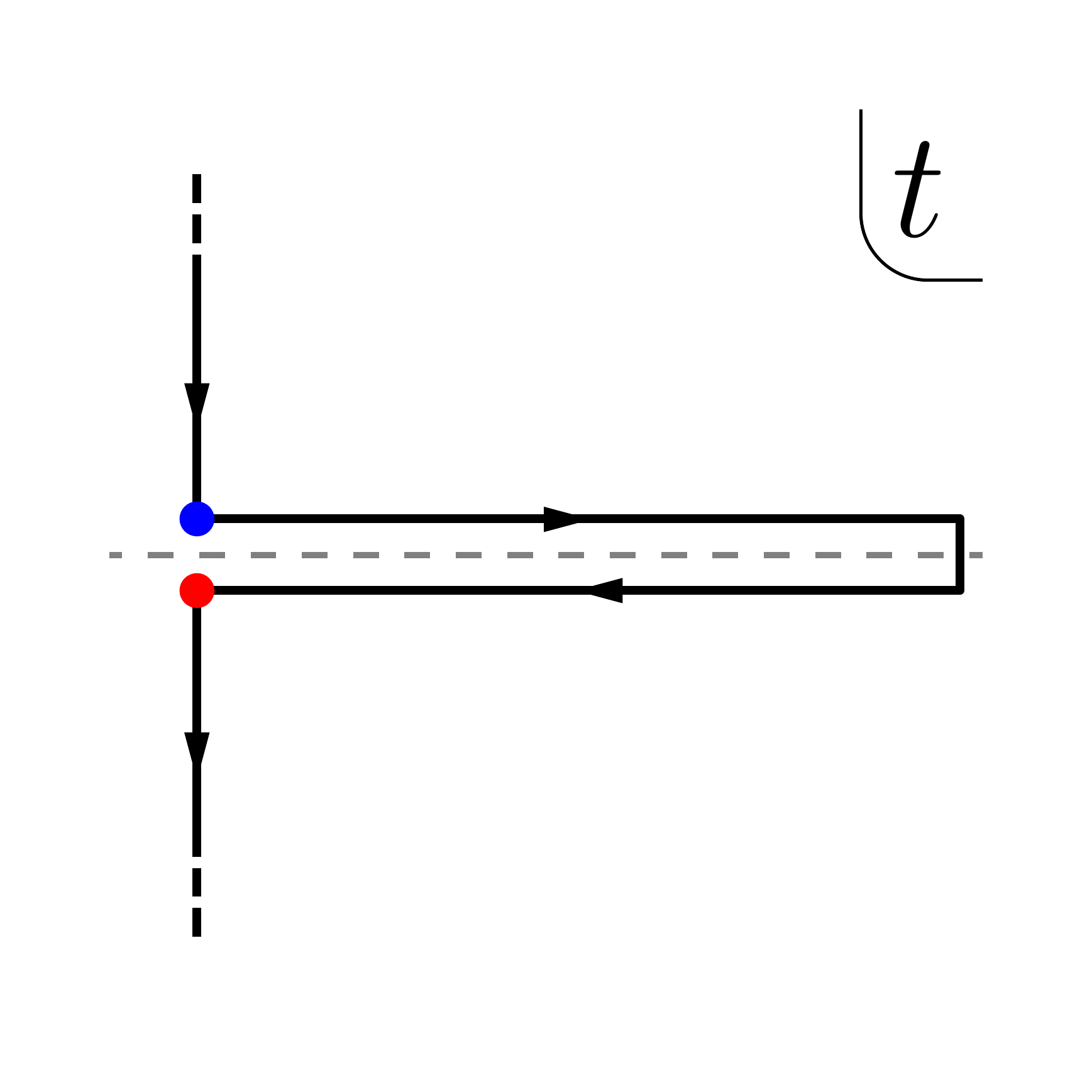}
\caption{}
\end{subfigure}
\begin{subfigure}{0.49\textwidth}\centering
\includegraphics[width=.9\linewidth] {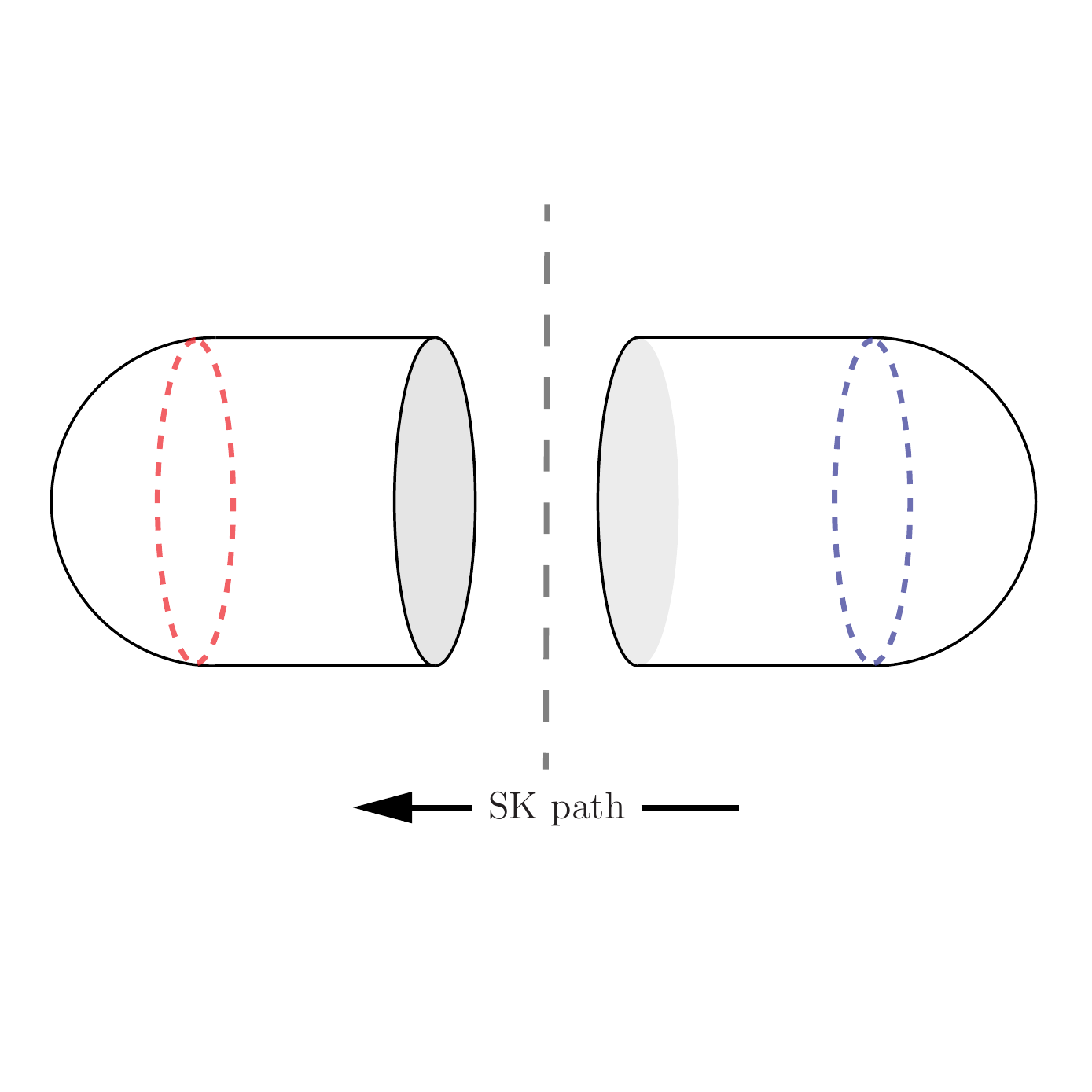}
\caption{}
\end{subfigure}
\caption{(a) In-In SK path. The state used for computing the expectation value is prepared by   Euclidean path integrals represented by the vertical segments. The upper horizontal real time segment corresponds to the system evolving in time, it is there that operators are inserted. The lower horizontal segment is usually associated to the environment DOFs in interaction with the system. The dashed grey line emphasizes a ``mirror" symmetry in the complex plane that helps build the geomtric dual.  (b) Bulk dual for the In-In SK path. We retain the grey dashed line to ease its interpretation.}
\label{AdS-Geom}
\end{figure}

We take this section as a warm up to set notation and showcase the discussion in Sec. \ref{Sec2}, including the construction of the mixed signature manifolds involved. We will  profit from 2- and 3-pt functions being exactly known in pure AdS. Thus, rather than giving a complete set of examples, we will focus on computations that are manageable with analytic expressions. The starting action and metrics are (setting $R_{AdS}^2=1$)
\begin{equation}\label{AdS-Action}
S[\Phi]=\frac 12\int \sqrt{g} \left((\partial_\mu \Phi)^2+m^2\Phi^2\right)+\frac{\lambda}{3}\int \sqrt{g}\;\Phi^3\;,\qquad\qquad \lambda\sim 1/N\ll1\;,
\end{equation}
\begin{equation}\label{AdS-Metrics}
ds^2=\left.(r^2+1)\times\begin{cases}-dt^2\\+d\tau^2\end{cases}\hspace{-3mm}\right\}+\frac{dr^2}{r^2+1}+r^2d\varphi^2\;.
\end{equation}
The action $S[\Phi]$ is understood to be defined over the manifold associated to an SK contour which we now introduce. For the zero temperature scenario, the correct SK path to study the evolution of expectation values is the so called In-In path presented in Fig. \ref{AdS-Geom}(a). The name emphasizes that both bra and ket in the path are prepared at $t=0$, cf. with In-Out path in Fig. \ref{Fig:In-Out}(a). Vertical pieces prepare the wavefunction (at $t=0$) via Euclidean path integrals.
The upper horizontal real time segment (forward in time) represents the physical system time evolution.
The lower horizontal real time segment (backward in time) is usually interpreted as the environment degrees of freedom (a $T=0$ reservoir in this case) interacting with the system \cite{Umezawa}, but this will not be relevant for our discussion. The bulk dual is shown in Fig. \ref{AdS-Geom}(b). It is built via a piece-wise map between pure imaginary (real) segments of the SK path and Euclidean (Lorentzian) AdS manifolds, ${C}^1$ glued across constant time surfaces in \eqref{AdS-Metrics}. For details on the gluing see \cite{SvRL,us1}. The resulting manifold is a natural zero temperature saddle possessing only asymptotic boundaries. This means that the variational problem following from \eqref{AdS-Action} has a unique solution in terms of boundary data which has direct CFT interpretation. 

We now proceed to construct the classical perturbative solution for the scalar field  \eqref{AdS-Action} in the mixed signature geometry. 
A recursive  expression for the solution to the EOMs can then be formally written as,
\begin{equation}\label{Scalar-GeneralSol}
\left( \square - m^2 \right)\Phi=\lambda \Phi^2 \qquad \Rightarrow \qquad   \Phi = \int_{\cal C} K \cdot \phi + \lambda \int_{\cal C} G\cdot \Phi^2
\end{equation}
where the first and second terms are the homogeneous and inhomogeneous solutions to the EOM. The functions $K$ and $G$ are the {\it complexified} bulk to boundary and bulk to bulk propagators respectively and we have denoted with $\phi$ the prescribed asymptotic data, irrespectively of their insertion in Euclidean or Lorentzian segments. 
We will use ${\cal C}$ to indicate that all time integrals involved should be taken ordered according to the relevant SK path. 

The solution constructed in this way is unique, with the gluing along the SK path in Fig.\ref{AdS-Geom} fixing  N-modes at $t=0$ in terms of the Euclidean sources. The precise way in which the SK path fixes the ordering of all real-time correlators both in QFT and AdS/CFT has been thoroughly studied  previously in the literature and will not be detailed here. The interested reader can see \cite{SvRL,Umezawa,Kamenev,Kamenev2,Kamenev3} for further details. Moreover, since observables $F_n$ have all real-time operators inserted at equal times, issues related to ordering along SK-path will not be relevant for our analysis.

For pure AdS, the propagators $K,G$ in \eqref{Scalar-GeneralSol} are known. Writing $\sf t$ to denote $\{t,-i\tau\}$, we have
\begin{equation}\label{Scalar-GeneralSol2}
    \Phi({\sf t},r,\varphi) = \int_{\cal C} K({\sf t},r,\varphi;{\sf t}',\varphi')  \phi({\sf t}',\varphi') + \lambda \int_{\cal C}
    G({\sf t},r,\varphi;{\sf t}',r',\varphi') \Phi({\sf t}',r',\varphi')^2
\end{equation}
 with\footnote{An $i\epsilon$ prescription is expected to appear in \eqref{KK} whenever $\sf t,t'$ are lightlike separated in the Lorentzian sections. For concreteness take ${\sf t}=t>t'={\sf t}'$ both in the upper Lorenztian section in \ref{AdS-Geom}(a). In that scenario the $\omega$ integral is
\begin{align*}
    K({\sf t},r,\varphi;{\sf t}',\varphi')&=\frac{1}{4\pi ^2}\sum_l \int_{\cal F}  d\omega\; e^{-i\omega (t-t')+i l (\varphi-\varphi')} f_{\omega l}(r)= \frac{2^{d/2-\Delta}(\Delta-d/2)^2\pi} {\left[\sqrt{r^2+1}\cos((t-t')(1-i\epsilon) )- r \cos(\varphi-\varphi')\right]^{\Delta}} \;, 
\end{align*}
where ${\cal F}$ stands for the Feynman integration path. On the other hand, in the lower Lorentzian section, the reverse flow of time ends up imposing an anti-Feynman correlator. Notice that points in different Lorentzian sections are never lightlike separated. See \cite{us1} and references within for details.} 
\begin{align}
\label{KK}
    K({\sf t},r,\varphi;{\sf t}',\varphi')&=\frac{1}{4\pi ^2}\sum_l \int  d\omega\; e^{-i\omega ({\sf t}-{\sf t}')+i l (\varphi-\varphi')} f_{\omega l}(r)\nonumber \\
    &= \frac{2^{d/2-\Delta}(\Delta-d/2)^2\pi} {\left[\sqrt{r^2+1}\cos({\sf t}-{\sf t}' )- r \cos(\varphi-\varphi')\right]^{\Delta}} \;,
\end{align}
We stress that correlators between any segments of the SK path are uniquely determined by the SK path ordering.
The function $f_{\omega l}(r)$ in the first line carries the radial profile, i.e. it is the regular homogeneous solution to the EOM properly normalized to give a delta-function at the boundary. Its precise form \cite{SvRC,us1} will be of no concern to us. It will suffice to say that it contains the information on AdS N-modes through poles in the 
complex $\omega$-plane. 
For pure AdS$_3$ they are located at 
\begin{equation}
\label{exOme}
\omega=\pm \omega_{nl} 
\qquad\text{where}\qquad \omega_{nl}=2n+\Delta +|l|\in\mathbb{R} 
\end{equation}
with $n\in\mathbb{N}$. The bulk to bulk Green function $G$ for AdS$_{d+1}$ is
\begin{equation}\label{GG}
    G(\zeta)=\frac{2^{-\Delta}\Gamma(\Delta)}{\pi^{\frac d2}\Gamma(\Delta-d/2)\Gamma(2\Delta-d)}(1-\zeta)^\Delta \,_2F_1\left(\frac \Delta 2,\frac{\Delta+1}{2};\Delta-\frac d2+1;1-\zeta^2\right),
\end{equation}
we should set  $d=2$ for AdS$_3$. The complexified AdS-invariant function $\zeta$ is defined as,
\begin{equation}
  1 -  \zeta =\frac{1}{\sqrt{r^2+1} \sqrt{r'^2+1} \cos({\sf t}-{\sf t}' )-r r' \cos (\varphi -\varphi' )}
\end{equation}
With these functions at hand one can readily compute 2- and 3-pt CFT correlation functions via holography. The 2-pt function reads
\begin{align}
\langle  {\cal O}_\Delta({\sf t},\varphi){\cal O}_{\Delta'}({\sf t}',\varphi')  \rangle &= \frac{\delta_{\Delta,\Delta'}}{4\pi i}\sum_l \int d\omega\; e^{-i\omega ({\sf t}-{\sf t}')+i l (\varphi-\varphi')}\alpha(\omega,l,\Delta)\beta(\omega,l,\Delta)\nonumber\\
&=  \frac{\left(2\Delta-d\right)^{2}}{2^{\Delta+2} \pi} \frac{\delta_{\Delta,\Delta'}}{[\cos({\sf t}-{\sf t}')-\cos(\varphi-\varphi')]^{\Delta}}\label{2p}
\end{align}
\begin{equation}
\alpha(\omega,l,\Delta)=\frac{1}{\Gamma(\Delta+1)\Gamma(\Delta-d/2+1)}\left(\frac{\omega + l + d-\Delta}{2}\right)_{\Delta-d/2}\left(\frac{\omega - l + d-\Delta}{2}\right)_{\Delta-d/2}
\label{alfa}
\end{equation}
\begin{equation}
\label{beta}
\beta(\omega,l,\Delta)=-\psi\left(\frac{\omega + l +\Delta}{2}\right)-\psi\left(\frac{-\omega + l + d-\Delta}{2}\right)
\end{equation}
with $(a)_b$ Pochhammer symbols and $\psi(x)=\Gamma'(x)/\Gamma(x)$ the Digamma function. $\alpha$ and $\beta$ are defined so that all $\omega$-poles in the first line of \eqref{2p} are contained in $\beta$ with Res$[\beta]=1$ irrespectively of $ {n,l}$; whilst $\alpha$ is a regular polynomial  throughout its domain, see \cite{SvRC,SvRL,us1} for details. The 3-pt function results
\begin{align}\label{3p}
\langle {\cal O}_{1} {\cal O}_{2}{\cal O}_{3}\rangle_{{\sf t}\varphi} &= \int_{\cal C}  \sqrt{g} K(1,b) K(2,b) K(3,b)\nonumber\\
&= [\cos({\sf t}_1-{\sf t}_2)-\cos(\varphi_1-\varphi_2)]^{\frac{\Delta_3-\Delta_1-\Delta_2}{2}}[2-3]_{{\sf t}\varphi}^{\frac{\Delta_1-\Delta_2-\Delta_3}{2}}[3-1]_{{\sf t}\varphi}^{\frac{\Delta_2-\Delta_3-\Delta_1}{2}} 
\end{align}
where the point $b$ is integrated on the bulk shown in Fig. \ref{AdS-Geom}(b) and a compact notation has been used for the sake of brevity. The expression in the first line is just the convolution of three bulk to boundary functions \eqref{KK}. The last expression in \eqref{3p}  can be obtained from the well known Poincaré result \cite{Rastelli} through a change of coordinates. Once again, the correct correlator ordering is uniquely fixed by SK path ordering. 

\paragraph{Remarks on holographic computations of $F_n$:}\;

\begin{figure}[t]\centering
\begin{subfigure}{0.49\textwidth}\centering
\includegraphics[width=.9\linewidth] {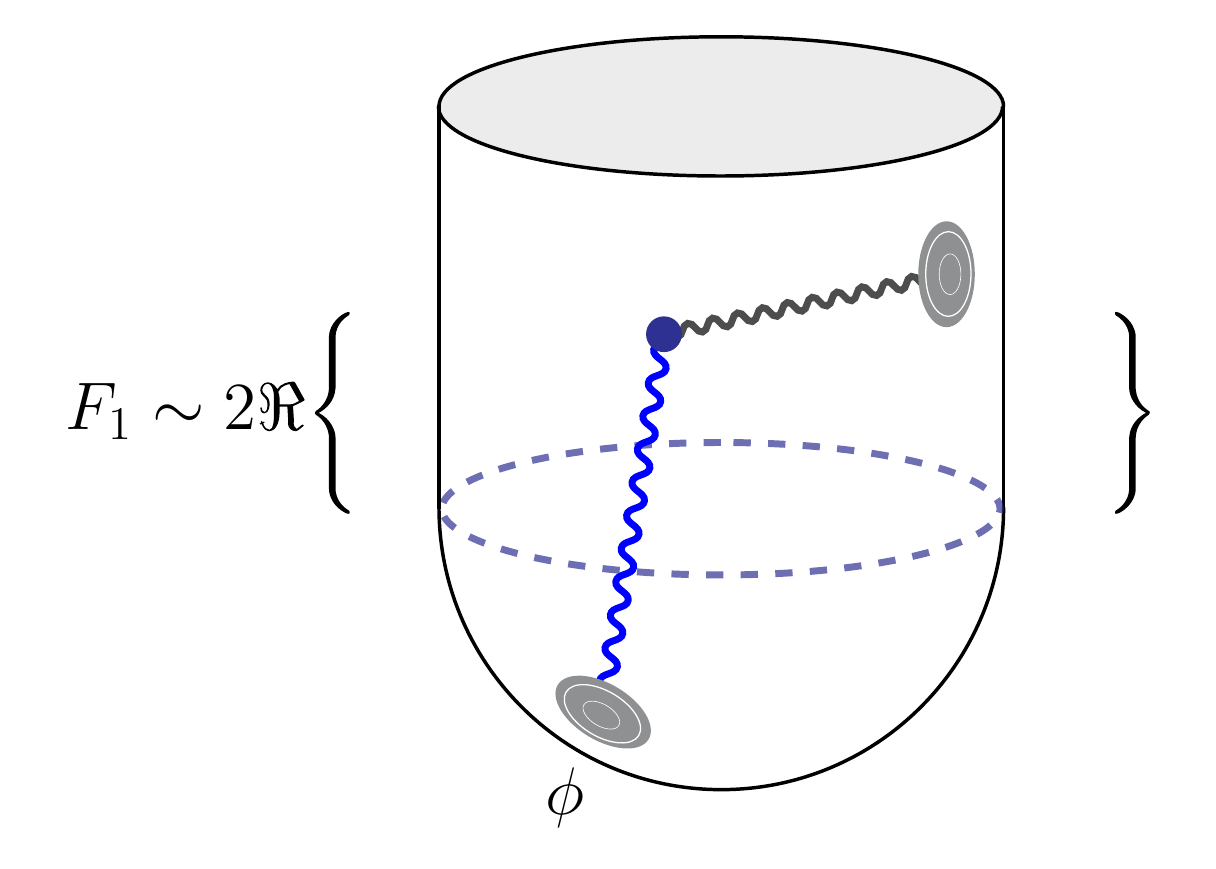}
\caption{}
\end{subfigure}
\begin{subfigure}{0.49\textwidth}\centering
\includegraphics[width=.9\linewidth] {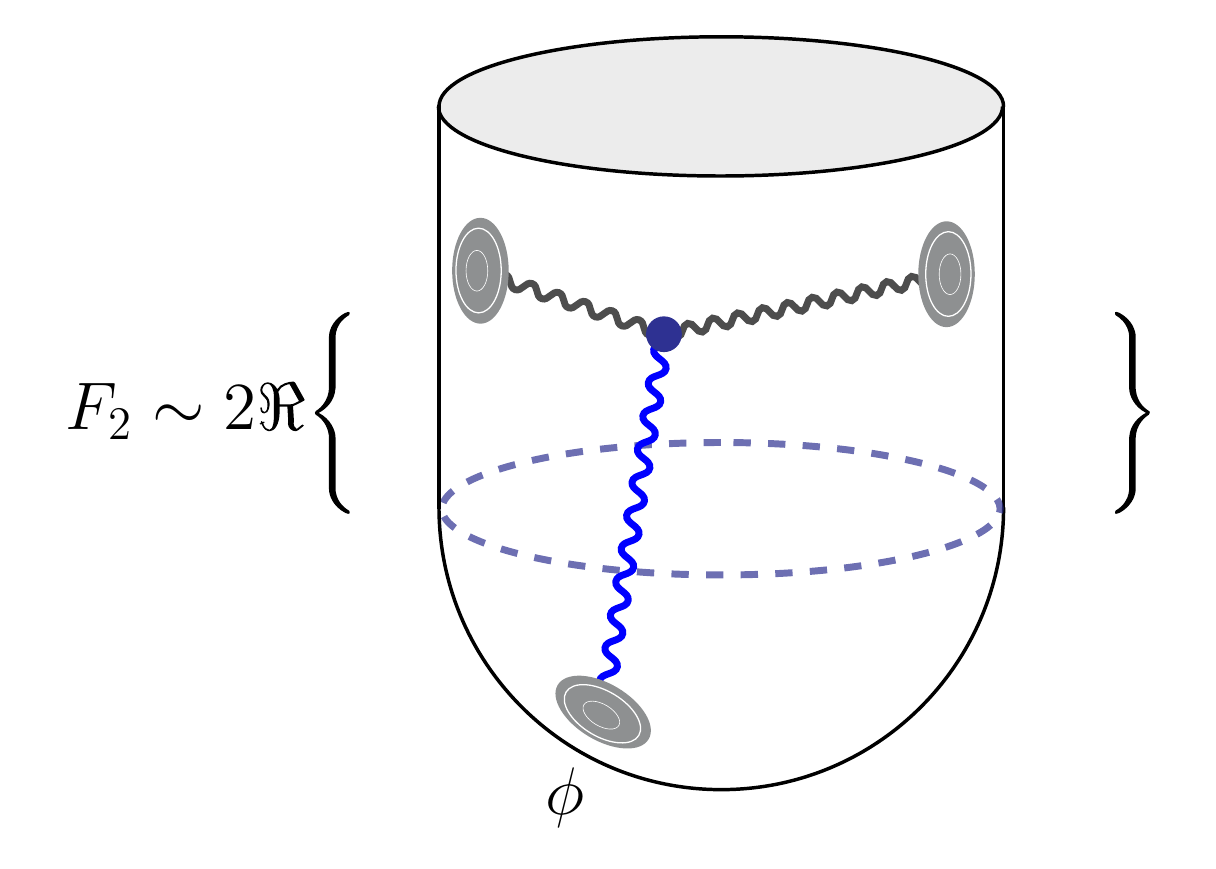}
\caption{}
\end{subfigure}
\caption{Diagrammatic representation of $F_1$ and $F_2$ to leading order in $\phi$ as Witten diagrams in pure AdS complexified manifolds. Analogous leading order contributions to the $F_n$ family of observables and other more complex geometries are straightforward to build.}
\label{Fig:AdS-Exact}
\end{figure}

We now comment on some mathematical properties of the bulk expressions computing the $F_n$ observables. In Fig. \ref{AdS-Geom} we explicitly displayed a dashed line dividing the SK path/bulk dual in two mirrored halves. We will see that all In-In SK paths appropriate for studying thermalization will present this symmetry. This property
manifests in the leading order $\phi$-contribution in $F_n$ as the two terms in the first line of \eqref{Fn2} being complex conjugates of each other. Hence, it is enough to compute any of the terms in the first line to get the leading physical response. Of these, in the expressions \eqref{Fn2},\eqref{F1},\eqref{F2} we have singled out those  in which all correlator insertions  lie in the first half of the SK-path (upper half plane). An analogous structure between terms should appear at all orders in order to guarantee a real result for $F_n$. 
Strictly speaking, a correlator computation involves the complete information on the bulk dual. However, to first order in $\phi$ and in the geodesic approximation, the leading order $F_n$'s can be computed using only the first half of the bulk dual\footnote{Profiting from the leading order in $\phi$ mirror symmetry, one could equivalently consider dropping the second half of the geometry and impose reflective boundary conditions on the dashed line in Fig. \ref{AdS-Geom}(b).}. We represent this recipe for  $F_1$ and $F_2$ in Fig. \ref{Fig:AdS-Exact}. Subleading corrections, i.e.  $O(\phi^2)$, will  generally involve the complete bulk geometry.

\subsection{Study of $F_1$}

In this section, given that the required correlators are known analytically known for pure AdS, we explore the most salient aspects of our construction for $F_1$. Computing $F_1$ amounts to compute the convolution of the 2-pt correlator and source. Furthermore, pure AdS geometry is simple enough so that the geodesic approximation turns out to be exact and matches the 2-pt function result \eqref{2p}, so that there is no relevant comparison to do between the exact and geodesic approx results. A more interesting study arises in the BTZ scenario, this will be done in Sec. \ref{BTZ}.

In the next section we show explicitly that  sources of the form \eqref{singlemode2}  correctly skip N-modes. This is   done transparently in the mode expansion representation of the correlator, an unsual situation since one does not often have access to it for the case of more general manifolds. We will also be able to check the result in configuration space representation. Finally, we will explore the properties of some specific Euclidean source profiles.

\subsubsection{Skipping N-modes}

As we have discussed, skipping a particular  N-mode  requires a quite cumbersome source. This stems from the fact that N-modes are degenerate. 
However, we will consider the lowest possible excitation $\omega_{00}=\Delta$ for which no degeneracy exists, so \eqref{singlemode2} is then adequate. On the other hand, avoiding the first excited state of a system is also probably the most realistic scenario for applying of our framework. The observable $F_1$ is computed to leading order as 
\begin{align}
\label{f1adsp}
F_1(t,\varphi)&=-2\Re \left\{\int_{-\infty}^0d\tau'\int_0^{2\pi} d\varphi' \,\langle  {\cal O}(t,\varphi) {\cal O}_E(\tau',\varphi') \rangle\,\phi(\tau',\varphi')\right\},
\end{align}
and to avoid the first excited state we can insert $\tilde \omega=\omega_{00}=\Delta$ in \eqref{singlemode2}, with $f_l=e^{il\varphi_0}$ 
$$\phi(\tau',\varphi')= \left(-\partial_\epsilon^2+\Delta^2\right) \delta(\tau'+\epsilon)\delta(\varphi'-\varphi_0)\,$$ 
Inserting the Fourier mode expansion \eqref{2p} with $t'=i\tau'$ in \eqref{f1adsp} results in
\begin{align} 
F_1(t,\varphi)
&=-\Re \left\{\frac{1}{2\pi i} \left(-\partial_\epsilon^2+\Delta^2\right)  \sum_{l} \int d\omega\; e^{-\omega \epsilon}e^{-i\omega t} e^{i l (\varphi-\varphi_0)}  
\alpha(\omega,l,\Delta)\beta(\omega,l,\Delta)\right\}\nonumber\\\label{f1o}
&=-2\Re \left\{\sum_{nl} e^{-i\omega_{nl} t}e^{i l (\varphi-\varphi_0)}  \left(\Delta^2-\omega_{nl}^2\right) \text{Res}_{\omega_{nl}}\left[\alpha(\omega,l,\Delta)\beta(\omega,l,\Delta)\right]\right\}
\end{align}
where in the last line we computed the $\omega$-integral using residues theorem. 
This expression explictly shows no component in the $\omega_{00}$-mode due to the vanishing of the parentheses in the second line. 

Writing $F_1$ as \eqref{f1o} is in general, not available, so it is perhaps more illuminating to see how convolution \eqref{f1adsp} works in configuration space. Consider a very narrow Gaussian peaked at some particular value of $\varphi_0$ and $\tau<0$ as a source, and take $\Delta\varphi=0$, according to \eqref{2p} we get 
\begin{align}
\label{dsou}
F_1(t,\varphi_0)&\sim 2\Re \left\{ \frac{1}{\left( \cos(t+i\tau)-1 \right)^{\Delta}}\right\}
\sim \cos(t \Delta )e^{\tau \Delta}c_{00} + \cos(t(\Delta+1)) e^{\tau(\Delta+1 )} c_{01} + \dots
\end{align} 
where the rhs follows from the first line in \eqref{2p} after performing the $\omega$-integral by residues.  
The result displays a linear combination of all N-modes of the system, as one would expect for a generic source profile. In the expression above the $c_{nl} $ coefficients represent the amplitude of the  $\omega_{nl}$ mode and are generically non-zero. Amusingly, if we now takes \eqref{singlemode2} as our source, we get
\begin{align*}
F_1(t,\varphi_0)&\sim 2\Re \left\{ (-\partial_\epsilon^2+\Delta^2) \left(\frac{1}{\left( \cos(t-i\epsilon)-1 \right)^{\Delta}}\right) \right\}=\Re \left\{\frac{-2\Delta  (1+2 \Delta )}{(\cos (t-i\epsilon)-1)^{\Delta+1 }}\right\}
\sim \cos(t(\Delta+1))e^{-\epsilon (\Delta+1)} \tilde c_{00} + \dots
\end{align*}
which explicitly shows the absence of the first excited state $\omega_{00}=\Delta$ in the response of the system. The first mode appearing in the expansion is the one immediate above $\omega_{01}=\omega_{10}=\Delta+1$.

\begin{figure}[t]
\begin{subfigure}{0.49\textwidth}\centering
\includegraphics[width=.9\linewidth] {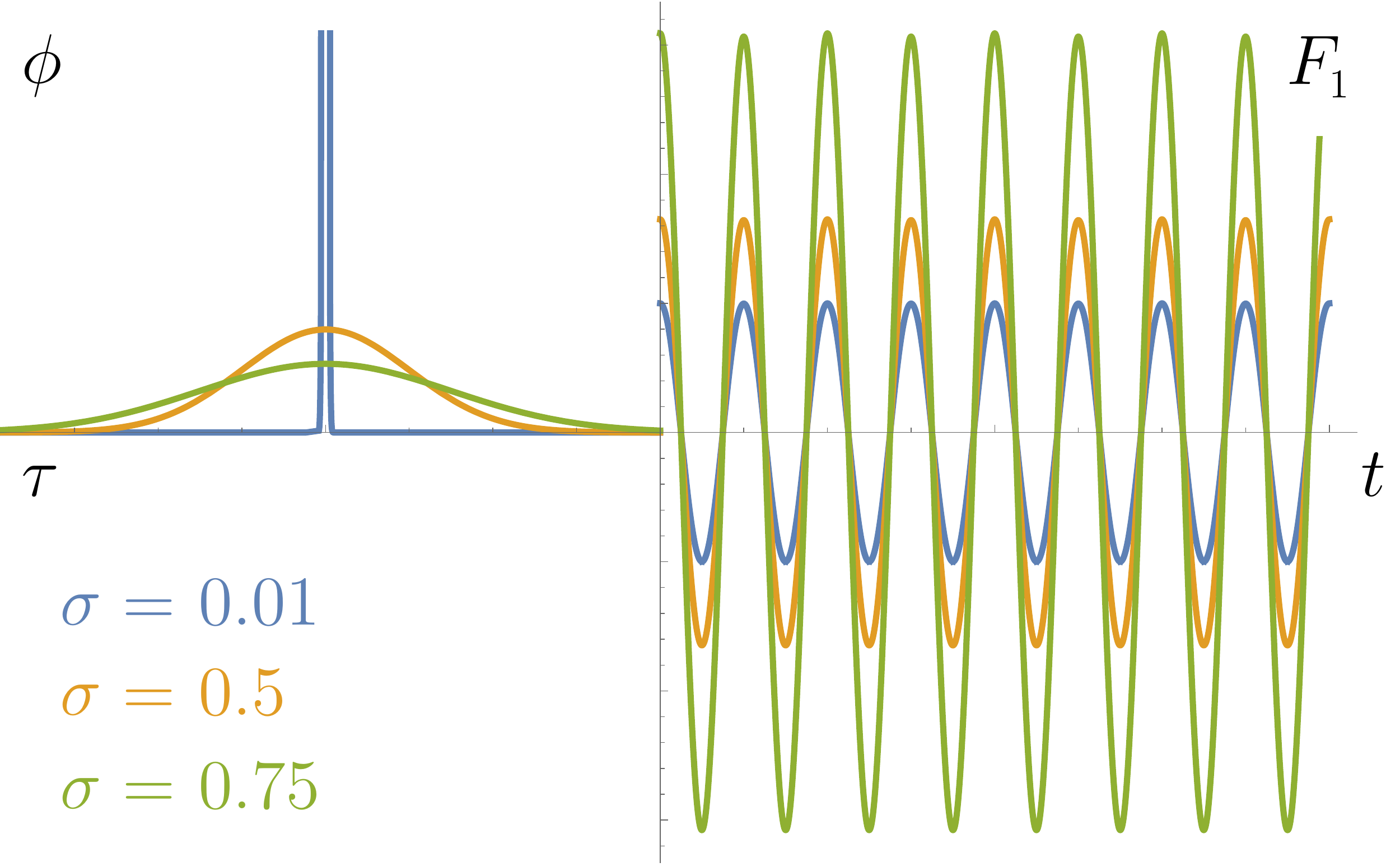}
\caption{}
\end{subfigure}
\begin{subfigure}{0.49\textwidth}\centering
\includegraphics[width=.9\linewidth] {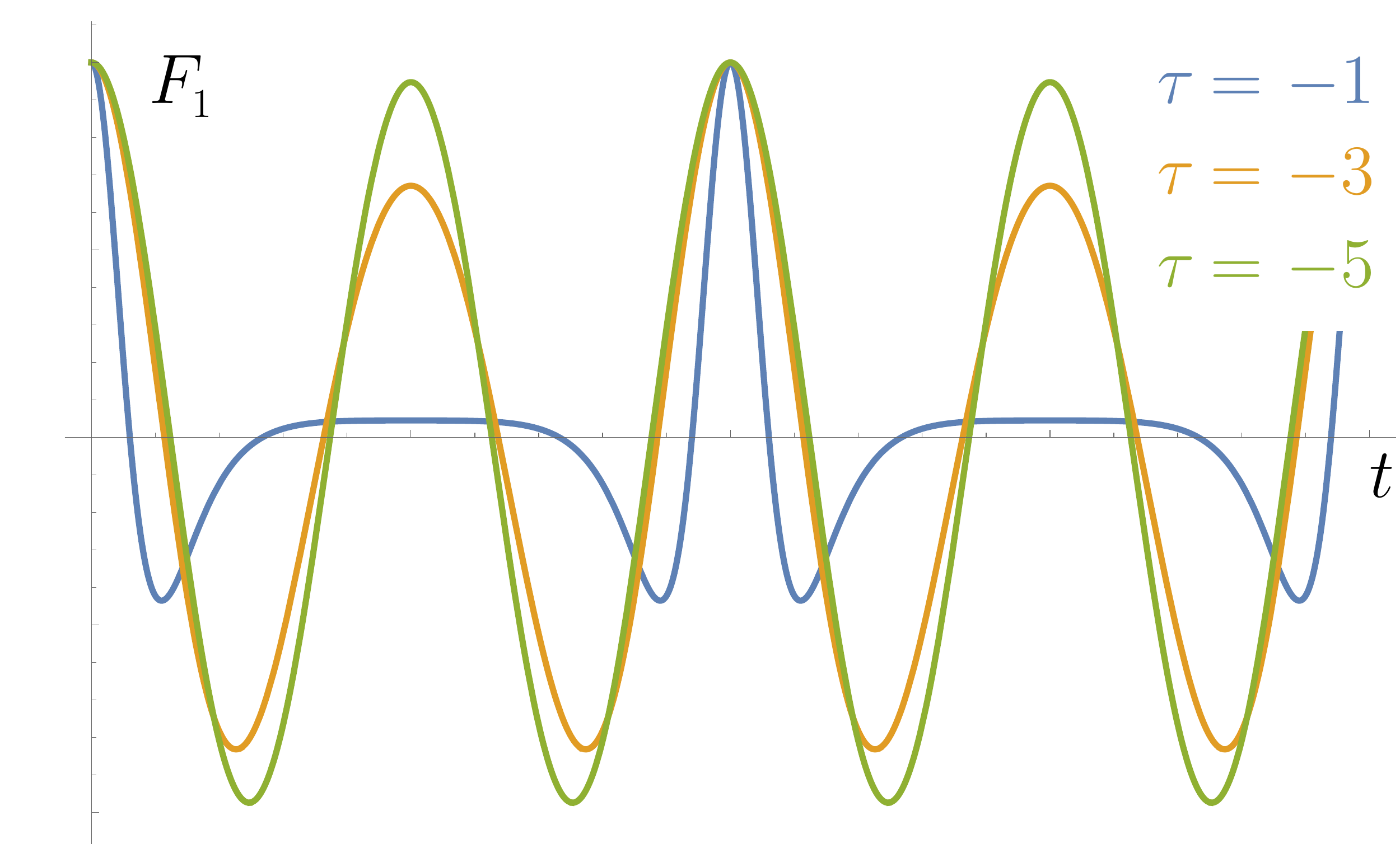}
\caption{}
\end{subfigure}    
\caption{$F_1(t)$ for localized sources. (a)  Gaussian sources on the Euclidean section with varying widths $\sigma$ for a sample value of $\tau$. 
The limit $\sigma\to0$ for which the source becomes a  Dirac delta gives a finite response. Although drawn horizontally, the negative  horizontal axis corresponds to imaginary time $\tau$. (b) Sample responses for Dirac delta sources at different Euclidean locations $\tau<0$. In the $\tau\to-\infty$ limit, the response reduces to the  (fundamental) frequency $\omega=\Delta$. This is consistent with the fact that the insertion of a source at $\tau=-\infty$ corresponds to a particle state in its fundamental state $\omega_{00}=\Delta$, i.e. at rest in the center of AdS.}
\label{Fig:Sources}
\end{figure}

As our final example consider an operator insertion displaced from the source as $\Delta\varphi=\pi/2$, then
\begin{align}
F_1(t,\varphi_0+\pi/2)&\sim 2\Re\left\{ (-\partial_\epsilon^2+\Delta^2) \left(\frac{1}{ \cos(t-i\epsilon) ^{\Delta}}\right)\right\}=\Re\left\{\frac{2\Delta  (\Delta+1 )}{\cos (t-i\epsilon)^{\Delta +2}}\right\}\sim \cos(t(\Delta+2))e^{-\epsilon (\Delta+2)} \hat c_{00}+\dots
\end{align}
Notice that for this particular case the first excited mode $\omega_{01}=\Delta+1$ is also absent. This feature is due to the choice\footnote{This is easy to see in AdS$_{2+1}$, spherical harmonics are simple exponentials $e^{il\phi}$. Since $\alpha,\beta$ in \eqref{alfa}-\eqref{beta} are insensitive to the sign of $l$, $l=\pm1$ modes contribute as $(e^{i\Delta \varphi}+e^{-i\Delta \varphi})\sim \cos(\Delta\varphi)$ which vanishes for $\Delta\varphi=\pi/2$.} $\Delta\varphi=\pi/2$. This last example aims to show that we can assure that the skipped modes will not be present but, depending of the precise observable, other modes might be absent as well.

\subsubsection{Paradigmatic simple sources}
\label{Sec:RelevantSources}

To gain intuition, before moving  to the study of $F_2$, we want to discuss some specific simple sources. For ease of computations we focus on   $\Delta=2$ for this section.

Our first  choice is a localized  $\delta$-source at particular $\varphi_0$ and $\tau<0$ values, much like \eqref{dsou}. We can regularize it by considering a Gaussian located at $\tau,\varphi_0$ of width $\sigma\to0$. We have already checked in the previous subsection that the limit is smooth, i.e. the excitations generated are of finite amplitude. We can ask how do the excitations behaves as we vary $\tau$. One can see that each $\omega_{nl}$ mode's amplitude behaves as $\sim e^{(\Delta+n) \tau}$ ($\tau<0$), so the leading contribution as $\tau\to-\infty$ comes from the fundamental $N$-mode $\omega= \Delta$. This is well known: a localized source at $\tau\to-\infty$  creates a single-particle excitation in the lowest fundamental state. Further comment on this will be made in the upcoming HHH subsection. This discussion is summarized in Fig. \ref{Fig:Sources}.

\begin{figure}[t]
\begin{subfigure}{0.49\textwidth}\centering
\includegraphics[width=.9\linewidth] {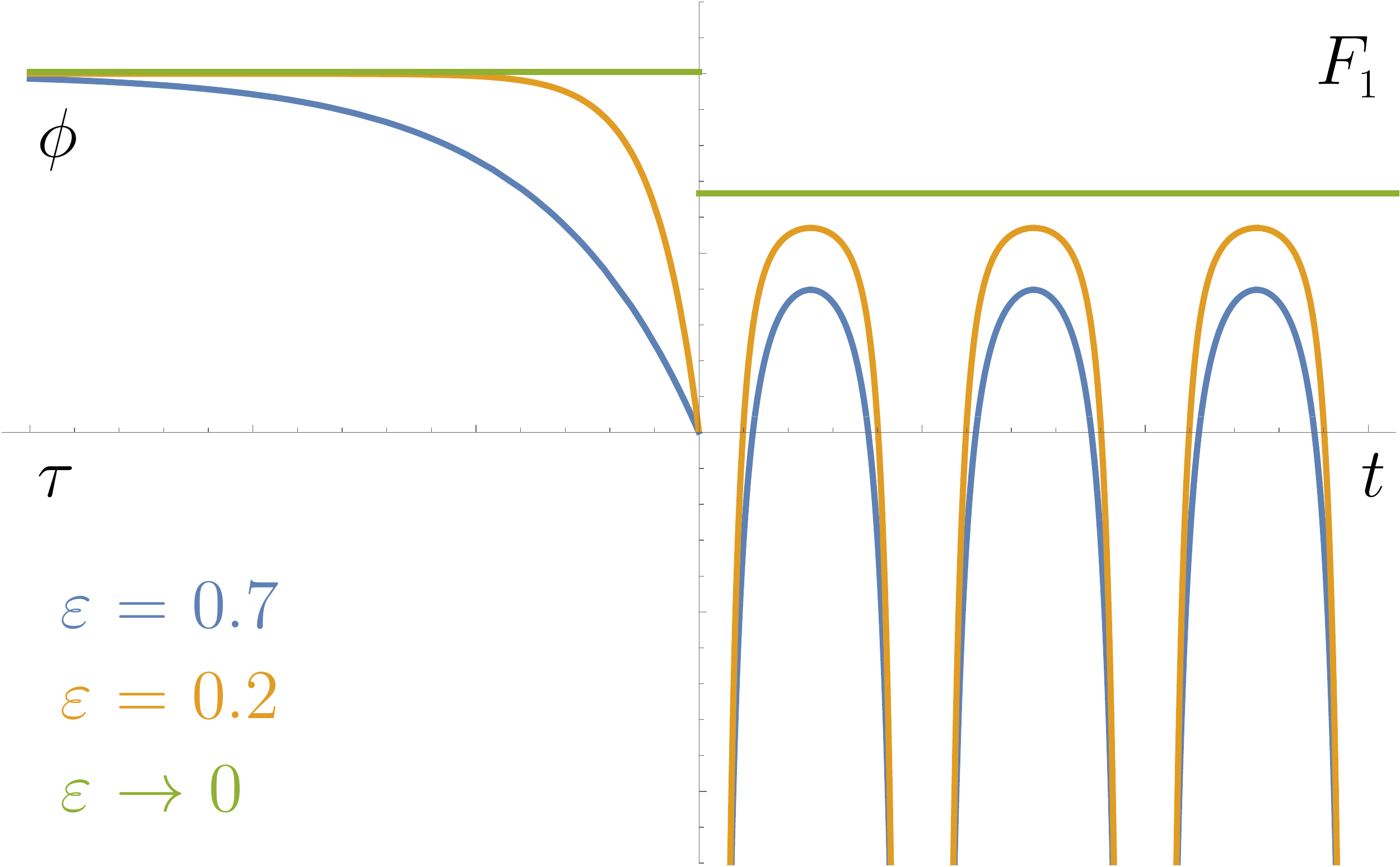}
\caption{}
\end{subfigure}
\begin{subfigure}{0.49\textwidth}\centering
\includegraphics[width=.9\linewidth] {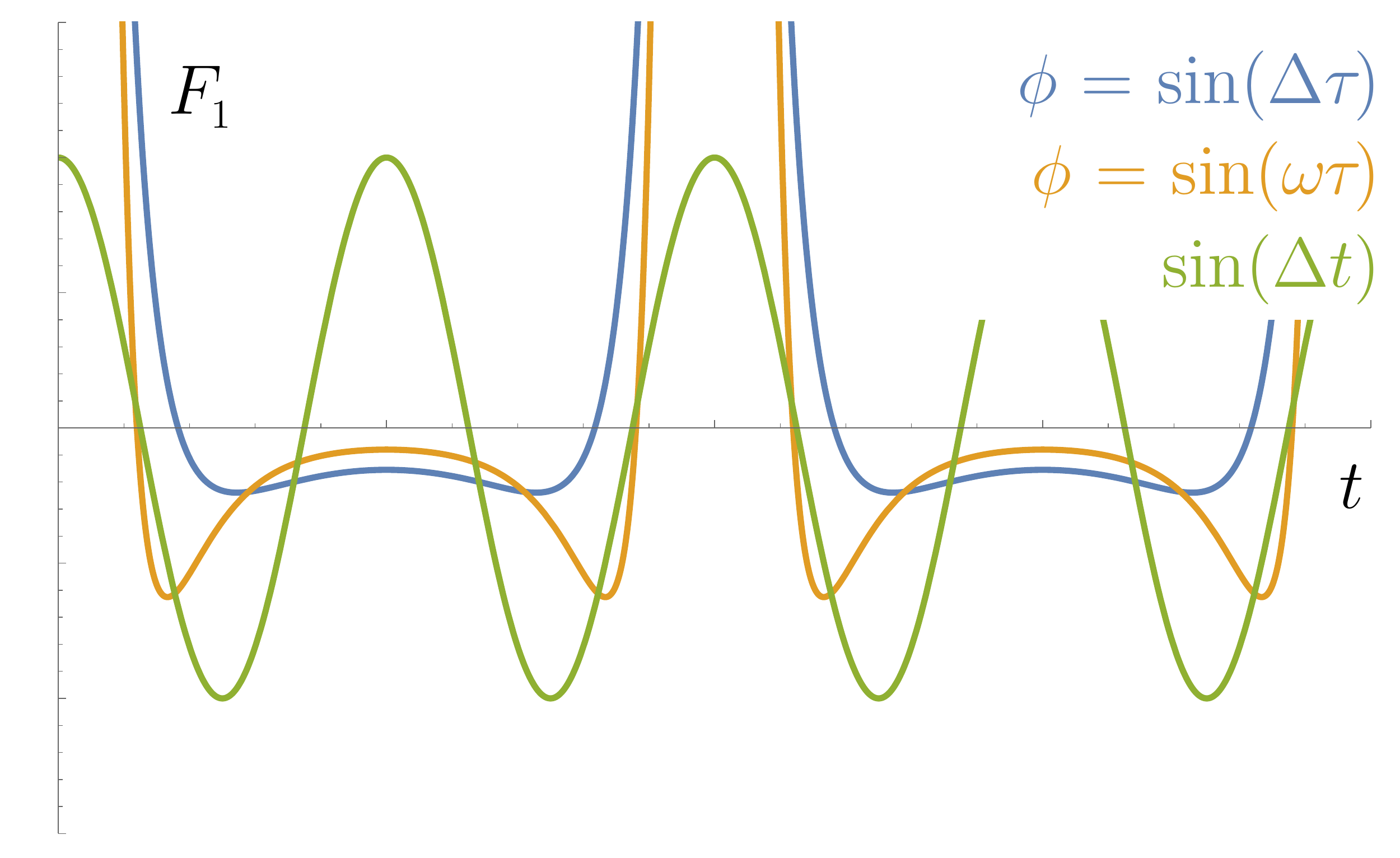}
\caption{}
\end{subfigure} 
\caption{$F_1(t)$ for special sources.  (a) we describe a limiting procedure to understand the constant $\phi$ configuration. One can see that these kind of sources generate periodic divergences as a response, but upon taking the limit the response stabilizes at a constant value. This is understood as the vacuum of a deformed theory as described in this one, as explained in more detail in the main text. One should regard the negative piece of the horizontal axis as $\tau$ and the positive piece as $t$. In (b) we perform a sanity check of our claims in the sense that we should not be able to select modes using single mode sources. We test this with two single mode sources with frequencies both corresponding to N modes of the system (blue) and for a generic frequency $\omega$. In green we show the physical response one should get for the lowest physical mode. Neither source is able to produce a single mode state.}
    \label{Fig:Sources2}
\end{figure}

Our second example corresponds to a delocalized (constant) source $\phi$, albeit one should be careful with its interpretation. This source is suspicious for two reasons: it has no compact support, and it does not vanish at $t=\tau=0$. Inspecting prescription \eqref{exc-state}, a constant source should actually be interpreted as a deformation of the original CFT. By definition then, we are preparing the vacuum state of a deformed theory rather than an excited state of the original CFT, i.e. the generated wavefunction does not properly belong to the Hilbert space of the theory. One can approach the problem by considering a source $\phi \sim (1-e^{\tau/\epsilon})$, where $\epsilon$ is a regulator taken to be zero at the end of the computations. This regulated source meets $\phi(\tau=0)=0$ so at each step in the limit we abide the rules of our excited state creation mechanism, disregarding the compact support. The result of this limit yields a $t$-independent shift in $F_1$ which is also the expected result for this scenario. We present these results in Fig. \ref{Fig:Sources2}(a).

A final test would be that of a sanity check considering single mode Euclidean sources $\phi\sim \sin(\omega \tau)$ for $\omega$ both contained and not in the set of N modes $\omega_{nl}$. One can directly see that there is no qualitative change in $F_1$ for neither type of $\omega$ and that they both generate a response containing (in principle ) all modes, rather than keeping only a single mode $\omega$, see Fig. \ref{Fig:Sources2}(b). This was mandatory, for a thermalizing system is only able to oscillate in its natural frequencies $\omega_{nl}$ independently of the initial condition.

\subsection{Study of $F_2$}\label{AdSF2}

An exhaustive study of $F_2$ in the pure AdS scenario would be redundant after our $F_1$ analysis above, since the 3p functions are also known analytically. In this section we aim first to showcase the HHL regime of conformal dimensions $\Delta\gg1$ but $\Delta_E\ll1$ present in $F_2$, but absent in $F_1$ that is only useful if the full bulk to boundary expression is available. This makes this regime somewhat restrictive but physically interesting nevertheless. 
Finally, we present the HHH regime in which all conformal dimensions are taken to be heavy. This is probably the most interesting problem to study in our set-up, as it involves finding a saddle point approximation on a complexified geometry. Interestingly, we will find that the correct saddle geodesics generically lead to a complexification of its proper time. Thus, they may not be able to be reinterpreted in terms of a curve in the complexified spacetime. This geometric reinterpretation, however, is not necessary for our purposes.

We remind the reader that the fact that $F_2$ relying on 3p functions, which are not diagonal in conformal dimensions, allows to consider $\Delta_E$ as a new parameter. Being a deformation on a 2p function in real time, we still take the both operator's conformal dimension to be equal $\Delta_1=\Delta_2=\Delta$. More concretely, we will always discuss and are naturally interested in $\Delta_E < 2 \Delta$ in order to consider the excitation still as a deformation close to the vacuum.

\subsubsection{HHL regime: half-bred geodesics}

\begin{figure}[t]\centering
\begin{subfigure}{0.49\textwidth}\centering
\includegraphics[width=.9\linewidth] {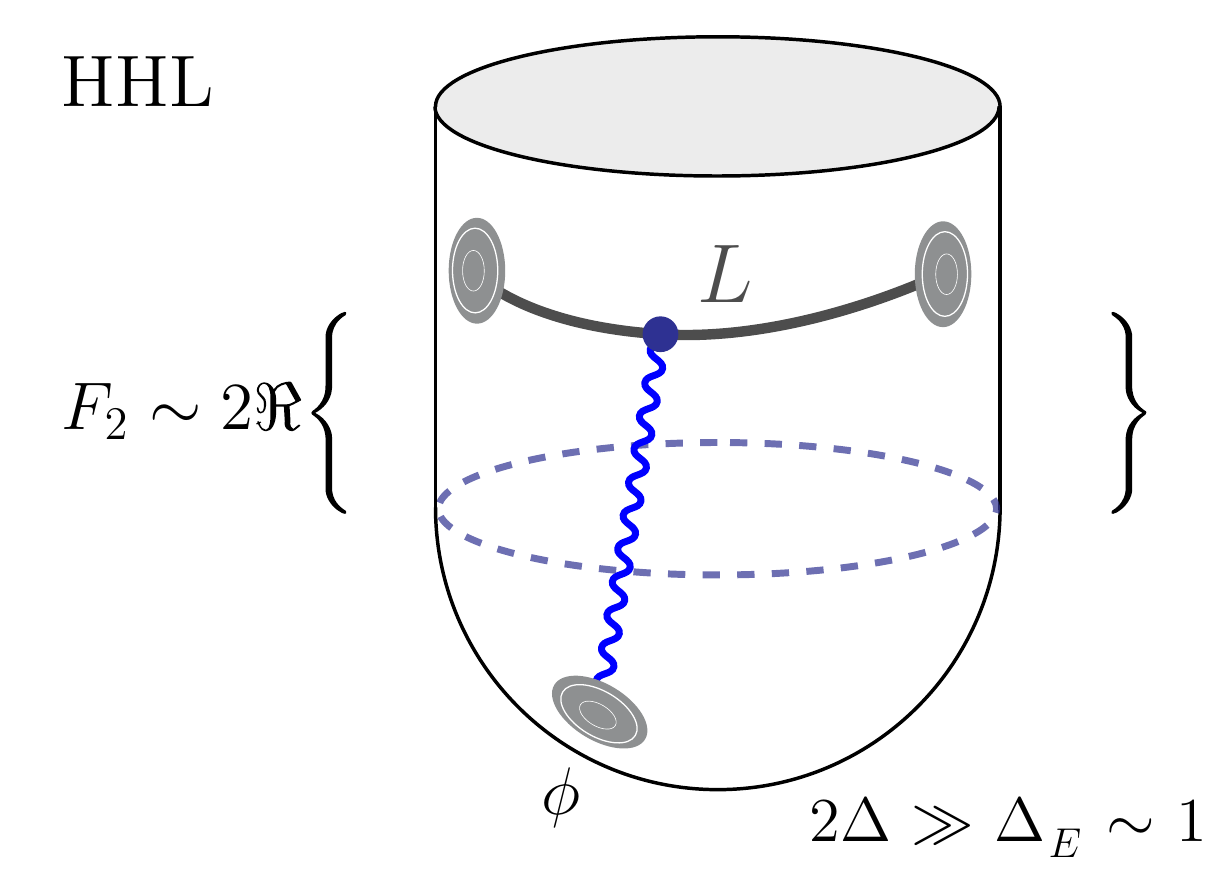}
\caption{}
\end{subfigure}
\begin{subfigure}{0.49\textwidth}\centering
\includegraphics[width=.9\linewidth] {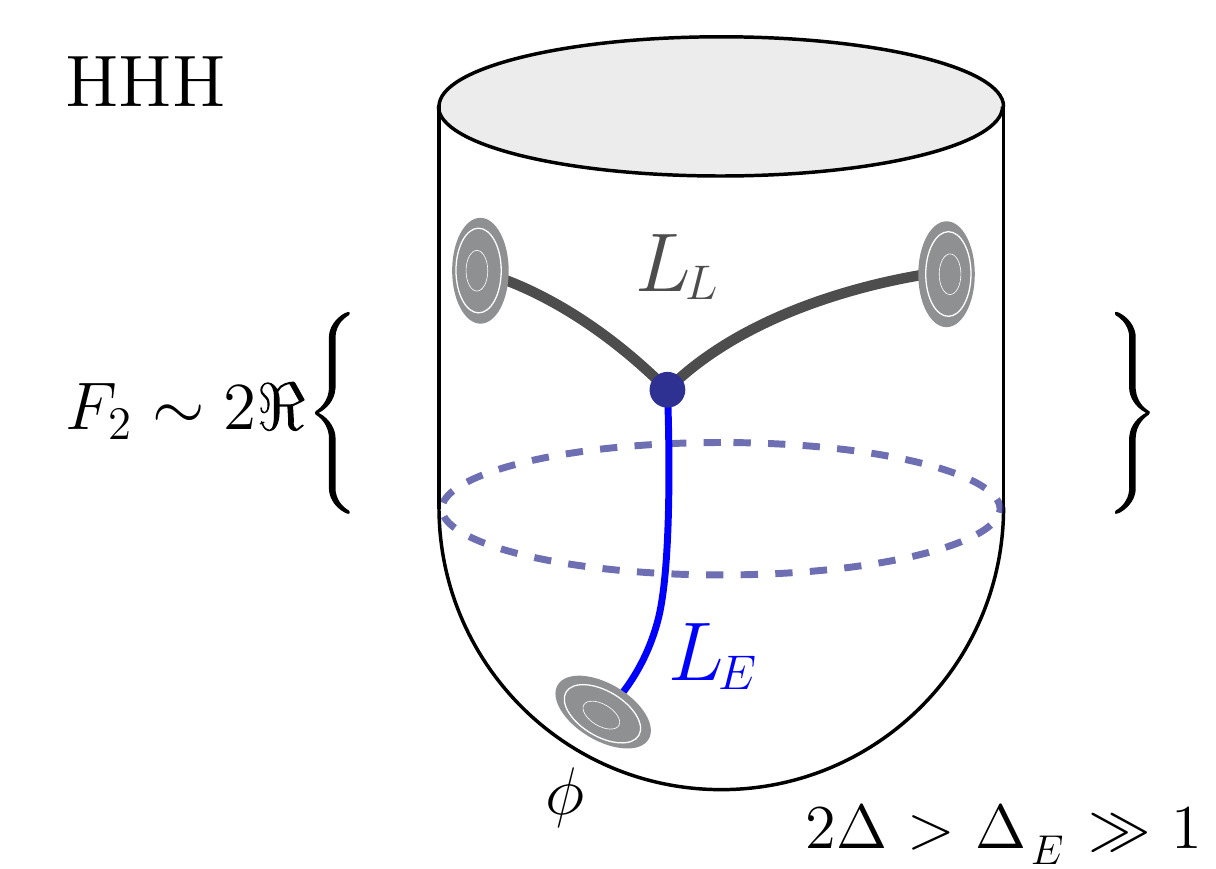}
\caption{}
\end{subfigure}
\caption{(a) We show a representation of $F_2$ in the HHL limit, when $2\Delta\gg\Delta_E\sim1$. In this limit, one computes the geodesic between the Lorentzian points and computes the convolution of its curve with the profile $\phi$ having the bulk to boundary as kernel. (b) When $2\Delta>\Delta_E\gg1$ one can also approximate the Euclidean leg by a geodesic. One then must compute a saddle between spacelike and time-like/Euclidean geodesics. The correct way to perform this computation usually involves loosing a ``curve on complex spacetime'' representation of the correlator, so the Figure should be taken as a pedagogical representation of the computation rather than an actual geodesics plot. }
\label{Fig:HHL}
\end{figure}

In this context, one could consider the regime $\Delta \gg 1$ but $\Delta_E\ll1$ so that the real time 2p function can be safely approximated by a geodesic, but the excitation must still be treated with the exact bulk to boundary correlator. This intermediate regime was explored in \cite{ZaremboHHL,CostaHHL} in Euclidean signature. The prescription for the relevant 3p function in this limit is
\begin{align}\label{Zarembo}
\langle {\cal O}_{\Delta}(t,\varphi_1){\cal O}_{\Delta}(t,\varphi_2){\cal O}_{\Delta_E}(\tau,\varphi_E) \rangle &\sim \langle {\cal O}_{\Delta}(t,\varphi_1){\cal O}_{\Delta}(t,\varphi_2) \rangle \times \int d\sigma K_{\Delta_E}(t(\sigma),r(\sigma),\varphi(\sigma);\tau,\varphi_E)
\end{align}
where $K_{\Delta_E}$ is the bulk to boundary correlator of a field of conformal dimension $\Delta_E$, connecting the asymptotic Euclidean boundary $\phi_E$ at $\{\tau,\varphi_E\}$ with all points in the bulk corresponding to the geodesic that approximates the heavy 2p function, i.e. all points $\{t(\sigma),r(\sigma),\varphi(\sigma)\}$, parametrized by $\sigma$. We show this schematically in Fig. \ref{Fig:HHL}(a).

The first step is to derive the spacelike geodesic equations from the Lorentzian metric in \eqref{AdS-Metrics}. 
In proper time $\sigma $ parametrization,
\begin{equation}
    +1=-(r^2+1)\dot t^2+\frac{\dot r^2}{r^2+1}+r^2 \dot \varphi^2 \qquad\qquad E=(r^2+1)\dot t\qquad\qquad J=r^2\dot \varphi\;.
\end{equation}
where $\dot \;$ denotes derivation wrt the proper parameter of the geodesic $\sigma$.
Aiming at a comparison with BH results as well as to simplify computations, we will consider a particular configuration of the real time insertions $\varphi_2=\varphi_1+\pi$. From the exact correlator \eqref{2p} one can readily see that $\langle {\cal O}_{\Delta} (t,\varphi_1) {\cal O}_{\Delta} (t,\varphi_1+\pi)\rangle$ is finite and constant. We now find the geodesic that connects these points and check that this is the case. Notice that $\varphi_2=\varphi_1+\pi$ can also be interpreted as a geodesic that has $J=0$ and that passes through $r=0$ and arrives at the other boundary. Since there is also no time evolution one could also propose $E=0$. We get
\begin{equation}\label{GeodEqAdS}
    \dot r^2 - r^2=1\;,\qquad E=J=0\qquad\Rightarrow\qquad r(\sigma)=\sinh(\sigma) \qquad \sigma\in\mathbb{R}
\end{equation}

The 2p function in this approximation is computed as $\langle {\cal O}_1 {\cal O}_2 \rangle \sim e^{-L_{reg}}$ where $L_{reg}$ is the regularized length of the geodesic above. The naive computation of the geodesic's length $\Delta\sigma=\int_{-\infty}^{\infty}d\sigma$ is naturally infinite, as the curve connects two points which are infinitely far apart in the bulk. A standard way to regularize these geodesics is to put a radial cut-off $r<R_c$, at a $\sigma_c$ and define $L_{reg}$ as the finite piece of this distance, i.e.
\begin{equation}
\Delta\sigma = 2\int_{0}^{\sigma_c}d\sigma=2\sigma_c=2\sinh^{-1}(R_c) \qquad\Rightarrow\qquad L_{reg} \equiv \Delta\sigma-2\sinh^{-1}(R_c)=0
\end{equation}
where we have noticed that the geodesic is symmetric with respect to $\sigma=0$ and thus its length can be also computed as 2 times its length up to $\sigma=0$. Since all geodesics computed in this work would be formally divergent, we will drop the $reg$ sub-index for the ease of notation from now on.
The correlator is thus $\langle{\cal O}_1{\cal O}_2\rangle \sim e^{-0}=1$ i.e. we have checked that the correlator is regular and constant for these boundary points. The fact that $L_{reg}=0$ in this regularization is just a matter of convention.

At this point we are ready to compute \eqref{Zarembo}, which in this case using \eqref{KK} and \eqref{GeodEqAdS} reduces to,
\begin{align}
F_2(t) &\sim 2\Re\left\{ \int d\tau d\varphi_E \int_{-\infty}^{\infty} d\sigma  \frac{\phi(\tau,\varphi_E)}{\left[\cosh(\sigma)\cos(t+i\tau)- \sinh(\sigma) \cos(\varphi_E)\right]^{\Delta_E}} \right\}
\end{align}
Notice that by integrating from $\sigma\in(-\infty,\infty)$ we are essentially capturing both $\varphi=0$ and $\varphi=\pi$ pieces of the geodesic. The $\sigma$ integral above can be analytically done for general $\Delta_E$, giving
\begin{align}
F_2(t)
&= 2\Re\left\{ \int d\tau d\varphi_E \frac{\phi(\tau,\varphi_E)}{[\cos(t+i \tau)^2-\cos(\varphi_E)^2]^{\Delta_E}}\right\}
\end{align}
which matches with \eqref{3p} for our points of interest. This reflects the fact that for pure AdS the geodesic approximation becomes exact. From this expression we can readily compute directly $F_2$ for any source profile we find of interest as we did in Sec. \ref{Sec:RelevantSources}.

\subsubsection{HHH: a geodesic warm up}
\label{Sec:AdSHHH}

As a final example in pure AdS, we take $1\ll\Delta_E<2\Delta$ limit, in which all contributions can be approximated by geodesics which meet at a point in the geometry as shown schematically in Fig \ref{Fig:HHL}(b). In this regime, also studied in Euclidean signature in \cite{CostaHHL}, the problem reduces to extremize with respect to the locus point location. Interestingly, we will see that the complex-signature nature of our set-up will make the locus point and the geodesic's proper length complex numbers. Thus we must understand Fig \ref{Fig:HHL}(b) more as a pedagogical drawing or a starting point for a quantity that may in fact loose a geodesic-as-a-curve geometric picture on its own with no detriment in its physical interpretation. 

Now, the problem of analytically finding the equilibrium locus in a given manifold is very hard in general (even for a pure Euclidean AdS) and it is beyond the scope of this work. 
We will thus pick again equal Lorentzian times $t$ and $\Delta\varphi=\pi$ but also a single delta-like insertion at $\tau=-\infty$. It is standard in AdS/CFT at zero temperature that this configuration corresponds to a single particle in the fundamental state. We will find that $F_2$ is defined in such a way that it can isolate the effect of this particle-like excitation. By symmetry, this problem should have an equilibrium point in the $r=0$ axis, so that the point is solely determined by the (perhaps complex) time $t_e$ at which they meet. The most interesting aspect of our approach is that, as shown in Fig \ref{Fig:HHL}(b), one should look for the intersection point between space-like and time-like (i.e the segment coming from the Euclidean piece after traversing $t=\tau=0$ surface) geodesics. In this sense, this is a rather unusual problem to solve in order to compute an observable. To be concrete, the minimization problem is 
\begin{equation}
    F_2\sim 2\Re\{\langle {\cal O}_{\Delta}(t){\cal O}_{\Delta}(t){\cal O}_{\Delta_E}(-\infty)\rangle\} \sim 2\Re\{ e^{- \Omega[t_e]}\}\qquad \Omega[z]\equiv \Delta \; L_{L} + \Delta_E\; L_{E}\qquad \Omega'[t_e]=0 
\end{equation}
where $L_{L/E}$ are the regulated lengths of the Lorentzian and Euclidean pieces which come from their respective boundary points up to $z\in\mathbb{C}$, see Fig. \ref{Fig:HHL}(b), and $z=t_e$ is the point that extremizes $\Omega[z]$. We stress that we denote by $L_L$ the total length of the two Lorentzian pieces summed. This notation is motivated by comparison with the $F_1$ computations in which $L$ denoted the full geodesic's length, see Fig. \ref{Fig:HHL}.

We begin by computing the Lorentzian geodesics that meet at $r=\varphi=0$ coming from the points $t$ in the boundary. These are geodesics similar to the ones studied in the subsection above, but now they have $E\neq0$ so that they arrive at $r=0$ at time $t_e\neq t$. The geodesics glued in this way meet at a cusp at $t=t_e$ as shown in see Fig. \ref{Fig:HHL}(b).
For a particle with energy $E$, we see that the radial solution is
\begin{equation}\label{AdSGeod}
    r(\sigma)=\sqrt{1+E^2}\sinh(\sigma)\qquad\qquad t(\sigma)-t=\arctan\left( \frac{E(1+\tanh(\sigma))}{1+E^2\tanh(\sigma)} \right)
\end{equation} 
such that the regulated length becomes
\begin{equation}
     L_{L}=\lim_{R_c\to\infty}\left( \Delta \sigma- 2\sinh^{-1}(R_c)\right) =-\ln \left(1+E^2\right)
\end{equation}
For these type of geodesics, and in general in this work, we will find that it will be more convenient to write the regulated length in terms of the energy $E$ instead of their meeting point $t_e$. 

The geodesic coming from the Euclidean segment requires some interpretation. By symmetry, it can be seen to consistently be sitting at $r=\varphi=0$. The metric with these restrictions becomes simply $ds^2=d\tau^2$, but one should begin at $\tau=-\infty$ and end at $\tau=it_e$. To make sense out of this problem, one should consider a holomorphic complexification of the metric in terms of a single complex variable $z=\tau+it$, such that
\begin{equation}\label{AdSEleg}
    ds^2=d\tau^2 \qquad\to\qquad ds^2=d(\tau+it)^2=dz^2 \qquad\Rightarrow\qquad \Delta\sigma = \Delta z = \Delta\tau+i\Delta t
\end{equation}
Notice that the proper distance of the geodesic has become complex, which immediately generates a tension with its interpretation as a curve. Notice however that this extension correctly reproduces pure spacelike/timelike geodeseic nature when $\Delta t=0$ or $\Delta \tau=0$ respectively. One could try to envision the result as two separate geodesics, one purely Euclidean and the other pure Lorentzian whose lengths are summed, but this alternative interpretation is not needed, nor guaranteed to be always possible in a more general scenario.
In App. \ref{CGeod} we review some arguments in favour of this analytic extension of the metric.

For our concrete example notice that $\Delta\tau$ generically comes from $\tau=-\infty$ up to $\tau=0$ so is actually infinite. This is just another manifestation of the asymptotic boundary being infinitely far away. Furthermore, this $\tau=-\infty$ up to $\tau=0$ geodesic that just falls to the AdS center is entirely equivalent to the $E=0$ spacelike geodesics that we computed in the last subsection, and thus in our regularization we get that this contribution is completely removed.
We are left with a finite contribution to $L_E=i\Delta t$ which is a geodesic beginning at the geometry's initial time (which can be taken to be zero without loss of generality) and ending at $t_e$, i.e.
\begin{equation}
L_{E}=i t_e= i(t+\arctan(E) )
\end{equation}
where we used eq. \eqref{AdSGeod} to relate $t_e$ to the energy of the $L$ geodesics.
The function to extremize becomes
\begin{equation}
    \Omega[E]=\Delta \; L_L+\Delta_E \; L_E=-\Delta \ln \left(1+E^2\right)+i\Delta_E(t+\arctan(E) )
\end{equation}
from where now we can find an extremum with respect to $E$, being a single (complex) variable problem,
\begin{equation}
    \Omega'[E]=0 \qquad\Rightarrow\qquad E=i \frac{\Delta_E}{2\Delta} \qquad\Rightarrow\qquad t_e = t +i \arctan\left(\frac{\Delta_E}{2\Delta}\right)
\end{equation}
Before studying the resulting $F_2$ some comments are due. 
Notice first that our solution is consistent with our analysis in the previous section, since we know that for $\Delta_E=0$ the geodesic follows an $E=0$, $t(\sigma)=t$ geodesic.
Notice that $t_e\in\mathbb{C}$ makes both the $L_L$ and $L_E$ to become complex by themselves. We emphasize this point because it is not only the Euclidean length that must be extended analytically to make sense, but also the Lorentzian legs become extended. We stress that this is no longer necessarily a set of 3 curves that meet at a point in a complexified bulk, even if for some cases there is a compatible reinterpretation of them as such. 
Our final result is,
\begin{equation}
    F_2(t)\sim2\Re\Bigg\{\frac{(1-\Delta_E/2\Delta)^{\Delta_E/2-\Delta}}{(1+\Delta_E/2\Delta)^{\Delta_E/2+\Delta}} e^{i \Delta_E t } \Bigg\}=A_{\Delta,\Delta_E} \cos(\Delta_E t)
\end{equation}
where $A_{\Delta,\Delta_E}$ is shorthand for the amplitude. Recall that we are always taking $\Delta_E<2\Delta$, so that the amplitude is free from singularities in our scenario.

Notice that $F_2$ is non-trivially able to capture and isolate exactly the excitation produced by a scalar particle of conformal dimension $\Delta_E$ in its fundamental state $\omega_{00}=\Delta_E$. This also matches the $\Delta_E\gg1$ limit of the HHL scenario obtained before, which we take as a check of our analytic extension of the  geodesics.

Notice also that $\delta$-like sources does not allow one to go further building all one-particle excited states semi-classically unless one is willing to compute higher point functions for operators of the form $\partial_\mu\dots\partial_\nu {\cal O}$. In that sense, our mode-skipping sources solve this complication by systematically avoiding poles from the 2p function at a semi-classical level. One could in principle also consider $n$-particle states by inserting more Euclidean legs and considering higher point vertexes in the bulk. These contributions should however be always subleading with respect to these, see \eqref{Fn2}.

Going back to our results in Sec. \ref{Sec:RelevantSources}, a $\delta$-like source at $\tau\to-\infty$ in $F_1$ also produce a single mode $\Delta$ of oscillation, but in that scenario one can only sense the excitation via an operator insertion of the exact same conformal dimension. Our result here shows that using $F_2$ one can study excited state effects on a thermalizing system using any operator in the theory.

This concludes the set of examples we wanted to present in pure AdS and mostly cover the full power of the presented formalism, albeit in a geometry were computations are simple enough. In our next sections, we apply the built intuition to less simple and physically more interesting scenarios.

\section{Case Study II: BTZ}
\label{BTZ}

In this section we consider the BTZ geometry \cite{BTZ} as a second case study in which analytic correlators are available. However, beyond exact computations, in this section we want to emphasize some aspects that we have not covered in detail above, and that are actually the main tasks in tackling a realistic scenario in our formalism. 

We note that different systems will require in general different SK paths consisting of many segments in order to study the $F_n$ observables. In terms of bulk duals, these segments manifests as the number of asymptotic boundaries of the complexified manifold. However, the problem of finding the topology of the manifold's interior is far from trivial in general.  
We will see that the two sided BH geometry being dual to a set of 2 entangled systems, requires a variation of the standard Thermal SK path \cite{HS} we present below. The BH geometry should be thought of as a high temperature dual of a finite temperature CFT, in the sense of the holographic Hawking-Page transition \cite{HP,WT}. This high temperature scenario manifests as a couple of highly entangled CFT systems and as wormholes connecting the entangled theories in the bulk. An analogous study to the one we will present in this section using a low temperature bulk dual in the fashion of a Thermal AdS geometry can also be carried out, but this would be mostly redundant after Sec. \ref{Sec:AdS}, the BH geometry will present a more interesting set-up.
From now on we will mainly focus on the geodesic approximation of the correlators, which is usually the only regime available in most scenarios.  

\subsection{SK path and Geometry}

The first step is to build the adequate SK path and geometry in which to study this scenario. Before doing so explicitly we make some comments. First, notice that we are now dealing with finite temperature $T\sim \beta^{-1}$ systems, so now the total Euclidean time evolution must be periodic $\tau\sim\tau+\beta$. In this sense, we are no longer in the In-In scenario. Moreover, we are also not in the standard Thermal path used in \cite{HS,VanRees}. This is because the standard Thermal scenario has only a single forward/backward time segment related to a boost-like time evolution \cite{VanRees,Rangamani19} rather than the global (Kruskal-like) time evolution that we are interested in studying.
Another set of SK paths filled up with BH pieces was given in \cite{us3,us4}, but their associated geometries also drop the BH interiors and consider TFD evolution, so they are not of our interest here. This discussion is intended to raise awareness of the plethora of possible SK paths at hand adequate to study different set-ups, but all of them concerning BHs. 

\begin{figure}[t]\centering
\begin{subfigure}{0.49\textwidth}\centering
\includegraphics[width=.9\linewidth] {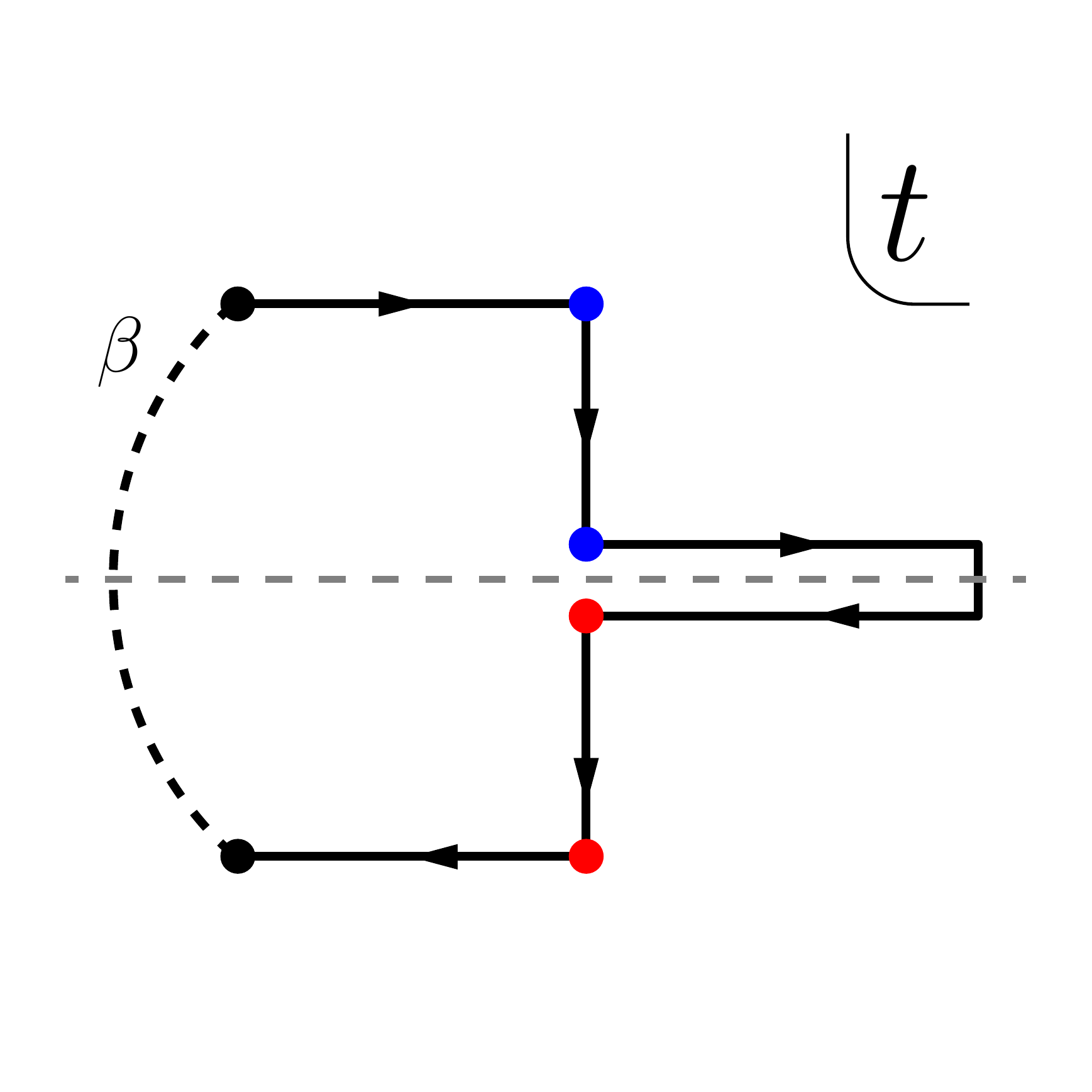}
\caption{}
\end{subfigure}
\begin{subfigure}{0.49\textwidth}\centering
\includegraphics[width=.9\linewidth] {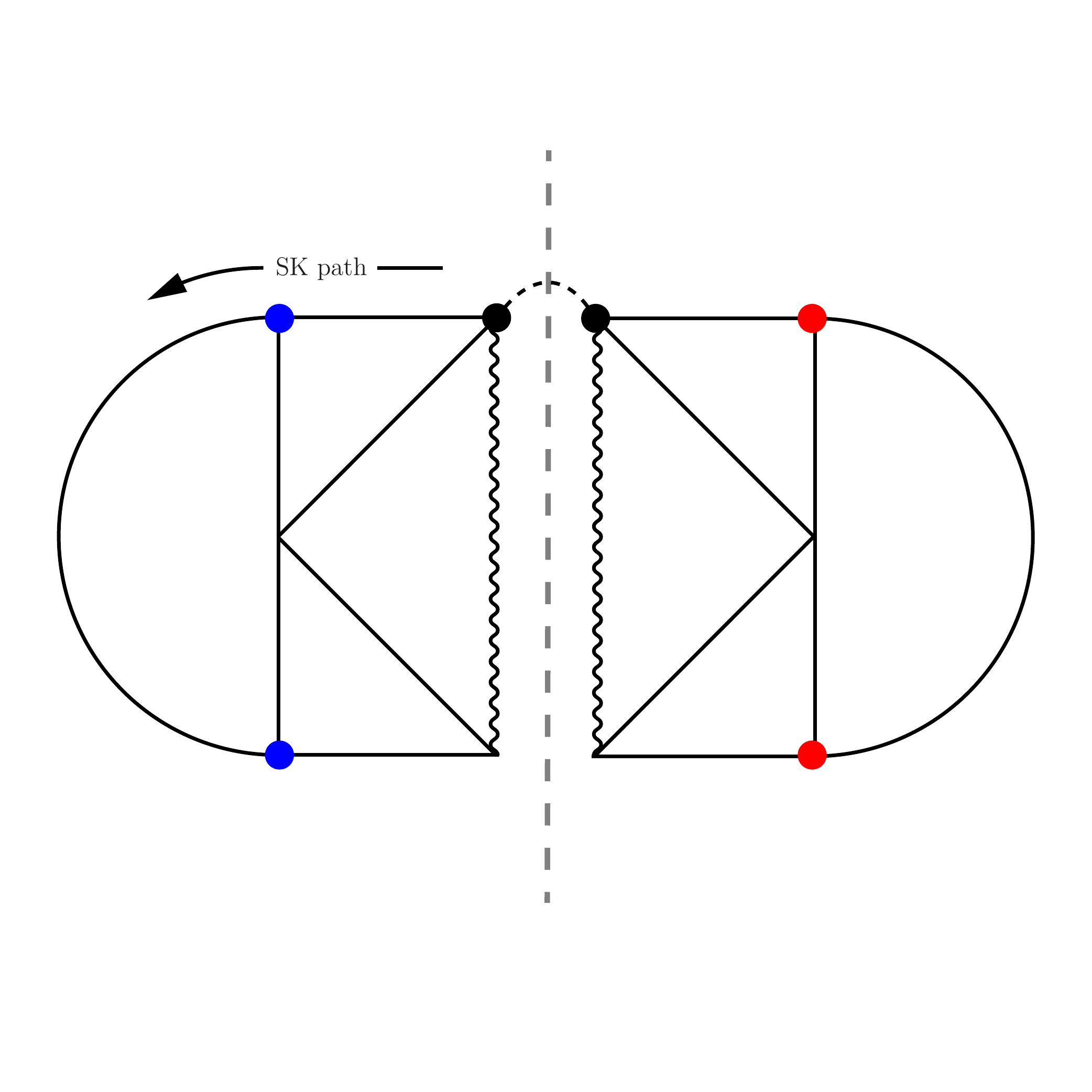}
\caption{}
\end{subfigure}
\caption{(a) The adequate SK path for a finite high temperature scenario is presented. This is essentially a standard Thermal path duplicated such that it can fit the number of asymptotic boundaries for a two-sided BH holographic dual. (b) The BTZ dual to the SK path on the left. The high temperature scenario is reflected in the bulk as wormholes connecting asymptotic boundaries through the bulk.}
\label{Fig:BH1}
\end{figure}

We focus on the path presented in \cite{SvRL}, which better fits our needs, shown in Fig. \ref{Fig:BH1}(a). Essentially, this path is a duplication of the standard Thermal scenario in order to adapt to the duplicated number of asymptotic boundaries in a two-sided BH geometry. Notice that, as in Fig. \ref{AdS-Geom}(a), the path can be made symmetric with respect to the real axis. The bulk dual of this path is built as follows. First, take the upper half of the path in Fig. \ref{Fig:BH1}(a). We start by associating a half Euclidean BH to the vertical piece. We will assign to both real time segments (both moving forward in time) a single upper half of a maximally extended BH geometry, ending up with a geometry much like the one in \cite{eternal}. This association is justified due to the local times in each exterior running in opposite directions, such that the asymptotic boundary of this geometry is seen to be consistent with the ordering in the SK path. 
Notice that the entanglement between the theories on each side makes itself manifest not only through the Euclidean segment but also through the wormhole using the holographic coordinate and that this connection is not present in the SK path. 
We assign to the second half of the path a mirror copy of the bulk we just described.
Finally, we need to glue each of these geometries between them to close the path, which we need to do at both final (global) time on each piece. One can think this gluing in two ways, both having the same limiting manifold. The first is to take a finite global time, gluing both copies and this gluing surface all the way to the singularity. The other works directly in exterior Schwarzschild patches and glues the copies across a finite (timelike) $r$ surface in the interior of the BH. One can then take the $r\to0$ limit. 
Both gluings lead to the geometry shown in Fig. \ref{Fig:BH1}(b) which explicitly contains regions behind the horizons and contains only asymptotic boundaries. 

With this manifold at hand, we can now proceed to study our family of observables $F_n$. To be concrete, we will foliate our manifold with exterior coordinates,
\begin{equation}\label{BTZ-metric}
    ds^2=\left.(r^2-r_s^2)\times\begin{cases}-dt^2\\+d\tau^2\end{cases}\hspace{-3mm}\right\}+\frac{dr^2}{r^2-r_s^2}+r^2d\varphi^2 \qquad r\in[r_s,\infty)\quad t\in\mathbb{R}\quad \{\varphi, r_s \tau\}\in[-\pi,\pi)
\end{equation}
with $r_s$ the Schwarzschild radius, and once again consider a massive scalar field \eqref{AdS-Action} over this fixed metric. For this geometry, both 2 and 3 point functions of the dual CFT can be also exactly computed both in Fourier and configuration space. For our purposes we would only need the 2p function, which is, for two operators on the same boundary, $t>t'>0$, \cite{Shenker02,us3}
\begin{align}\label{GRR}
\langle 0|{\cal O}_R (t,\varphi) {\cal O}_R (t', \varphi')|0\rangle=\frac{(\Delta-1)^2}{2^{\Delta-1}\pi}\sum_{j\in\mathbb{Z}}\left[\cosh((\varphi-\varphi')+2\pi r_s j)-\cosh(( t-t')(1-i\epsilon))  \right]^{-\Delta}\;,
\end{align}
Albeit known, we will not write the mode expansion of this expression, see \cite{SvRL,us3}. It suffices to say that it has single poles at the QN modes of the system which lie at all 4 possible signs of $\omega_{nl} = \pm i r_s (\Delta + 2n) \pm r_s |l| $. Since our manifold is analytic, all other correlators can be obtained from this one via analytic continuation following the SK path in Fig. \ref{Fig:BH1}. For example, an Euclidean-Lorentzian correlator can be obtained by moving $t'\to 0 - i\tau'$, the sign on the imaginary piece mandated by the SK path
\begin{align}\label{GRE}
\langle 0|{\cal O}_R (t,\varphi) {\cal O}_E (\tau', \varphi')|0\rangle=\frac{(\Delta-1)^2}{2^{\Delta-1}\pi}\sum_{j\in\mathbb{Z}}\left[\cosh((\varphi-\varphi')+2\pi r_s j)-\cosh( t+i\tau')  \right]^{-\Delta}\;,
\end{align}
where the $i\epsilon$ regulator is no longer necessary, and the correlator between two boundaries connected through a wormhole is obtained by performing a full $t'\to t'-i\pi$ 
\begin{align}\label{GRL}
\langle 0|{\cal O}_R (t,\varphi) {\cal O}_L (t', \varphi')|0\rangle=\frac{(\Delta-1)^2}{2^{\Delta-1}\pi}\sum_{j\in\mathbb{Z}}\left[\cosh((\varphi-\varphi')+2\pi r_s j)+\cosh( t-t')  \right]^{-\Delta}\;,
\end{align}
Notice that this last correlator does actually represent entanglement between the DOFs at R and L rather than travelling information since the points are always space-like separated and so, for example, light-cone singularities are no longer present. On this regard, notice that $t>0$ and $t'<0$ from the SK path so that $t-t'\neq0$ unless both are zero. A correlator at $t=|t'|$ has $\Delta t=2t$, a relation that will be useful in what follows.

A comment on N modes and QN modes in BTZ and general BH geometries might be clarifying. Given an initial state built from a HH Euclidean path integral, interpreted as initial data at global time $T=0$, one can then evolve the system with respect to 2 different Hamiltonians $H_\pm = H_R \pm H_L$. Here, $H_+$ corresponds to global time evolution whilst $H_-$ corresponds to a boost-like time evolution. The former, which is the one we are interested in this work, generates a global evolution and is not a Killing vector field, since all information ends up at the singularity for an eternal BH. In this scenario, of course, no N modes exists and the physical system relaxes via QN modes that decay in time. On the other hand, the latter is in fact a Killing vector whose action leaves the HH state invariant. Under this second evolution, one can in fact build its corresponding set of N modes as in \cite{Kenmoku,us4}. As we stated, though, the HH state can be thought of as an initial state configuration for both physical scenarios. The bottom line of this discussion is that in our current set-up we are studying the HH state as an initial condition for the $H_+$ Hamiltonian, whose Hilbert space contains only QN modes. 

Before moving on, we reiterate that the geometry in Fig. \ref{Fig:BH1}(b) corresponds to a high temperature geometric dual of our SK path in the light of a holographic HP transition \cite{WT}, and that a low temperature dual would correspond to an SK path filled with pure AdS segments, analogous to the one explored in \cite{us3}. The analysis of this saddle is very similar to the pure AdS example studied in the previous section, and thus less interesting, albeit some interesting discussion arises in comparing the correlators obtained in both saddles, see \cite{us3}. 

\subsection{Study of $F_1$}

For this discussion we will first make use of full correlator \eqref{GRE} to study some relevant source profiles along the lines of Sec. \ref{Sec:RelevantSources} and then we will concentrate on the geodesic approximation of $F_1$.
The discussion on mode skipping in BTZ in the geodesic approximation will allow for an interesting analysis on its nature as an orbifold of pure AdS$_3$.

\subsubsection{Relevant sources}

The only major change with respect to the pure AdS scenario is that the Euclidean sections are no longer infinite but rather have length $r_s\tau\in[-\pi,0]$ and $r_s\tau\in[0,\pi]$ respectively. Notice also no N modes are present, for no (global) $n$-particle states exist in this set-up\footnote{In the boost-like time evolution, a delta-like source at $r_s \tau=-\pi/2$ should actually represent a single particle of that real time system related to $H_-$. That analysis, however is beyond the point of this work and requires careful regularization near the horizon to correctly define the physical set of N modes, see \cite{Kenmoku}}, see discussion in the paragraph below eq. \eqref{GRL}. 

\begin{figure}[t]
\begin{subfigure}{0.49\textwidth}\centering
\includegraphics[width=.9\linewidth] {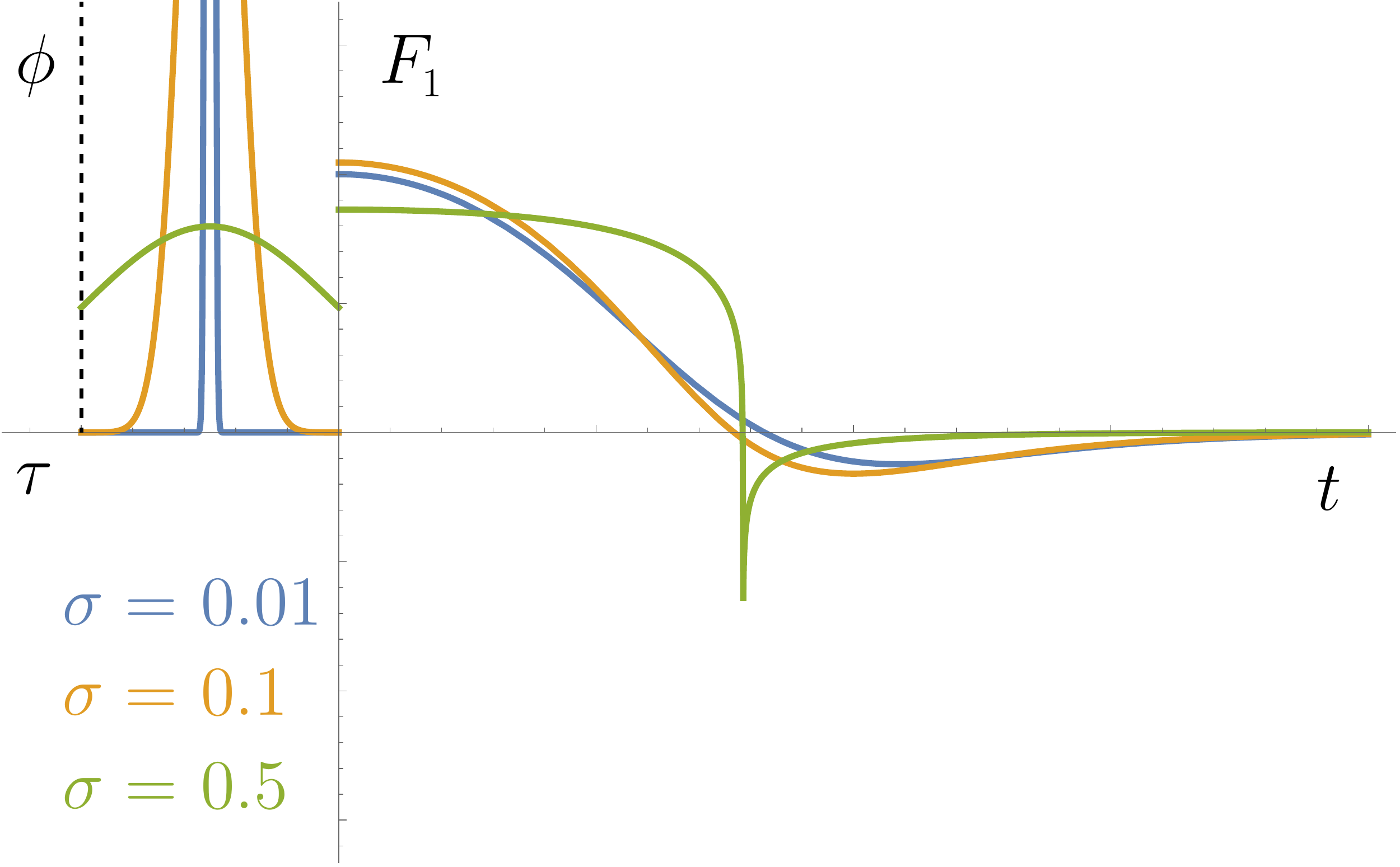}
\caption{}
\end{subfigure}
\begin{subfigure}{0.49\textwidth}\centering
\includegraphics[width=.9\linewidth] {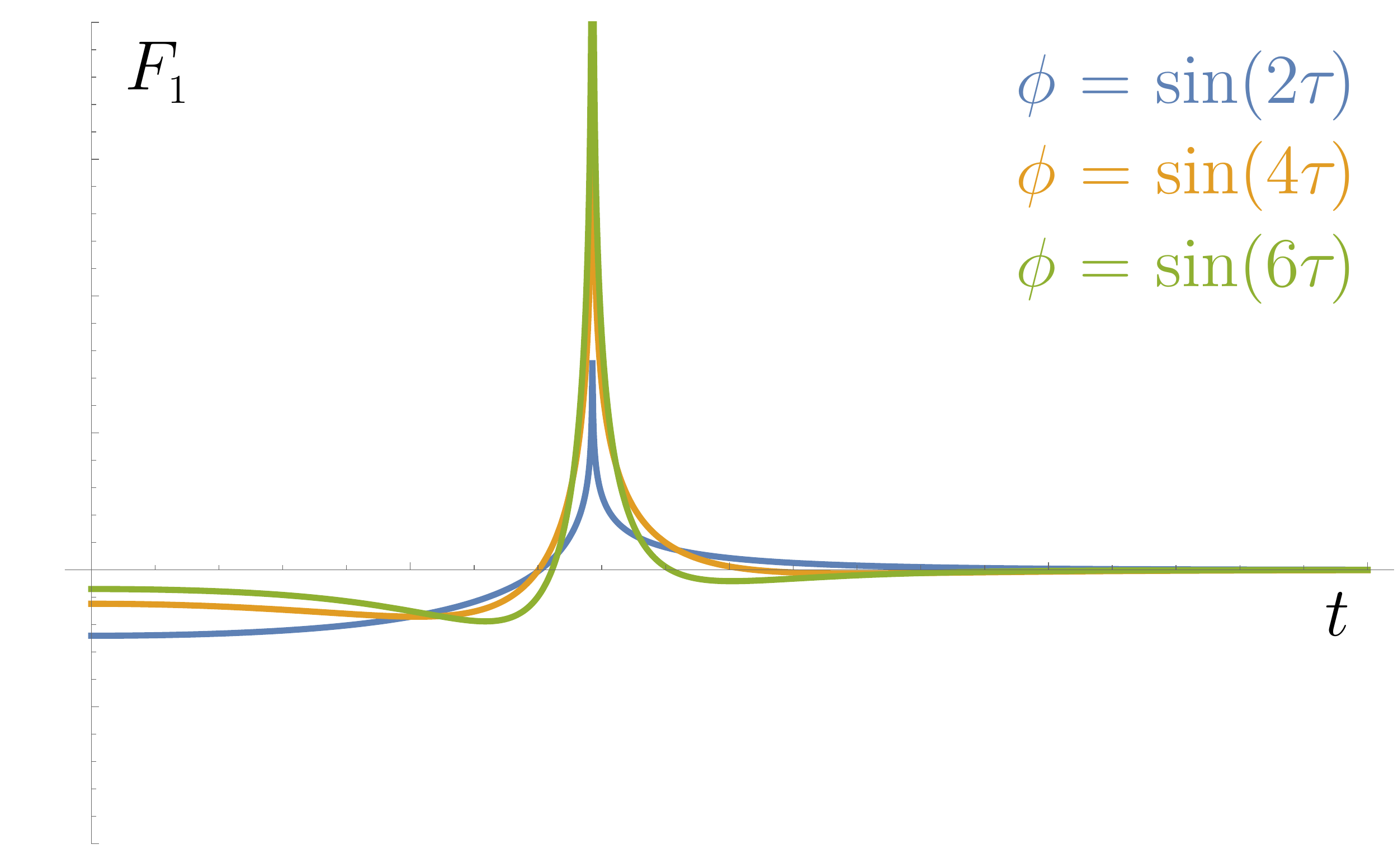}
\caption{}
\end{subfigure}
\caption{We plot $F_1$ for some relevant sources profiles for the BTZ scenario. In (a) we take Gaussians of different width shown to the left and plot $F_1$ in the positive axis. We consider the limit in which the source is a Dirac delta $\sigma\to0$. The limit can be seen to be finite. One should regard the negative piece of the horizontal axis as $\tau$ and the positive piece as $t$. Notice that there is only a finite domain for $r_s\tau\in[-\pi/2,0]$ and that if the Gaussian is too fat, $\phi(0)\neq0$ and the system responds with divergences in finite time. In (b) we different single frequency modes as sources. Despite meeting $\phi(0)=0$ these modes not only do not produce single mode responses, which matches expectation, but also produce divergences at finite time in the response function.}
\label{Fig:sourcesBTZ}
\end{figure}

Besides the exponential decaying amplitude of the response due to the QN modes, most of the lessons we learnt from relevant sources in the pure AdS examples extrapolate to this scenario also. For starters, a Gaussian profile can be made arbitrarily thin and still yield a finite response from the system. The source can be moved around producing also finite responses, albeit no analog of the $\tau\to-\infty$ limit in Sec. \ref{Sec:RelevantSources}, as we explained above. One can also see that the relevant domain in which to insert delta-like excitations is $r_s\tau\in[-\pi/2,0]$ since going beyond $-\pi/2$ can be reproduced by exchanging the L and R wedges.
Some examples are presented in Fig. \ref{Fig:sourcesBTZ}(a), where one sees that it is important for the source to turn off at $\tau=0$ to avoid singularities in the response function in real time. 

As for single mode profiles, we must consider $\phi(\tau)\sim\sin(n\tau)$ with integer $n$ such that $\phi(0)=\phi(-\pi)=0$. 
The response of the system for some frequencies can be seen in Fig. \ref{Fig:sourcesBTZ}(b). 
Finally, a constant source once again lifts the initial field configuration by a constant, indicating a vacuum state of a different theory. 

\subsubsection{Geodesic approximation}

As we said before, the BTZ geometry still allows for exact computation of $F_1$ in this semiclassical limit, since the relevant propagator \eqref{GRE} is known. However, we are more interested on showing the general approach to build the $F_1$ so that we will focus on the geodesic approximation. The lesson to take here is that, as we learnt in Sec. \ref{Sec:AdSHHH}, our SK complex signature demands we take all our parameters as complex variables. Specifically, we will find that the correct way to find these geodesics that begin in an Euclidean point and travel to a real time requires the energy of the geodesic to be complex. We will once again find that the interpretation of the resulting geodesic as a curve on a complexified manifold may not be possible.

We will now solve for a geodesic starting in an Euclidean point $\tau$ and fix the solution parameters such that travels up until a real time $t$. For simplicity, we will restrict ourselves to geodesics with no angular momentum. The geodesic equations with $J=0$ for the Euclidean metric \eqref{BTZ-metric} are
\begin{equation}\label{GeodEqBTZ}
    +1 = \frac{E^2}{r^2-r_s^2}  + \frac{\dot r^2}{r^2-r_s^2}\qquad\qquad E=(r^2-r_s^2)\dot \tau\qquad\qquad J=r^2\dot \varphi=0
\end{equation}
with solutions
\begin{equation}\label{GeodSolBTZ}
    r(\sigma)=\sqrt{r_s^2+E^2}\cosh \sigma \qquad\qquad r_s\left(\tau(\sigma)-\tau\right)=  \arctan \left( \frac{r_s}{E}  \right) + \arctan\left( \frac{r_s}{E}\tanh\sigma \right) 
\end{equation}
where $\sigma\in\mathbb{R}$ is the affine parameter and we have already fixed the geodesic such that $\tau(-\infty)=\tau$. All is left to do now is to fix our only free parameter $E$ to be such that $\tau(+\infty)=0 + i t$, i.e. a point in the R wedge according to the SK path. A correlator to the L wedge requires $\tau(+\infty)=-\pi r_s + i t$.
The easiest way to do this is to relate first the energy with $\Delta t= t+i\tau $ by forcing $\tau(+\infty)=0 + i t$ which leads to
\begin{equation}
    r_s(i t-\tau) =i r_s\Delta t = 2 \arctan \left( \frac{r_s}{E}  \right) \qquad\Rightarrow\qquad E=-i r_s \coth \left(\frac{r_s\Delta t}{2}  \right)\in\mathbb{C}
\end{equation}
As a consistency check, notice that for $t=0$ the energy becomes real again and for $\Delta t\to0$, which is a geodesic returning to the original point, the energy diverges as it should for such a process. 

The correlator now is given by the exponential of the (regulated) proper length of this complex geodesic, written in terms of $\Delta t$. In a similar fashion as in Sec. \ref{AdSF2}, this proper length can be computed considering that $r(\sigma)$ in \eqref{GeodSolBTZ} is an even function of $\sigma$, so that the regulated length is just twice the length 
up until $\sigma=0$, i.e.
\begin{equation}
    L = \Delta\sigma - 2\ln(R_c)\sim  \log\left(\frac{4R_c^2}{E^2+r_s^2}\right)- 2\ln(R_c) = \log\left(\frac{4}{E^2+r_s^2}\right)
\end{equation}
where $R_c$ is once again a regulator distance in the asymtotic AdS boundary and we have chosen to subtract $2\ln(R_c)$, i.e. only the divergent piece in the $\sigma(R_c)$, $R_c\gg1$. This leads to the correlator
\begin{equation}\label{BTZ2pgeod}
    \langle {\cal O}(t,\varphi){\cal O}(\tau,\varphi)\rangle\sim e^{-\Delta \;L} \sim \frac{1}{[1-\cosh(r_s(t+i\tau))]^{\Delta}}
\end{equation} 
which matches the leading term in \eqref{GRE}.
In this approximation, we can put sources spread in $\tau$ but fixed at $\Delta\varphi=0$. A more general $J\neq0$ study of geodesics can more generally recover the full $\Delta\varphi$ analysis \cite{Shenker02}. We finally get from \eqref{BTZ2pgeod} an $F_1$ of the form
\begin{equation}
    F_1 (t)\sim 2\Re \int_{-\infty}^0 d\tau d\varphi_E  \phi(\tau,\varphi_E)  \langle {\cal O}(t,\varphi){\cal O}(\tau,\varphi_E)\rangle 
    =2\Re \int_{-\infty}^0 d\tau  \frac{\phi(\tau)}{[1-\cosh(r_s(t+i\tau))]^{\Delta}}
\end{equation}
where we have made $\phi(\tau,\varphi_E) =  \phi(\tau)\delta(\varphi_E-\varphi)$, i.e. fixed at the same angular position of $ {\cal O}(t,\varphi)$, and kept a general $\phi(\tau)$ profile. 

An interesting analysis comes from skipping the first mode $\omega_{00}=i r_s \Delta$ with the source 
\begin{equation}
    \phi(\omega)= (\omega^2+(i r_s \Delta)^2)e^{i\omega \epsilon} \qquad\Rightarrow\qquad \phi(\tau)= (-\partial_\epsilon^2- r_s^2 \Delta^2) \delta(\tau+\epsilon)
\end{equation}
which successfully avoids the first $e^{-r_s(\Delta+1) t}$ mode but leads to
\begin{equation}
    F_1 \sim 2\Re \int_{-\infty}^0 d\tau  \frac{\phi(\tau)}{[1-\cosh(r_s(t+i\tau))]^{\Delta}}\sim e^{-r_s(\Delta+1) t}+\dots
\end{equation}
which does not match with any other BTZ QN mode $\omega_{nl}=\pm i r_s (\Delta+2n)\pm r_s |l| $.
The reason for this lies in the geodesic approximation and the close relation between BTZ and AdS$_3$. Notice that as it stands, eq. \eqref{BTZ2pgeod} can be reinterpreted in itself as a Wick rotation in $\tau$ of the pure AdS$_3$ with $\Delta\varphi=\pi$. This mathematical identity arising in this approximation of the BTZ correlator result in spurious modes appearing rather than the physical BTZ modes. Notice that the pure AdS Wick rotated correlator has poles at $\omega=\pm i r_s(\Delta+2n+|l|)$, which our method satisfactorily reproduces and in this computation have effectively replaced the BTZ QN modes.
Fortunately, in this scenario we can explicitly perform the integral in the mode expansion of the exact correlator \eqref{GRR} to check that our prescribed source actually skips the first QN mode successfully and does not introduce any unwanted modes. As stated, this is a problem with the specific BTZ example being deeply connected with the pure AdS$_3$ geometry and should not arise in higher dimension examples.  

In a more pragmatical approach, suppose one has skipped the fundamental mode and has a leading mode $\tilde \omega$ in $F_1$ in a certain approximation. If in doubt if $\tilde \omega$ is an actual QN mode of the system, we propose the following. Recall that all information on $\phi$ but its zeroes on the actual QN modes only modifies the initial condition on the specific coefficients in front of each mode, but not it presence or absence. If an independent method is available to check whether a candidate QN frequency is actually a good QN for a given system or not, one may take the leading $\tilde \omega$ mode as a candidate QN mode and test it. If the test fails, one can then add an ad-hoc correction to the source such that removes both the previous mode and $\tilde \omega$ as well. In our BTZ example,  $\phi(\omega) \sim (\omega^2 - r_s^2 \Delta^2) (\omega^2 - r_s^2(\Delta+1)^2) $ would do, removing the unphysical behaviour and falling onto a physical mode $\omega=r_s(\Delta+2)$. Given that the $\tilde \omega$ mode is unphysical, the $ (\omega^2 - \tilde \omega^2) $ zero is guaranteed only to affect the precise coefficients in front of each mode, but will not avoid nor create any other mode. In this sense, we are profiting from the ambiguity in the sources $\phi$ in providing a particular set of initial conditions. This is in line with the ideas in \cite{Belin20}.

A more detailed analysis on this is beyond the scope of this work as we emphasize that the heart of our approach is being able to skip particular sets of QN modes, which we have done successfully.

\subsection{Study of $F_2$}

\begin{figure}[t]\centering
\includegraphics[width=.55\linewidth] {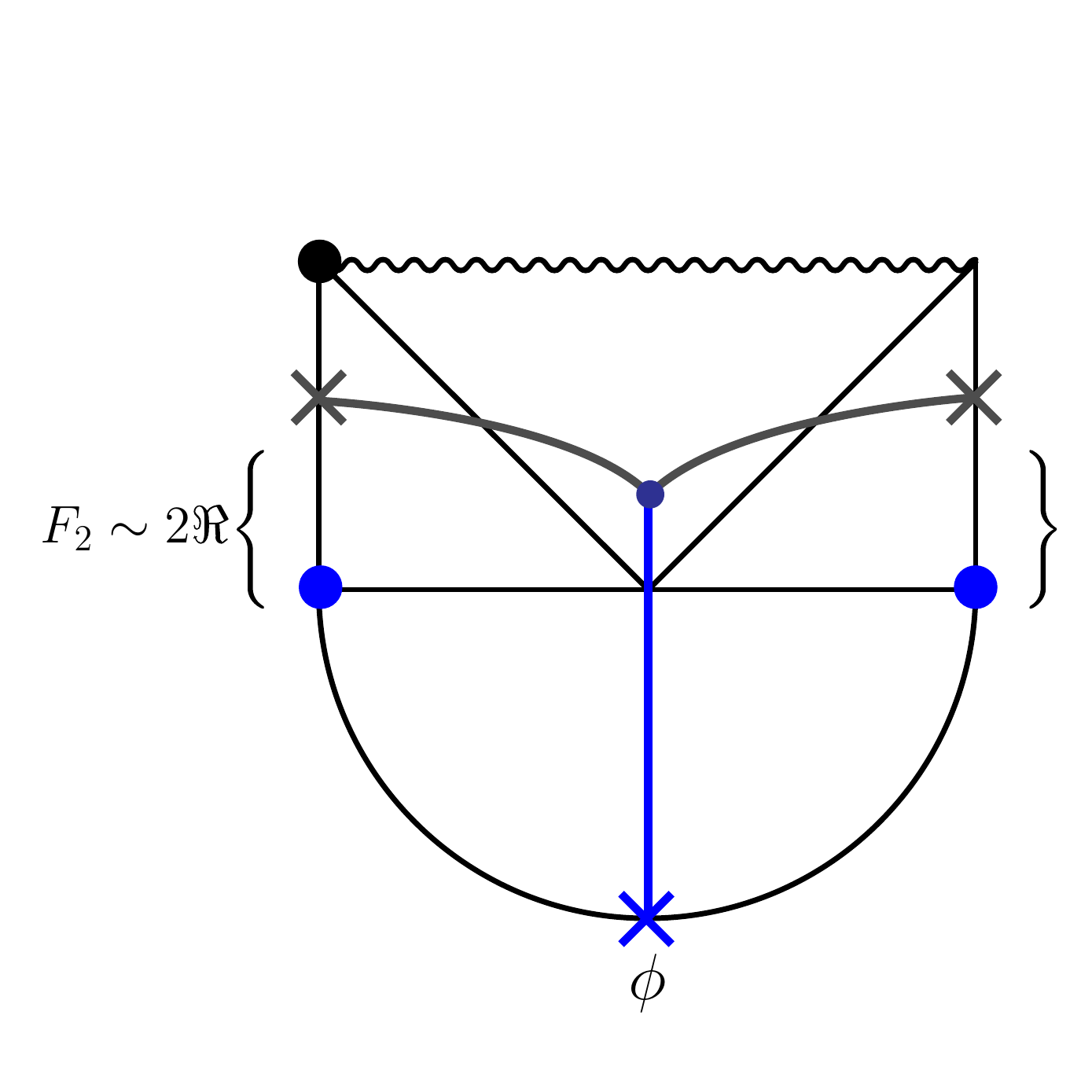}
\caption{We show a schematic representation of the $F_2$ observable in the geodesic approximation for the BTZ scenario. The effect of the Euclidean leg in blue will ultimately deform the geodesics away from the diagram. Furthermore, in general, a geodesic interpretation of the correlator in this limit may not allow a ``curve in complex spacetime'' interpretation. The representation must then be taken more as a pedagogical drawing rather than a representation of the correlator as geodesics in spacetime.}
\label{Fig:ShenkerBTZ}
\end{figure}

We now proceed to study $F_2$ in a geodesic approximation. As in Sec. \ref{Sec:AdSHHH}, we will not consider insertions of sources $\phi(\tau)$ at generic points $\tau$. This is because finding the intersecting point in the bulk for the geodesics can become a quite non trivial problem, albeit always possible to solve, at least numerically. We will thus consider only a delta like $\phi(\tau)$ source at $r_s\tau=\pi/2$ and Lorentzian points in symmetric points at $\pm t$ in the R/L boundaries respectively, $\Delta t=2t$, all angular points $\varphi$ being identical. Notice that we need to find the intersection point between two spacelike and one timelike geodesic, which will inevitably lead to a complex result. A representation for the geodesic we are after is shown in Fig. \ref{Fig:ShenkerBTZ}.

We begin our study of these geodesics with the Euclidean one, which is the simplest. Looking again at \eqref{GeodSolBTZ} and noticing that the geodesic we are after has $\dot\tau=\dot t=0$ by symmetry, so we see that we can express its length $L_{E}$ as, see metric \eqref{BTZ-metric},
\begin{equation}\label{BTZF2E}
    L_E 
    = \int_{R_c}^{r_e} \frac{dr}{\sqrt{r^2-r_s^2}}-\ln(R_c)
    = \ln(2) +i \arccos(r_e)
\end{equation}
where we defined $r_e<r_s$ as the intersection point lying in principle inside the horizon.

\begin{figure}[t]\centering
\begin{subfigure}{0.49\textwidth}\centering
\includegraphics[width=.9\linewidth] {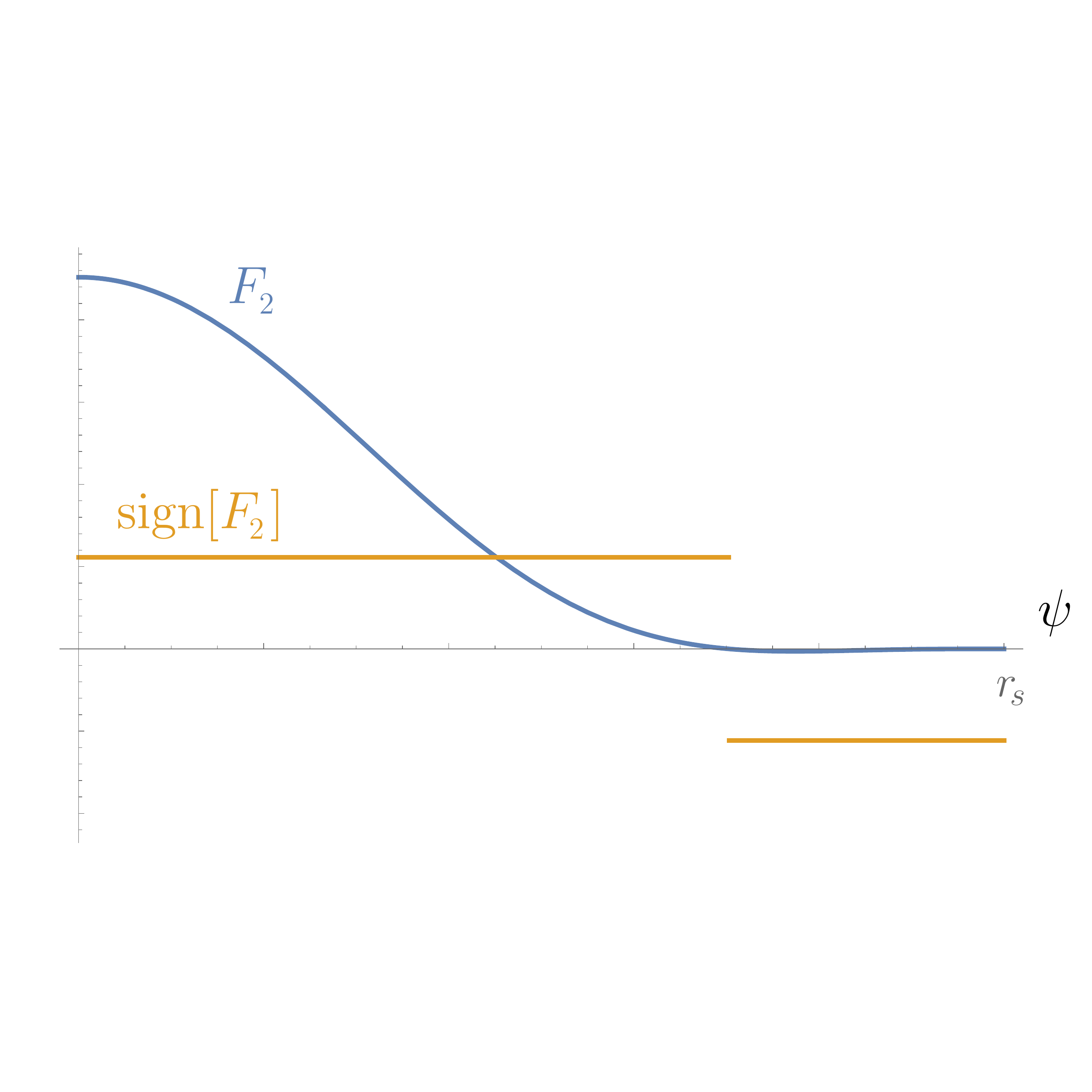}
\caption{}
\end{subfigure}
\begin{subfigure}{0.49\textwidth}\centering
\includegraphics[width=.9\linewidth] {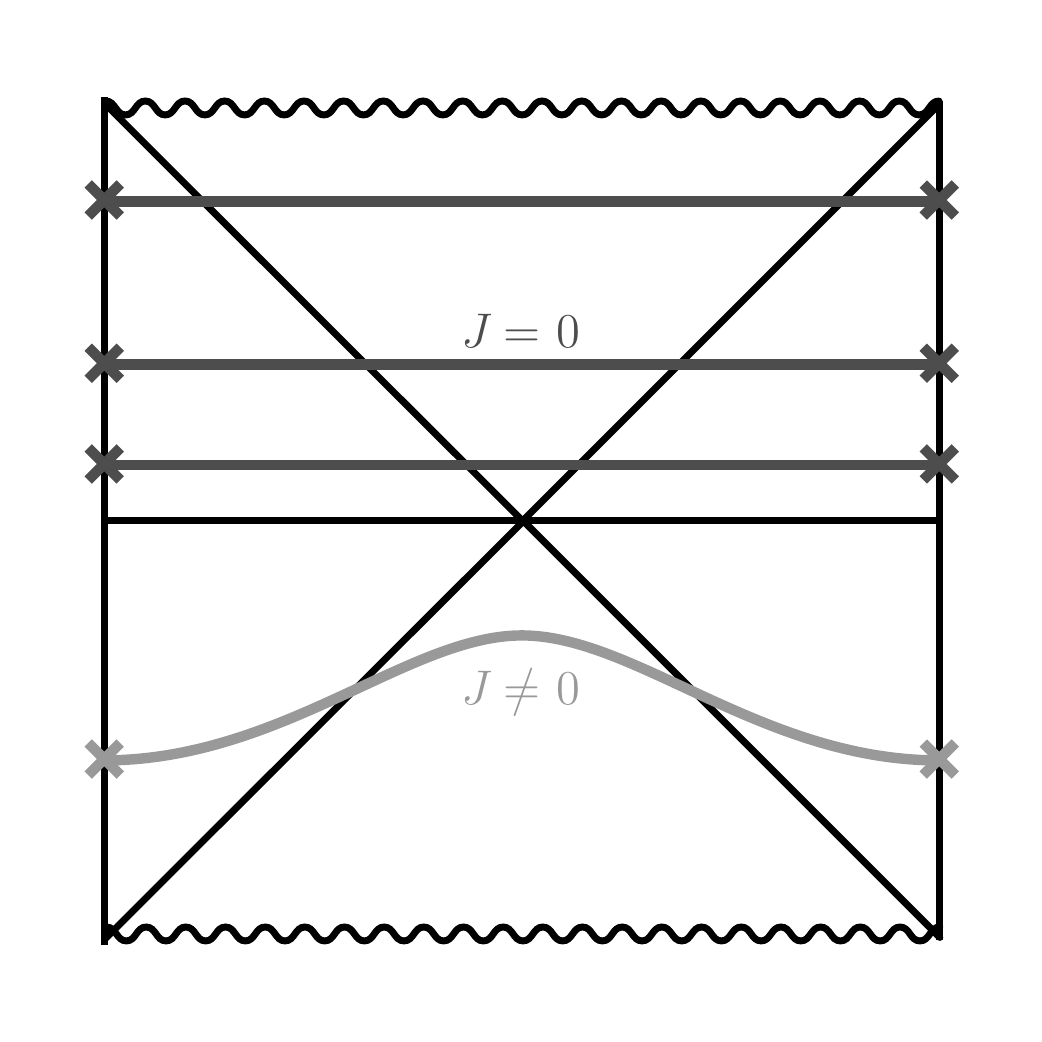}
\caption{}
\end{subfigure}
\caption{(a) A plot of a generic result for $F_2$ in BTZ is presented. Notice that the fact that the first QN mode is pure imaginary results in the system only admitting a single absorption and re-emission process throughout the BHs history. Notice that since the horizontal axis is $\psi\in[0,r_s]$ and describes the complete BH history. (b) We plot some geodesics in the BTZ BH in dark grey to show that the $\dot t=0$ are also $\dot T=0$ curves at  $J=0$. Of course, for $J\neq0$ the geodesics bend, as we show in a lighter tone of grey below.}
\label{Fig:BTZHHH}
\end{figure}

The symmetric spacelike geodesics fixed at boundary time $t$ must have equal length by symmetry, so we can focus only on the one in R. The BTZ spacelike geodesics for the real time metric \eqref{BTZ-metric}, are
\begin{equation}
      r(\sigma)=\sqrt{r_s^2-E^2}\cosh \sigma \qquad\qquad r_s\left(t(\sigma)-t\right)= - \tanh^{-1} \left( \frac{r_s}{E}  \right) - \tanh^{-1}\left( \frac{r_s}{E}\tanh\sigma \right)
\end{equation}
It is important for consistency to notice that we have chosen the $t$ coordinate on the R wedge (which we have glued to the $\tau=0$ surface) to have no imaginary piece, i.e. $t(-\infty) = t \in \mathbb{R}$. Having fixed this, notice that the BH interior has $\Im[t]=\tau=-\pi/(2 r_s)$ and the L wedge has $\Im[t]=\tau=-\pi/r_s$, which is consistent with gluing it to the $\tau=-\pi/r_s$ on the other half of the Euclidean BH as we have done, cf. with the SK path in Fig. \ref{Fig:BH1}(a). As this contributions are fixed, we will disregard them when writing, for example $\Delta t=2t$ for the R to L correlators. To compute this geodesic's length, we must now find which is the correct $\sigma_0$ such that $\Re[t(\sigma_0)]=0$, leading to
\begin{equation}\label{psi}
    \sigma_0=-\tanh ^{-1}\left(\frac{E (\psi-E )}{r_s^2-E \psi }\right) \qquad\qquad \psi\equiv r_s \tanh(t)\in[0,r_s]
\end{equation}
where we found convenient to reparametrize our initial time so that we have a parameter in a finite domain.
One can quickly check the expression is correct by considering the cases $E=0$, $\Delta t=0$ and $E=\psi$, $\Delta t=2t$ which are the only scenarios in which $\sigma_0=0$, i.e. the geodesics is symmetric in the vertical axis. We will come back to this geodesics after we estimate $F_2$. The regulated geodesic length is
\begin{equation}
    L_{L}=\Delta \sigma -\ln(R_c)=\ln \left(\frac{4r_s^2}{r_s^2-E^2}\right)-2\tanh ^{-1}\left(\frac{E
   (E-\psi )}{r_s^2-E \psi }\right)
\end{equation}
The point at which the geodesics intersect can be also rewritten in terms of the energy as
\begin{equation}
    r_e=\frac{r_s^2-E \psi }{\sqrt{E^2-2 E \psi +r_s^2}} \qquad\Rightarrow\qquad L_{E}=\ln(2) +i \arccos\left(\frac{r_s^2-E \psi }{\sqrt{E^2-2 E \psi +r_s^2}}\right)
\end{equation}
Putting everything together and disregarding the $\ln2$ in $L_E$ which plays no physical role, one finds that the quantity to minimize is
\begin{equation}
    \Omega(E)=\Delta L_{L}+\Delta_E L_{E}=  \Delta  \left(\ln \left(\frac{4r_s^2}{r_s^2-E^2}\right)-2\tanh ^{-1}\left(\frac{E
   (E-\psi )}{r_s^2-E \psi }\right)\right)+ i\Delta _3 
   \arccos\left(\frac{r_s^2-E \psi }{\sqrt{E^2-2 E \psi +r_s^2}}\right)
\end{equation}
which has an extremum at
\begin{equation}
    \Omega'[E_0]=0 \qquad\Rightarrow\qquad E_0=\psi - i\frac{ \Delta _3 }{2 \Delta } \sqrt{r_s^2-\psi ^2}
\end{equation}
Below, we will check that the limit $\Delta _3\to0$ leading to $E=\psi$ is the correct saddle in the vacuum. We get for $F_2$,
\begin{equation}
    F_2\sim 2\Re\{\langle {\cal O}_{\Delta}(t){\cal O}_{\Delta}(t){\cal O}_{\Delta_E}(-\pi/2)\rangle\} \sim 2\Re\{ e^{- \Omega[E_0]}\}
\end{equation}
whose generic profile is shown in Fig. \ref{Fig:BTZHHH}(a). 
The response of the system to an excited profile shows both absorption and emission up until the system finally relaxes completely at $\psi=r_s$, $t=+\infty$. This single absorption and re-emission process is related to the fact that the first BTZ QN mode is pure imaginary $\omega_{00}=ir_s\Delta$. This will not be the case for higher dimensional BHs below. Notice that since we have kept the background fixed, the BH is not allowed to grow during the process, and this computation should be complemented with a backreaction analysis if one is to make precise predictions, which is beyond the scope of this work. 

\paragraph{Dire Straights:} \;

Before moving on to BHs in $d+1>3$, we make a small comment on these geodesics in the $\Delta_E\to0$ limit, i.e. simple spacelike geodesics that cross from R to L at opposite times as in \cite{Shenker02}. Albeit perhaps minor, the authors have not found this observation stressed enough in the literature. The solution to the geodesic equations for symmetric geodesics $\Delta t=2t$, $J=0$, are
\begin{equation}
    r(\sigma)=\sqrt{r_s^2-E^2}\cosh \sigma \qquad\qquad r_s t(\sigma)=- \tanh^{-1}\left( \frac{r_s}{E}\tanh\sigma \right)+i\frac \pi 2
\end{equation}
where much like in \eqref{psi} the $+i\frac \pi 2$ factor fixes the quantity $r_s t(\sigma)$ to be real on R, $\sigma\to-\infty$. Now, notice that $E$ and the initial/final boundary times are related, and actually using again the definition in \eqref{psi} it can be seen that the symmetric geodesics meet $E=\psi$. Albeit not clear in these coordinates, these geodesics have an interesting property exclusive of this 2+1 set-up. Upon mapping them to Kruskal $X,T$ coordinates, one finds that they are actually $\dot T=0$ curves, i.e. straight lines, on the Penrose diagram. This is perhaps surprising since one does not expect $\partial_T$ to be a Killing vector in the metric due to the singularity. This gets clarified by rewriting \eqref{BTZ-metric} in standard Kruskal coordinates \cite{eternal},
\begin{equation}
ds^2 
= \frac{- dT^2+dX^2}{\cos(X)^2}+\frac{\cos(T)^2}{\cos(X)^2}d\varphi^2 
\end{equation}
where one can see that $\partial_T$ is, as expected, not a Killing vector of the geometry. However, we also see that only for BTZ the $g_{XX},g_{TT}$ do not depend on $T$, so in the $J=0$ scenario all dependence in $T$ is lost, and $\dot T=0$ curves become geodesics, as long as $T\in[-\pi/2,\pi/2]$. These are shown in Fig. \ref{Fig:BTZHHH}(b) alongside a $J\neq0$ geodesic which naturally does bend in the diagram, i.e. has no longer $\dot T=0$.

\section{ Case Study III: AdS$_{4+1}$ BH }
\label{Sec:BH5}

\begin{figure}[t]\centering
\begin{subfigure}{0.49\textwidth}\centering
\includegraphics[width=.9\linewidth] {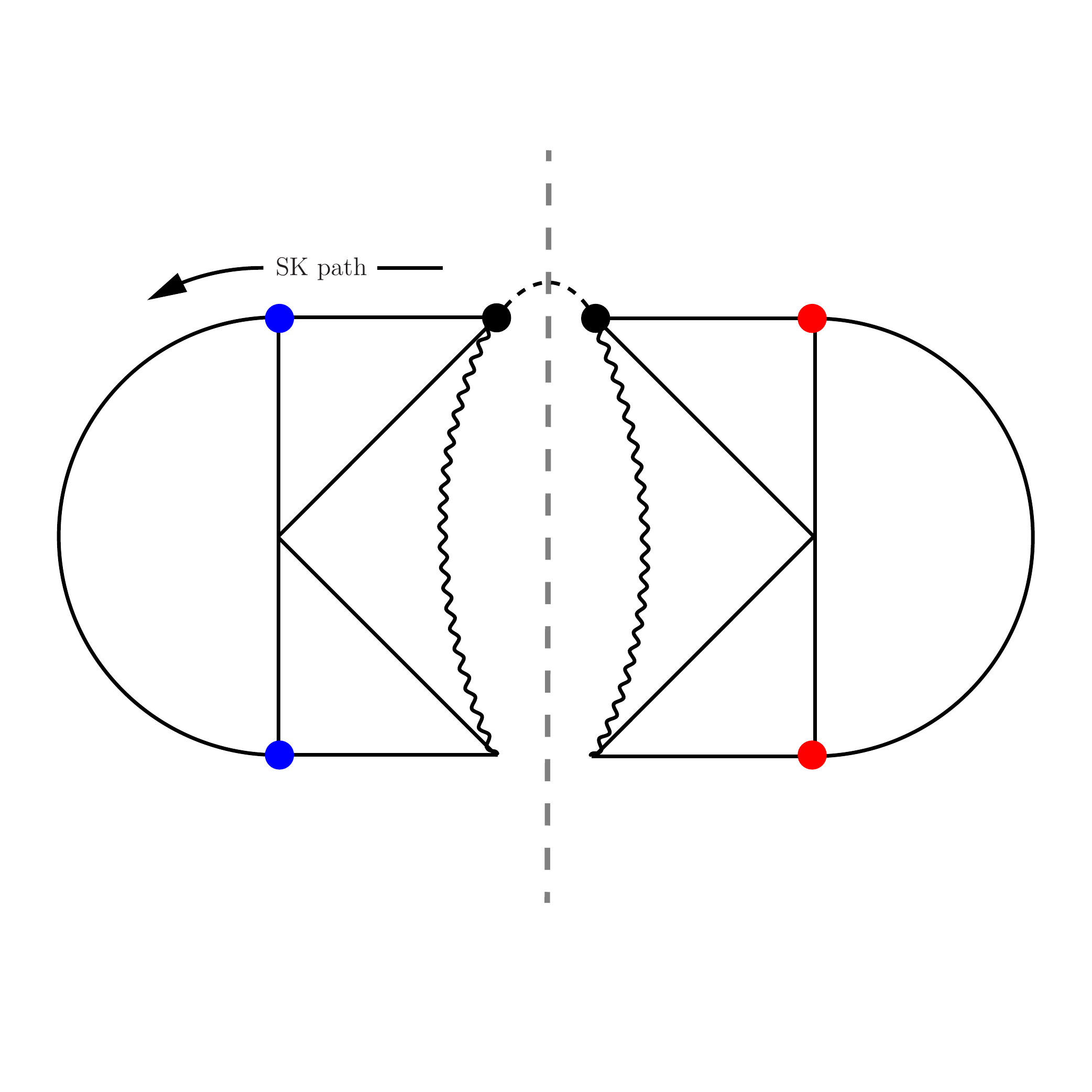}
\caption{}
\end{subfigure}
\begin{subfigure}{0.49\textwidth}\centering
\includegraphics[width=.9\linewidth] {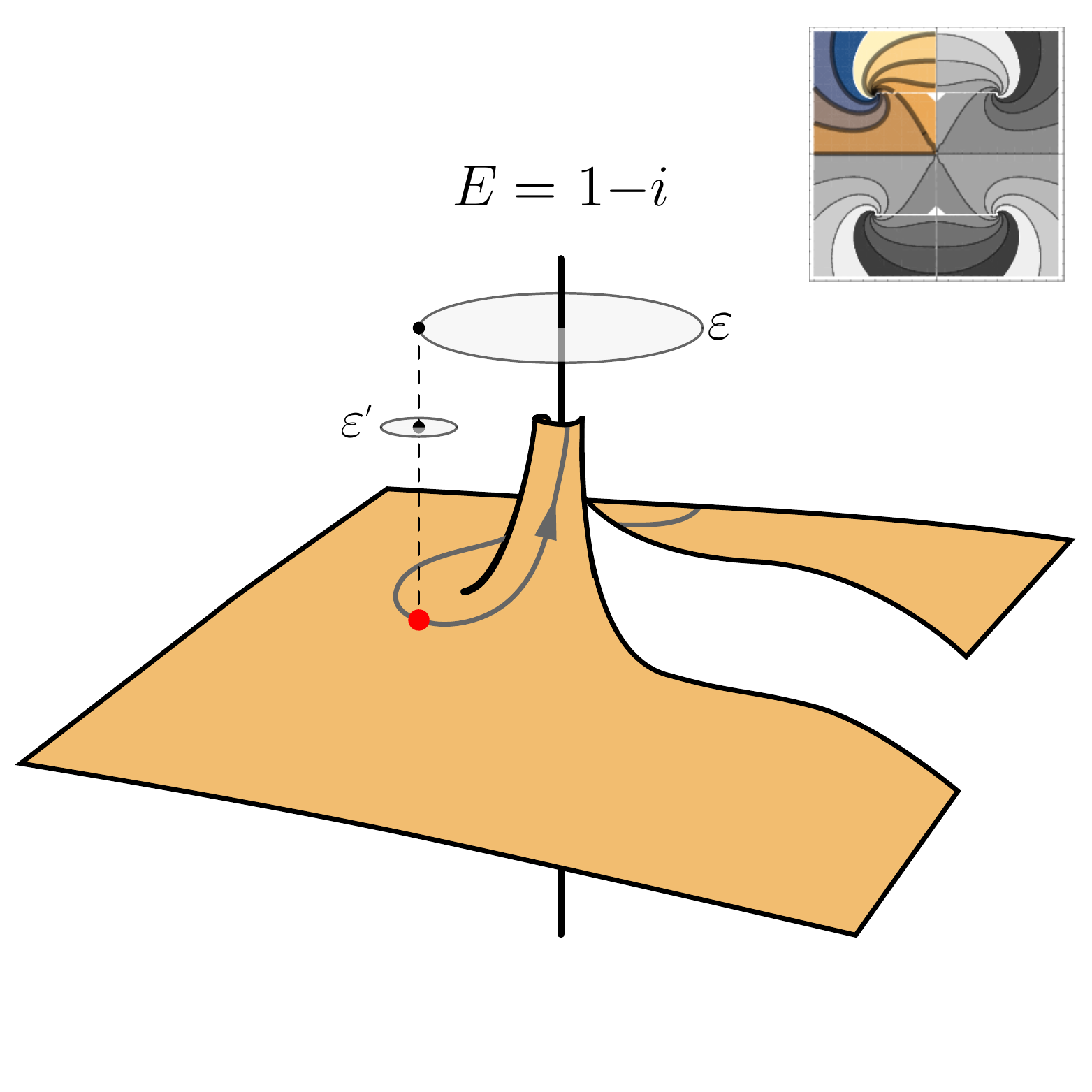}
\caption{}
\end{subfigure}
\caption{(a) We show the higher dimensional bulk dual to the SK path in Fig. \ref{Fig:BH1}(a). The bulk is quite similar to the BTZ one, but the singularities are necessarily bent inwards if the asymptotic boundary is to be kept a straight line. (b) The real part of \eqref{BH5-Deltat} is presented in a countour plot in the upper right, and the colored piece, near $E=1-i$, is represented in 3D in the main part of the Figure. The grey curve represents the path of the energy saddles as $t\gg t_c$. At the red point, the saddle is at a radius $\epsilon\sim e^{-4(1+i)t}$ away from $E=1-i$. A second, smaller $\epsilon'\ll\epsilon$ deformation is also represented, that will be useful in our computation of $F_2$. }
\label{Fig:BH5}
\end{figure}

In this last example, we tackle a more realistic scenario of a $4+1$ BH in the geodesic approximation and study $F_1$ and $F_2$ in the geodesic approximation.
As in \cite{Shenker03}, we will pick the infinitely massive two sided BH in AdS$_5$, i.e. 
\begin{equation}\label{BH5metric}
    ds^2=\left. f(r) \times\begin{cases}-dt^2\\+d\tau^2\end{cases}\hspace{-3mm}\right\} + \frac{dr^2}{f(r)} + r^2 d\Omega_3^2 \qquad f(r)=r^2-\frac{1}{r^2} \qquad \tau\in[-\pi/2,\pi/2]
\end{equation}
Notice that in the standard notation that we have used, this BH has $\beta=\pi$ so that $\tau\sim\tau + \pi$ is not a good angle around the origin of the Euclidean disk. 
The relevant SK path is the same as in \ref{Fig:BH1}(a). The dual geometry is still similar to \ref{Fig:BH1}(b) but the $d+1>2+1$ BH Penrose diagram is not a square \cite{Shenker03} and the singularity bends over inwards if the asymptotic boundaries are taken to be straight lines as in Fig. \ref{Fig:BH5}(a).

Interestingly, for this geometry, and generically at higher dimensional BHs, the saddles corresponding to spacelike geodesics between the asymptotic boundaries already sit at complex values of the energy. More concretely, a pair of complex conjugate saddles provide the correct (real) CFT correlator. A naive, but subleading saddle sitting at real energies also exists but it predicts a singularity at a certain time $t_c$ once it hits the singularity in the conformal diagram, which is unphysical from the CFT point of view \cite{Shenker03}.
This complex saddles structure, however, is quite subtle to unveil from pure Lorentzian computations.  
We begin this section by studying $F_1$, where complex energy values are naturally expected, and these complex saddles arise more naturally and unambiguously.
We then study $F_2$ which describe deformations of the geodesics connecting the asymptotic boundaries. Due to the more complicated geometry, full analytic control of the computations will not always be possible. We will see that the excited states under study affect the saddle energy in such a way that it effectively chooses only a single saddle of the pair of conjugated ones, effectively simplifying the problem.

\subsection{Study of $F_1$}

We begin our study of $F_1$ as in the BTZ geometry by computing a pure Euclidean geodesic with fixed initial time $\tau$, and then moving the final time to be pure imaginary $+i t$. This will force the Euclidean geodesic's energy to be complex and will provide in turn the geodesic's length and correlator.
The geodesic equations with no angular momentum in this geometry are 
\begin{equation}
    +f(r) = E^2 + \dot r^2 \qquad E=\dot \tau f(r)
\end{equation}
and the radial equation is explicitly solved by
\begin{equation}
    r(\sigma)=\sqrt{1+\frac{E^4}{4}} \cosh (2 \sigma )+\frac{E^2}{2}
\end{equation}
The boundary to boundary geodesic's length can be once again obtained by taking twice the distance from the asymptotic boundary up until $\sigma=0$, regardless of $E\in\mathbb{C}$. Taking an $R_c$ cutoff, we get
\begin{equation}
    L=\Delta\sigma-2\ln(R_c) 
    \sim\ln \left(\frac{4 R_c^2}{\sqrt{4+E^4}}\right)-2\ln(R_c)=\ln \left(\frac{4}{\sqrt{4+E^4}}\right) 
\end{equation}
The Euclidean time solution $\tau(\sigma)$ can also be found exactly, but need only the expression for $\tau(+\infty)=+it$ which is, $\omega_{0}\equiv 1+i$,
\begin{equation}\label{BH5-Deltat}
    t+i\tau =-\frac i4 \ln\left( 
       \frac{(E+\omega_0) (E+\omega_0^*)}{ (E-\omega_0) (E-\omega_0^*)}
    \frac{\left( (E-\omega_0)  (E+\omega_0^*)\right)^i}{\left(  (E+\omega_0) (E-\omega_0^*)\right)^i} \right)
\end{equation}
One can see that the rhs above has $\ln$ singularities at all four signs of $E=\pm 1\pm i$, i.e. at $\pm \omega_0$ and its conjugates. Our specific problem can be solved by looking for level surfaces of the real piece of \eqref{BH5-Deltat} and then fix the imaginary piece to match the precise $\tau\in[-\pi/2,0]$. A plot of the real piece of \eqref{BH5-Deltat} is presented in Fig. \ref{Fig:BH5}(b). The first thing to notice is that for a given fixed $t$ there are many possible solutions for the energy, especially below $t_c=-\pi/4$, $0<t<t_c$ \cite{Shenker03}. However, one can also see that for $t\gg t_c$ the candidate solutions reduce to two and lie near to $E=\pm 1 + i$. In this regime one can find a leading order relation between energy and time is $E= 1 \mp i +e^{-2(1 \pm i) (t+i \tau )} $. Up to this point, the discussion has come out pretty similar to that in \cite{Shenker03} in which two complex conjugate saddles reproduced the correct CFT correlator. However, in our case, we have an Euclidean initial time which breaks the equilibrium between the saddles and one can see that only one dominates, $E\sim 1-i$ in this case\footnote{Of course, there is nothing special in the $E\sim 1-i$ saddle and different singularities in \eqref{BH5-Deltat} will dominate depending on the signs of $t$ and $\tau$ we are trying to solve for.}. We get for the regulated length and $F_1$,
\begin{equation}\label{BH5-F1}
    L\sim (1+i)(t+i\tau)
    \qquad\Rightarrow\qquad
    F_1\sim 2\Re \left\{\int_{-\pi/2}^0 \phi(\tau) e^{- \Delta\; L}\right\} \sim \int_{-\pi/2}^0 \phi(\tau) \left( e^{-(1+i)(t+i\tau)\Delta} +  e^{-(1-i)(t-i\tau)\Delta} + \dots\right)
\end{equation}
which correctly reproduces the first quasinormal modes $\omega=\Delta (1\pm i)$. 
As a consistency check, notice that our solution was found using $t>0$ and we obtained a correlator decaying in time, as we should have. Recall that $\tau$'s domain is finite so that no divergences comes from it despite its contribution to $F_1$. 
As in \cite{Shenker03} one can look further in the expansion and discover all higher QN modes and our pole-skipping sources can once again be used. 

The convolution in \eqref{BH5-F1} between the correlator and different sources profiles $\phi(\tau)$ can be carried in a straightforward manner. Its analysis leads to a striking change with respect to the BTZ scenario in that the first QN mode here is already complex $\omega=\Delta (1\pm i)$ whilst in the BTZ scenario we only got exponentially decaying behavior. This suggests that BH relaxation actually goes through a series of absorption and re-emission of the excited state before fully absorbing its energy. We will more directly see this in our computation of $F_2$ below. 

An interesting property of this observable rather than the \cite{Shenker03} scenario is that there is only one unambiguous saddle that dominates the correlator, which provides a less subtle problem to solve. Since the set-up already requires a complex saddle, one can see that the difficulty of the problem has not increased, and that the problem is easier to solve.

\subsection{Study of $F_2$}

We now look at a sample computation of $F_2$ in this geometry. To this end, we must first find the geodesic crossing from one asymptotic boundary to the other $|\Delta t|=2t$ and then explore its deformations due to a third one coming from the Euclidean region. From our analysis on $F_1$ above, one can already see that the energy of the spacelike geodesics between the asymptotic boundaries will be complex. 
From the correlator implicit in \eqref{BH5-F1} one can safely extend the second one to the other boundary $\tau\to-\pi/2+it'$, which retains the complex saddles and readily discards the $E\in\mathbb{R}$ candidates.

The correlator between both sides of the BH can be more constructively found by solving the Lorentzian geodesic equations
\begin{equation}
    +f(r) = -E^2 + \dot r^2 \qquad E=\dot t f(r)
\end{equation}
whose solutions for $t(-\infty)=t$ are,
\begin{equation}
    r(\sigma)=\sqrt{1+\frac{E^4}{4}} \cosh (2 \sigma )-\frac{E^2}{2}
\end{equation}
\begin{equation}\label{5-realt}
  t(\sigma)-t= \frac 14 \ln\left( \frac{\left(\sqrt{4+E^4}+e^{2 \sigma } (i
   E-\omega_0) (i E+\omega_0^*)\right)
   \left(\sqrt{4+E^4}+e^{2 \sigma } (i
   E-\omega_0) (i
   E-\omega_0^*)\right)^i}{\left(\sqrt{4+E^4}+e^{2 \sigma } (i E+\omega_0) (i
   E-\omega_0^*)\right)
   \left(\sqrt{4+E^4}+e^{2 \sigma } (i
   E+\omega_0) (i E+\omega_0^*)\right)^i}\right)\;.
\end{equation}
By demanding $t(\infty)=-t$ above one gets the symmetric geodesics found in \cite{Shenker03}. For $t \gg t_c$ there are two complex conjugate saddles near $E_\pm\sim 1\pm i+e^{-4(1\pm i)t}$ whose length is given by. 
\begin{equation}
   L_{\pm}=\ln\left(\frac{4}{\sqrt{4+E_\pm^4}}\right) \sim 2(1\pm i)t
    \qquad \Rightarrow \qquad
    \langle {\cal O}_L{\cal O}_R \rangle \sim
    \sum_{\pm}e^{-\Delta \; L_{\pm}} 
    \sim e^{-\Delta(1+i)2t}+e^{-\Delta(1-i)2t} 
\end{equation}
where recall that $\Delta t=2t$ for these geodesics so that they reproduce the expected correlator with the lowest allowed QN modes $\omega=(1+i)\Delta$ and its conjugate.
We are ultimately interested here in deformations over these geodesics that meet at $\Re(t)=0$ but with deformed energy $E$ such that they would form a cusped curve, recall Fig. \ref{Fig:BH1}(b). As in the BTZ example we parametrize these deformations by keeping $t$ fixed and deforming the geodesics energy. The length of these deformed geodesics is
\begin{equation}\label{5length}
   L_\pm=\ln\left(\frac{4}{\sqrt{4+E_\pm^4}}\right) \qquad\to\qquad L_L=\ln\left(\frac{4}{\sqrt{4+E^4}}\right)
   + 2\sigma_0
\end{equation}
where $E$ are no longer $E_\pm$ and $\sigma_0$ takes into account that the geodesic no longer reaches $\Re(t)=0$ at $\sigma=0$. This $\sigma_0$ is to be obtained from \eqref{5-realt}. In the BTZ scenario, the $\sigma_0$ deformation could be obtained analytically, see \eqref{psi}, but this is not the case in more general set-ups. Much like in the $F_1$ scenario, we will see that the deformation coming from the excited state unbalances the complex conjugate saddles so that there is only one that dominates. In this case it will be $E\sim 1-i$ and we will continue working only around this saddle in the following. We denote $\epsilon=e^{-4(1 - i)t}$ and consider a deformation $\epsilon'$ of the energy such that $E\sim 1- i+e^{-4(1- i)t}+\epsilon'$, and $\epsilon' \ll \epsilon$, see Fig. \ref{Fig:BH5}(b). By construction, one expects that an expansion of $L_L$ in $0<\epsilon'<\epsilon$ should have no linear term in $\epsilon'$ since otherwise our starting curve would not have been a geodesic to begin with. This is exactly the case as we expand \eqref{5-realt} and \eqref{5length} to leading order we find
\begin{equation}
    \ln\left(\frac{4}{\sqrt{4+E^4}}\right)\sim L_--\frac{\epsilon'}{4 \epsilon }+\frac{\epsilon'^2}{8 \epsilon ^2}+\dots \qquad\qquad 2\sigma_0=\frac{\epsilon '}{4 \epsilon} -\frac{\epsilon '^2}{8
   \epsilon ^2} + \omega_0  \frac{ \epsilon '^2}{8\epsilon }+\dots
\end{equation}
such that 
\begin{equation}
    L_L\sim L_- +\omega_0\frac{ \epsilon '^2}{8\epsilon }
\end{equation}

\begin{figure}[t]\centering
\includegraphics[width=.55\linewidth] {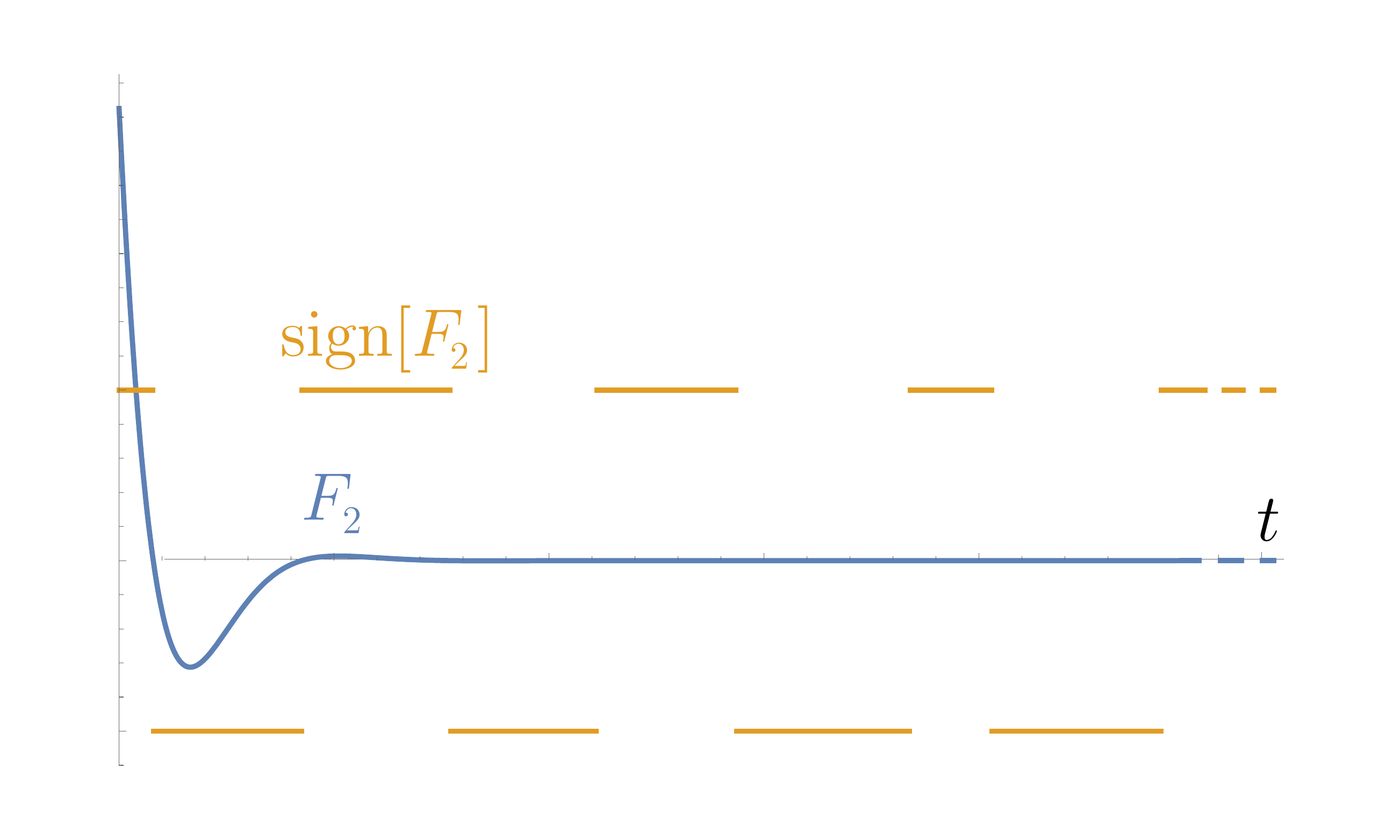}
\caption{Plot of a typical $F_2$ observable and its sign in AdS$_5$ BH. In contrast with the $F_2$ of the BTZ scenario in Fig. \ref{Fig:BTZHHH}(a), the excited state relaxes via a series of absorption and re-emission processes. This is directly related with the fact that the lowest QN state in any higher dimensional BH is already complex, as opposite to the BTZ case in which is pure imaginary.}
\label{Fig:BH5HHH}
\end{figure}

The next step in finding the deformed geodesic is to measure the leg coming from the excited state $L_E$. This can be exactly done in terms of the meeting point $r_e$, which in turn must be approximated in terms of $\epsilon'$ via the $\sigma_0$ found above. As should be standard by now, the geodesic travels in $\dot \tau=\dot t=0$ geodesics and the Euclidean piece only contributes as an $R_c$ regulator dependent constant which ultimately can be reabsorbed in the correlators normalization. The relevant pieces of $L_E$ are
\begin{equation}
    r_e^2=
    i-\left(\omega_0^*-i\sqrt{\frac{\omega_0}{2\epsilon
   }}\right)\epsilon ' +\dots \qquad\Rightarrow\qquad 
    L_E =  \frac{i}{2}  \arccos\left(r_e^2\right) 
    \sim i\frac{\pi}{4} + \frac{1}{2} \sinh
   ^{-1}(1)-i\left(\omega_0^*-i\sqrt{\frac{\omega_0}{2\epsilon
   }}\right) \frac{\epsilon '}{\sqrt{8}}+\dots
\end{equation}
and thus the minimization problem for the deformed geodesic becomes
\begin{equation}
    \Omega[\epsilon']=\Delta L_L + \Delta_E L_E 
    \sim
    \Delta 
   \left(L_-+\omega_0\frac{ \epsilon
   '^2}{8\epsilon }\right)+
    \Delta _E \left(i\frac{\pi}{4}+\frac{1}{2} \sinh
   ^{-1}(1)-i\left(\omega_0^*-i\sqrt{\frac{\omega_0}{2\epsilon
   }}\right) \frac{\epsilon '}{\sqrt{8}}\right)\;.
\end{equation}
The equilibrium between the complex conjugated geodesics $L_\pm$ can be explicitly seen to be broken by $+\Delta_E \;L_E\in\mathbb{C}$ and specifically by its $+$ sign. This sign is dictated by the SK path and thus it affects differently the real pieces on each geodesic. As we said above, at this point and for our configuration $t>0$ and $\tau=-\pi/4$ one can see that the $E\sim 1-i$ saddle dominates. The particular $\epsilon'$ that minimizes $\Omega$ is now straightforward to obtain,
\begin{equation}
    \Omega'[\epsilon'_0]=0 \qquad\Rightarrow\qquad \epsilon'_0 = \frac{\Delta _E }{2
   \Delta }\left(- i \omega
   _0^{3/2} \sqrt{\epsilon}+ \sqrt{8} \,\epsilon\right)\;.
\end{equation}
This also allows to compute $F_2$ in this approximation,
\begin{equation}
    F_2\sim 2\Re\{\langle {\cal O}_{\Delta}(t){\cal O}_{\Delta}(t){\cal O}_{\Delta_E}(-\pi/2)\rangle\} \sim 2\Re\{ e^{- \Omega[\epsilon'_0]}\}
\end{equation}
which concludes our computation. Beyond the precise expression for $\epsilon'_0$ of $F_2$, the most relevant result of our analysis can be seen in Fig. \ref{Fig:BH5HHH}. As typically the lowest QN mode in higher dimensions BHs are already complex, it turns our that the relaxation of the holographic excited states is done via a high number of absorption and re-emission processes. This is consistent with our results obtained above for $F_1$ and contrast with the BTZ results which only has a single absorption and re-emission process. 

A final technical comment should be made regarding our $\epsilon'\ll\epsilon$ approximation. Notice that since $|\epsilon|\sim e^{-4t}$, for $t\gg1$ one should take $\epsilon'$ exponentially small such that $\epsilon' e^{4t}\ll1$. In a naive $F_2$ plot like in Fig. \ref{Fig:BH5HHH} for a fixed $\epsilon'$ this will manifest as un-physical divergences for sufficient large times. The correct way to understand the calculations, at least to this leading approximation computation, is thus considering first an order of $t$ one is interested in and then fix $\epsilon'$ such that $\epsilon' e^{4t}\ll1$.

One can further, at least numerically, solve $F_2$ for insertions at different points of the Euclidean piece and find its convolution with a specific source $\phi(\tau)$ of interest. We will not pursue this computation in this work. This concludes our set of examples.

\section{Discussion and Conclusions}
\label{Sec:Conc}

In the present work, we have succeeded in two goals. We have enlarged the holographic map to include mode-skipping sources \eqref{singlemode} and we have presented a general framework in which to study real-time relaxation processes in thermal systems. We studied a particular class of excited states in terms of the family of observables $F_n$  defined in \eqref{Fn}.

In Sec. \ref{Sec2}, we extended our understanding on the excited states \eqref{exc-state} by analyzing the consequences on the initial wavefunction following from the particular (asymptotic) boundary conditions choosen in \eqref{singlemode}. We found that it is possible to fine tune boundary conditions in the Euclidean path integral to avoid the presence of any number of QN modes in the initial wavefunction. We call these ``mode-skipping sources''. The result in turn made possible to construct an initial state consisting on a single QN mode, and this seemed to contradict the general expectation which says that single modes should not have a simple geometric interpretation. As we elaborated, single QN mode wavefunctions can be described in the bulk  at the expense of an infinite superposition of geometric states, in a similar fashion as coherent states expand energy eigenfunctions.  This result is complimentary  to the traditional picture of the BH geometry as a TFD state \cite{eternal, VanRaamessay}, in which the former arises as  an emergent geometry for an infinite superposition of eigenstates each of which does not have  a smooth geometric description.

Secondly, the relaxation process of holographic excited states was  studied using the family of observables $F_n$ defined in \eqref{Fn}.  Using Skenderis-van Rees' holographic prescription, in Secs. \ref{Sec:AdS}, \ref{BTZ} and \ref{Sec:BH5}, we were able to extract the leading order contributions to $F_1$ and $F_2$ in the large $N$ limit.  This led to a discussion on the specific Schwinger-Keldysh path adequate to study each scenario and built bulk duals to each path. Specifically, we computed $F_1$ and $F_2$ to leading order in $1/N$ for scalar fields with cubic self-interactions in the bulk. We studied the profiles of initial conditions and its Lorentzian time evolution obtained from a set of physically motivated Euclidean sources. We also checked that our mode skipping sources can avoid particular QN modes in the initial wavefunction as long as their eigenfrequencies are known.

In the large conformal dimension limit, the geodesic approximation to the bulk correlators demanded a discussion on the nature of the analytic extension performed on the geodesics. We proposed that the system's physical response arises from a holomorphic complexification of the geodesics' parameters. This somehow obscures the visualization of the geodesics as paths in a complexified manifold and becomes especially manifest in that their proper lengths become complex numbers, i.e. not real nor pure imaginary.
In line with \cite{SvRL}, our framework makes manifest that naive analytic extensions of Euclidean correlators becomes increasingly cumbersome in practice, and that a direct real-time recipe, as the one we present, becomes handy and more tractable.

As a byproduct of our complex geodesics analysis in the context of excited states, we found that the computation of real time correlators simplifies. As discussed in \cite{Shenker03}, correlators between asymptotic boundaries in higher dimensional BHs are dominated by a pair of complex conjugated saddles, both contributing with the same weight to the path integral. Our setup typically breaks this symmetry, turning a problem with possibly many saddles into one that has a single dominating contribution, which is generally simpler.

Finally, comparing BTZ results with those for AdS$_5$ BH (cf. Figs. \ref{Fig:BTZHHH}(a) and \ref{Fig:BH5HHH}) shows that the relaxation process in BTZ develops a single absorption and re-emission process while the latter traverses a long series of absorption/re-emission processes before reaching equilibrium. This is a direct consequence of BTZ having a pure imaginary first QN mode whilst higher dimensional BHs have  generically complex QN modes.

All computations in this work were done on a fixed background geometry. As such, we could envisage extending them taking into account backreaction, we plan to pursue this line in future work. Another interesting avenue to consider is to include higher point interactions in the bulk. Keeping tree level computations in the bulk, one should note that additional sources both at Euclidean or Lorentzian regions must be taken into account. A simple example to consider is the $O(\phi^2)$ correction to $F_1$  with two insertions in the Euclidean region and a single insertion real time. As discussed in previous work \cite{us2}, these modify the coherent nature of the state. A computation of this kind should help to characterize the deformation from coherence. To conclude we mention that complexified geodesics have been used both in older \cite{HawkingOld} and  modern approaches \cite{Sundrum}  as tools to obtain information from regions beyond the singularity. Within the present formalism complex geodesics make sense as a direct observable of the system. Then, it may be interesting to revisit these ``beyond the singularity'' geodesics and try to interpret them in terms of physical phenomena in the CFT. 

\section*{Acknowledgments}

The authors want to thank Mark Van Raamsdonk and Jorge Russo for fruitful discussions. GS is supported by CONICET  and UNLP. PJM is supported by CONICET, CNEA and Universidad Nacional de Cuyo, Argentina. PJM is specially indebted to the Siembra-HoLAGrav collaboration.

\pagebreak
\appendix

\section{Finite energy of Holographic excited states}
\label{App:A}

In this Appendix, we review some standard considerations on the states \eqref{exc-state} regarding their normalization and finite energy conditions. More concretely, we want to specify which restrictions should be imposed on the asymptotic boundary conditions $\phi$ in \eqref{exc-state} such that the state can be actually thought of as a physical state in the Hilbert space. We apply this analysis directly to our AdS/CFT framework and specifically to the $2+1$ pure AdS set-up. Its generalization to higher dimensions and other geometries is straightforward.

Consider a fixed pure Lorentzian AdS$_{2+1}$ space with metric \eqref{AdS-Metrics} and consider a free massive scalar field $\Phi$. By standard methods one can see that a global energy can be defined in terms of the Dilatation symmetry $D$, written as
\begin{equation}
    D=\sum_{nl}\omega_{nl} a_{nl}^\dagger a_{nl}=\sum_{nl}(\Delta+2n+|l|) a_{nl}^\dagger a_{nl}
\end{equation}
This can be thought as both the AdS and CFT Hamiltonian in the strict large $N$ limit. 

Recall that for a QFT defined in a non-compact spatial manifold (take AdS-Poincare for example) the $n$-particle states do not strictly belong to the physical Hilbert space since they have infinite norm due to the infinite volume of space. For a theory defined on a compact space, however, this problem is alleviated. The finite nature of the volume leads to a discrete state basis which in turn lead to Kronecker rather than a Dirac deltas on the orthogonality relations, i.e.
\begin{equation}
    [ a_{nl},a_{n'l'}^\dagger]=\delta_{n n'}\delta_{l l'}
\end{equation}
so that the $n$-particle states can be nicely normalized and of course, provide finite energy $\omega_{nl}$,
\begin{equation}
    \langle n' l' | n l\rangle = \langle 0 |  a_{n'l'}a_{nl}^\dagger| 0\rangle=\langle 0 | [a_{n'l'},a_{nl}^\dagger]| 0\rangle=\delta_{n n'}\delta_{l l'} \qquad\qquad \langle n' l' |D| n l\rangle =\omega_{nl}\delta_{n n'}\delta_{l l'}
\end{equation}

Our states defined in \eqref{exc-state} are associated with coherent states in the large $N$ limit. By BDHM \cite{BDHM} prescription, we can even directly compute its eigenvalues, see \cite{us1}
\begin{equation}
    \lambda_{nl}=-\sqrt{4\pi(\Delta-1)\alpha(\omega_{nl},l)} \; \; \bar \phi_{}
    \qquad\qquad 
    \bar \phi_{}\equiv \frac{1}{2\pi}\int_{-\infty}^0 d\tau d\varphi e^{\omega_{nl}\tau+i l \varphi}\phi(\tau,\varphi)
\end{equation}
$$\alpha(\omega_{nl},l)=2(\Delta-1)\frac{\Gamma(\Delta+n+|l|)\Gamma(\Delta+n)}{n!\Gamma(\Delta)^2\Gamma(n+|l|+1)}$$
Normalization of our states is guaranteed by the fact that the inner product between them reduces to the original GKPW prescription with sources, i.e. essentially a Gaussian in $\phi(\tau,\varphi)$ with a positive definite kernel. A less trivial condition thus arises from the finite energy condition, namely $\langle \phi |D| \phi\rangle<+\infty$, since now for a generic source, all modes are turned on, giving
\begin{equation}
    \langle \phi |D| \phi\rangle=\sum_{nl}(\Delta+2n+|l|)  |\lambda_{nl}|^2
\end{equation}
so finite energy coherent states amount essentially to $\bar \phi_{}$ containing only a finite number of modes. This is easily met by Gaussians and general $L_2$ functions. We are, in particular, interested in Gaussian $\phi(\tau,\varphi)$ which can be seen to approximate \eqref{singlemodereal} as much as one like, since relaxing the delta to a Gaussian, the derivative can be easily applied and the source reduces to a sum of Gaussians. Our pole skipping procedure is thus seen as a limiting process on that precise finite energy Gaussian source.

\section{On complexified geodesics}
\label{CGeod}

In this short appendix, we would like to clarify some aspects of what we call ``complex geodesics'' in this work in order to avoid possible confusion.

The bottom line of the discussion is that one should not enforce a geodesic as a curve in complex spacetime in the sense that its proper length can be always be given by a real number. The complex geodesic problem requires instead a well defined physical set-up to start with, and then one performs a holomorphic analytic extension on a parameter (or set of parameters) that one should let to become complex. This somewhat blurs the interpretation given in our drawing of actual geodesics as curves meeting at a certain point in the bulk, see Fig. \ref{Fig:HHL} for example. In our examples we show that the summed proper length of these complex geodesics becomes a complex quantity in itself rather than a length $\sigma$ (or sum of lengths $\sigma_i$). Being the complex numbers a non-ordered set, one should necessarily struggle if would insist in thinking of these quantities as defining the length of a curve. The ``geodesic'' denomination is still justified however, in the sense that we first start from a well posed quantity defined in terms of expressions that do meet the geodesic equations. Upon starting from a well defined problem we proceed to perform a holomorphic complexification of the parameters involved. It is in this sense that we call these complex geodesics.

To conclude, we present a concrete example of the incorrect way to complexify the Euclidean leg in eq. \eqref{AdSEleg} in the simple AdS$_3$ set-up. Specifically if one would like to enforce the notion of geodesics as curves in spacetime, one could picture $L_E$ as a curve that travels through the complexified geometry. The problem is that by enforcing the geodesic to have a well defined "length" (i.e. its length being a real number) leads to an analytic extension that is incompatible with recovering pure imaginary/real distances when the signature is changed. Specifically, take the metric \eqref{AdS-Geom} at $r=0$ as before, starting from the Euclidean piece for the excited state geodesic
\begin{equation}
    ds^2 = d\tau^2 \qquad \rightarrow \qquad ds^2= (d\tau+idt) (d\tau-idt) \in \mathbb{R}\qquad\qquad\text{(real but unphysical)}
\end{equation}
This is tempting, but leads to unphysical results. This extension provides a clear picture of a geodesic travelling on a $t,\tau$ plane and provides also a real condition of extremization of the geodesic. However, this extension is clearly not the one we are after since going back to a scenario in which $d\tau=0$ one does not recover a time-like geodesic. To be precise, the adequate extension in this scenario is a holomorphic extension that one can envision by defining $z=t-i\tau$
\begin{equation}
    ds^2=d\tau^2 \qquad\rightarrow\qquad ds^2=-dz^2\in\mathbb{C}\qquad\qquad\text{(complex but physical)}
\end{equation}
where notice that no $z^*$ is involved and that we arrive at a simple solution for the proper distance $\Delta\sigma= i \Delta z=i \Delta t + \Delta \tau$ as we have used to reach correct physical results in Sec. \ref{Sec:AdS}. Albeit more complicated for general metrics, we will always follow this principle and one should consider this our guiding principle and definition of ``complexified geodesic'' throughout this work.


\end{document}